\documentclass{article}

\usepackage{arxiv}

\usepackage[utf8]{inputenc} % allow utf-8 input
\usepackage[T1]{fontenc}    % use 8-bit T1 fonts
\usepackage{mathptmx}
\usepackage{hyperref}       % hyperlinks
\usepackage{url}            % simple URL typesetting
\usepackage{booktabs}       % professional-quality tables
\usepackage{amsfonts, amssymb}       % blackboard math symbols
\usepackage{nicefrac}       % compact symbols for 1/2, etc.
\usepackage{microtype}      % microtypography
\usepackage[dvipsnames,table,xcdraw]{xcolor}

\usepackage[left]{lineno}

\usepackage{lipsum}
\usepackage{graphicx, subcaption}
\usepackage{amsmath}
\usepackage[font=small]{caption}
\usepackage{multirow}
\usepackage{makecell, makebox}
\usepackage{pifont}
\usepackage{colortbl}
\usepackage{authblk}
\usepackage{enumitem}

\title{A Machine Learning Benchmarking Framework for Lipid Nanoparticle Transfection Efficiency Prediction}

\author[1,$\dagger$,*]{Asal Mehradfar}
\author[1,$\dagger$,*]{Mohammad Shahab Sepehri}
\author[2]{Jose Miguel Hernandez-Lobato}
\author[3]{Glen S. Kwon}
\author[1]{Mahdi Soltanolkotabi}
\author[1]{Salman Avestimehr}
\author[3,4,*]{Morteza Rasoulianboroujeni}

\affil[1]{Department of Electrical and Computer Engineering, University of Southern California, Los Angeles, CA}
\affil[2]{Department of Engineering, University of Cambridge, Cambridge, UK}
\affil[3]{Pharmaceutical Sciences Division, School of Pharmacy, University of Wisconsin-Madison, Madison, WI}
\affil[4]{\vspace{0.5em}Department of Pharmaceutical Sciences, Gatton College of Pharmacy, East Tennessee State University, Johnson City, TN}
\affil[$\dagger$]{These authors contributed equally to this work.}
\affil[*]{Corresponding authors: \texttt{mehradfa@usc.edu}, \texttt{sepehri@usc.edu}, \texttt{rasoulianbor@etsu.edu}}

\begin{document}
% \linenumbers

\renewcommand{\thefootnote}{}
\footnotetext{Published in Communications AI \& Computing (Nature Portfolio), 2026.}

\maketitle

\def\thefootnote{\arabic{footnote}}
\newcommand{\rc}{\textcolor[rgb]{1,0,0}}
\newcommand{\bc}{\textcolor[rgb]{0,0,1}}
\newcommand{\gc}{\textcolor[rgb]{0,1,0}}
\newcommand{\oc}{\textcolor{orange}}
\newcommand{\method}{LANTERN}

\begin{abstract}

The discovery of new ionizable lipids for efficient lipid nanoparticle (LNP)–mediated RNA delivery remains a major bottleneck in RNA therapeutics development. Recent advances demonstrate the potential of machine learning (ML) models to predict transfection efficiency directly from lipid structure, enabling high-throughput virtual screening and accelerating lead identification. However, as new models for LNP transfection prediction continue to emerge, the lack of rigorous and standardized benchmarking poses a significant risk and may undermine confidence in their reliability for discovery. Here, we present a robust ML benchmarking framework for evaluating transfection prediction models based on ionizable lipid structures. This framework systematically benchmarks diverse molecular representations—including Morgan fingerprints, Expert RDKit descriptors, and Grover graph-based embeddings—paired with a broad range of ML architectures spanning traditional models, feedforward neural networks, and state-of-the-art graph-based methods. In addition, the presented framework supports assessment of model generalization through Murcko scaffold splitting and evaluates prediction reliability beyond standard regression metrics by incorporating analyses of relative error distributions and ranking accuracy. Using a curated dataset of 1,100 unique ionizable lipid structures derived from the HeLa transfection dataset originally reported by Xu et al. \cite{xu2024agile}, we show that within this framework, models leveraging explicit molecular substructure encoding—particularly multilayer perceptrons (MLPs) trained on Morgan fingerprints combined with Expert descriptors—consistently achieve the highest predictive accuracy and should serve as essential baselines for the development of new, more sophisticated models. In contrast, some current graph-based models, including AGILE, Chemprop, and KPGT, tend to show comparatively lower accuracy. The presented framework provides a standardized, transparent, and comprehensive benchmarking resource that enables meaningful comparison of emerging architectures and establishes strong baselines for future development of predictive models in lipid-based RNA delivery. Our code and dataset are publicly available for reproducibility at \url{https://github.com/AsalMehradfar/LANTERN}.

\end{abstract}

%%%%%%%%%%%%%%%%%%%%%%%%%%%%%%%%%%%%%%%%%%%%%%%%%%%%%%%%%%%%

\section{Introduction}

RNA therapy holds significant promise for treating diseases by introducing exogenous nucleic acids, such as mRNA, siRNA, and ASOs, to control gene expression in target cells. The FDA's approval of ONPATTRO™ in 2018 and Givosiran in 2019, followed by the success of lipid nanoparticle (LNP)-delivered mRNA vaccines against COVID-19, has firmly established the commercial and clinical viability of RNA therapeutics \cite{yan2022non}.

RNA therapies rely heavily on delivery systems to overcome biological barriers and reach specific tissues and cells to elicit a therapeutic response \cite{zhu2022rna}. LNPs represent a novel class of non-viral delivery vectors whose efficiency and versatility have been broadly validated by the clinical success of ONPATTRO™ and SARS-CoV-2 mRNA vaccines, including Moderna’s Spikevax and Pfizer/BioNTech’s Comirnaty. LNPs are now considered superior to viral vectors, which pose challenges in clinical applications due to immunogenicity, production costs, and biosafety concerns \cite{ibba2021advances}.

Among the four lipid components comprising LNPs, ionizable lipids play a central role in RNA encapsulation, protection against nuclease-mediated degradation, and facilitating intracellular delivery. Within the acidic environment of endosomes, ionizable lipids undergo protonation, leading to endosomal membrane disruption and promoting the cytosolic release of mRNA for subsequent protein synthesis \cite{kim2021engineered, wittrup2015visualizing, xu2021escaping}. While each lipid in the LNP formulation contributes to overall structural integrity and biological function \cite{albertsen2022role, cheng2020selective, ni2022piperazine, gan2020nanoparticles, qiu2022lung, zhu2022multi}, the molecular structure and physicochemical properties of ionizable lipids are particularly influential in determining transfection efficiency. 

The precise structure–function relationships governing ionizable lipid performance remain poorly understood, necessitating the empirical screening of large lipid libraries to identify high-performing candidates \cite{li2023combinatorial, liu2021membrane, liu2021zwitterionic, radmand2023transcriptional, radmand2024cationic}. The enormous chemical space, encompassing billions of potential ionizable lipid structures, renders exhaustive experimental synthesis and evaluation infeasible. While combinatorial chemistry, including multi-component reaction strategies, enables high-throughput synthesis of diverse lipid libraries \cite{li2023combinatorial,miao2019delivery,li2024accelerating}, these approaches remain time-consuming and resource-intensive \cite{han2021ionizable}. Moreover, the success rate in identifying highly effective formulations is low. For example, Li et al. \cite{li2023combinatorial} synthesized 720 ionizable lipids using a three-component reaction system combining 72 head groups and 10 lipid tails via a nitro ricinoleic acrylate (NRA) linker. Despite this diversity, more than 80\% of the formulations showed minimal or no transfection when delivering mLuc mRNA to A549 cells, highlighting the challenges in discovering functional ionizable lipid structures.

Given the challenges associated with experimentally designing and screening new ionizable lipids, virtual high-throughput screening (vHTS) using machine learning (ML) models trained on limited datasets has emerged as a promising alternative, enabling more efficient exploration of the vast chemical space relevant to lipid design. Ding et al. \cite{ding2023machine} approached LNP transfection prediction as a classification task using supervised ML models and achieved high accuracy on a dataset of 572 LNPs with luciferase expression in IGROV1 cells \cite{liu2021membrane}, demonstrating the potential of ML to accelerate lipid discovery. Building on this work, Moayedpour et al. \cite{moayedpour2024representations} evaluated diverse molecular representations—including RDKit descriptors, circular fingerprints, and embeddings from pre-trained graph-based and language-model architectures \cite{rong2020self, duvenaud2015convolutional, irwin2022chemformer}—and showed that large language model (LLM)–derived embeddings paired with gradient boosting achieved the strongest performance across binary and multi-class transfection prediction tasks. While such classifiers provide valuable structural insights and enable coarse-grained prediction of transfection efficiency, they lack the resolution required to reliably identify the highest-performing candidates for LNP discovery.

More recently, deep learning regressors have been developed for quantitative LNP transfection prediction. Xu et al. \cite{xu2024agile} introduced the AI-Guided Ionizable Lipid Engineering (AGILE) platform, a graph-based regression model designed to predict the transfection efficiency of ionizable lipids across large-scale chemical libraries. AGILE combines deep learning with combinatorial chemistry, employing a graph neural network (GNN) encoder based on the MolCLR architecture \cite{wang2022molecular}, which was initially pre-trained using contrastive learning on a dataset of 60,000 virtual lipid structures \cite{chen2020simple}. To fine-tune the model for quantitative transfection prediction, the authors synthesized a library of 1,200 ionizable lipids via a Ugi-based three-component reaction (3-CR) strategy \cite{miao2019delivery} and experimentally evaluated their mRNA transfection efficiencies in HeLa and RAW264.7 cell lines. The resulting labeled dataset was used to adapt the pre-trained GNN to the specific task of predicting LNP-mediated transfection efficiency. Building on this approach, Cui et al. \cite{cui2025lumi} developed the LUMI-model, which follows a three-stage training pipeline: (i) self-supervised pretraining on 13 million molecules and their 3D conformations using masked atom prediction and contrastive learning to capture general chemical and spatial features; (ii) continual pretraining on lipid-specific molecules to adapt the learned representations to the structural space of ionizable lipids; and (iii) supervised fine-tuning, where model predictions of mRNA transfection potential guide lipid synthesis and are iteratively refined through experimental feedback. 

Despite substantial progress in designing deep learning regressors for LNP transfection prediction, exemplified by models such as AGILE and LUMI, the field still lacks a robust benchmarking framework for rigorous and standardized evaluation of model performance. As increasingly complex models are introduced with the goal of improving both transfection efficiency prediction and generalizability, the absence of standardized evaluation criteria raises concerns about the validity of reported improvements. Without proper benchmarking, these models, regardless of their architectural sophistication, may fail to properly lead discovery efforts. Establishing such frameworks is essential to ensure reproducibility, facilitate model comparison, and enable the field to move toward more reliable and effective AI-driven therapeutic design.

In the present study, we introduce a computationally efficient and robust benchmarking platform designed to support the development of ML models for predicting LNP transfection efficiency based on the molecular structure of ionizable lipids. This framework provides a meticulously curated version of the ionizable lipid dataset originally published by Xu et al. \cite{xu2024agile}, and supports the training and evaluation of a wide range of model architectures across multiple chemically informative molecular representations that capture key structure–activity relationships. The framework supports both random and scaffold-based data splitting, enabling rigorous evaluation of model generalization to novel chemical scaffolds. It also integrates regression-based performance metrics and ranking-based evaluations to provide comprehensive, quantitative insights into model accuracy, robustness, and practical utility. By establishing a standardized evaluation pipeline, this framework addresses a critical methodological gap in the field, promoting reproducibility, transparency, and comparability across future studies and serves as a valuable reference for researchers developing predictive models in lipid-based RNA delivery and help accelerate the discovery of high-performing LNPs.

%%%%%%%%%%%%%%%%%%%%%%%%%%%%%%%%%%%%%%%%%%%%%%%%%%%%%%%%%%%%

\section{Results}

\subsection{Overview}
To develop a high-accuracy regression model for predicting LNP transfection efficiency from ionizable lipid structures, it is critical to optimize both the molecular representation and the model architecture. Here, we provide a robust benchmarking framework, established through systematic evaluation of key modeling components, including molecular input features, learning algorithms, and data splitting strategies, using a curated version of the experimental HeLa transfection dataset originally published by Xu et al. \cite{xu2024agile}. All the models included in this framework were either newly implemented or adapted specifically for the transfection prediction task, leveraging a unified dataset and standardized preprocessing pipeline. AGILE \cite{xu2024agile} was included as a baseline in this study and evaluated using its official implementation without architectural modifications, retrained on the curated dataset to ensure consistency. Full implementation details and configuration parameters for all models are provided in Section~\ref{sec:models}.

To facilitate future research, both the curated dataset and the models are publicly available. Although this study focuses on LNP transfection, the proposed framework is designed to be modular and extensible, enabling its application to a broad range of molecular property prediction tasks in drug delivery and beyond.

%%%%%%%%%%%%%%%%%%%%%%%%%%%%%%%%%%%%%%%%%%%%%%%%%%%%%%%%%%%%

\subsection{Dataset Curation}

To ensure consistency in evaluation, all models in this study were trained and tested using a dataset of 1,100 unique ionizable lipid structures derived from the HeLa cell transfection dataset originally reported by Xu et al. \cite{xu2024agile}, which was independently curated as part of this work.

Although a curated version of the Xu et al. dataset \cite{xu2024agile} is available on the AGILE GitHub repository (\href{https://github.com/bowang-lab/AGILE}{https://github.com/bowang-lab/AGILE}) and has been used to fine-tune the AGILE model, we hesitated to use this version due to multiple inconsistencies identified upon detailed inspection. Specifically, we found 235 label mismatches in the SMILES representations and 100 duplicate entries arising from geometric isomerism, particularly cis–trans variants. These discrepancies were identified by comparing the GitHub dataset against the “source data” file published with the article (\href{https://www.nature.com/articles/s41467-024-50619-z}{Nature Communications, 2024}). Additional details and representative examples of these inconsistencies, along with their impact on model performance, are provided in the Supplementary Information and Supplementary Figure~1.

%%%%%%%%%%%%%%%%%%%%%%%%%%%%%%%%%%%%%%%%%%%%%%%%%%%%%%%%%%%%

\subsection{Impact of Feature Representations on LNP Transfection Prediction}\label{sec:feature_rep}

The proposed framework enables systematic benchmarking across three distinct feature representations for ionizable lipids, as well as across all pairwise and three-way combinations of these feature sets. The feature sets included are: (i) count-based Morgan fingerprints, (ii) Expert RDKit-derived molecular descriptors, and (iii) Grover embeddings, a GNN–based representation learned via transformers.

Morgan fingerprints encode molecular substructures as fixed-length bit vectors by algorithmically fragmenting molecules into atomic neighborhoods and iteratively expanding circular substructures around each atom \cite{rogers2010extended}. Expert molecular descriptors capture predefined physicochemical and topological properties, such as molecular weight, logP, and topological polar surface area, derived using RDKit \cite{landrum2013rdkit, rdkitdesc}. Graph-based representations, such as Grover embeddings, employ self-supervised learning on large molecular datasets to extract expressive features directly from molecular graphs \cite{rong2020self}.

\begin{figure}[t]
	\centering
        \includegraphics[width=0.8\textwidth]{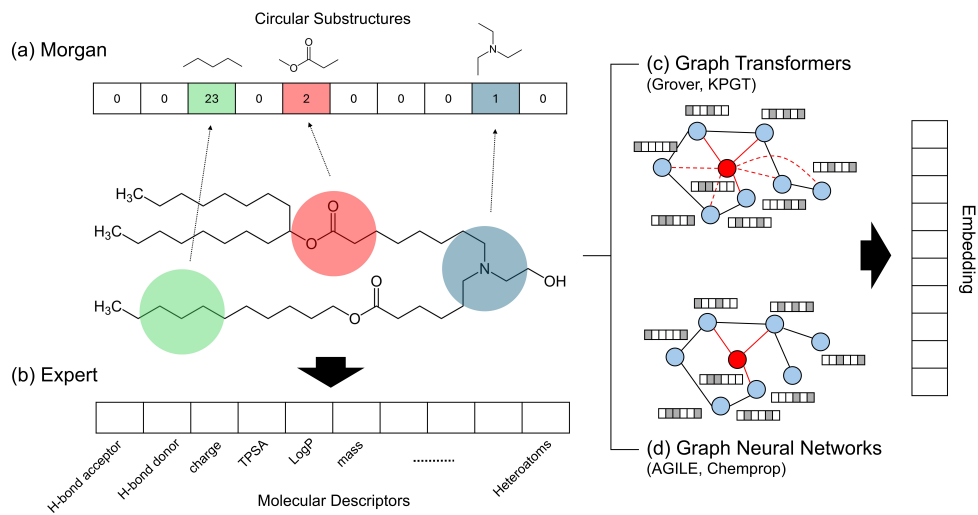}
        \caption{Overview of molecular representation techniques. (a) Morgan fingerprints encode molecular structures as binary vectors based on circular substructures. (b) Expert descriptors use predefined molecular properties, such as hydrogen bonding capacity, charge, and mass. (c) Graph transformers (e.g., Grover, KPGT) capture structural relationships through self-attention mechanisms, generating molecular embeddings. (d) Graph neural networks (GNNs) (e.g., AGILE, Chemprop) learn molecular representations by aggregating information from neighboring nodes in a molecular graph. The resulting embeddings from (c) and (d) can be used for downstream predictive modeling tasks.}
        % \caption{}
	\label{fig:feature_overview}
\end{figure}

Looking ahead, increasingly sophisticated models are expected to rely on end-to-end learned molecular representations, as demonstrated by recent architectures such as AGILE \cite{xu2024agile} and LUMI \cite{cui2025lumi}. Evaluating the performance of these learned representations relative to the fixed feature sets implemented in the proposed framework is essential, as the choice of representation exerts a substantial influence on predictive performance. Figure~\ref{fig:feature_overview} provides a schematic overview of different feature types considered in this study.

The proposed workflow enables quantification of the predictive power of each feature set, individually and in combination, through evaluation across multiple model architectures. We included two representative models: multi-layer perceptron (MLP), a deep learning architecture capable of capturing complex nonlinear relationships, and support vector regression (SVR) \cite{awad2015support}, a robust traditional machine learning method that leverages kernel functions for nonlinear modeling \cite{MIT04_Kernel}.

To further demonstrate the importance of benchmarking different feature representations, we incorporated the AGILE model for comparison. AGILE was evaluated in two forms: (i) its original configuration, which uses end-to-end learned embeddings supplemented with Mordred molecular descriptors \cite{moriwaki2018mordred} to refine lipid representations within the pre-trained graph encoder, and (ii) an extended version augmented with additional explicit molecular features. All models were trained and evaluated using our curated dataset to ensure fair and consistent comparison. A random data split was used, and performance was assessed on the held-out test set following a uniform evaluation protocol.

Table~\ref{tab:feature_analysis} summarizes the results. The choice of feature representation has a substantial impact on model performance across all architectures. In this dataset, feature sets that include Morgan fingerprints consistently achieve the highest predictive accuracy for both the MLP and SVR models. Augmenting Morgan fingerprints with Expert descriptors yields the highest performance for MLP (Supplementary Figure~2) and SVR (Supplementary Figure~3).

In contrast, the graph-based Grover embeddings underperform, indicating limitations of purely data-driven feature extraction methods that do not explicitly encode molecular substructures. Moreover, Grover’s pretraining corpus consists primarily of small molecules~\cite{rong2020self}, which may limit the relevance of its learned representations for structurally larger and more diverse lipid molecules.

\begin{table}[!t]
    \centering
    \small
    % \renewcommand{\arraystretch}{1.1} 
    % \resizebox{0.95\textwidth}{!}{
    \begin{tabular}{c|cccccccccc}
        \toprule
         && \multicolumn{4}{c}{\textbf{Feature Set}} && \multicolumn{4}{c}{\textbf{Metrics}}\\
        \cmidrule{3-6} \cmidrule{8-11}
        \textbf{Method} && \textbf{Morgan} & \textbf{Expert} & \textbf{Grover} & \textbf{End-to-End} && $\textbf{R}^\textbf{2}$ & \textbf{RMSE} & \textbf{MAE} & \textbf{r} \\
        \midrule
        \multirow{7}{*}{MLP} && \ding{52} & \ding{56} & \ding{56} & \ding{56} && \underline{0.7974} & \underline{1.5018} & \underline{1.1507} & \underline{0.8959}\\
        && \ding{56} & \ding{52} & \ding{56} & \ding{56} && 0.5815 & 2.1585 & 1.6922 & 0.7639\\
        && \ding{56} & \ding{56} & \ding{52} & \ding{56} && 0.5171 & 2.3187 & 1.8576 & 0.7215\\
        && \ding{52} & \ding{52} & \ding{56} & \ding{56} && \textbf{0.8161} & \textbf{1.4308} & \textbf{1.1003} & \textbf{0.9053}\\
        && \ding{52} & \ding{56} & \ding{52} & \ding{56} && 0.7489 & 1.6721 & 1.2311 & 0.8734\\
        && \ding{56} & \ding{52} & \ding{52} & \ding{56} && 0.5403 & 2.2622 & 1.7785 & 0.7398\\
        && \ding{52} & \ding{52} & \ding{52} & \ding{56} && 0.7449 & 1.6853 & 1.2162 & 0.8674\\
        \midrule
        \multirow{7}{*}{SVR} && \ding{52} & \ding{56} & \ding{56} & \ding{56} && \underline{0.7109} & \underline{1.7942} & \underline{1.4162} & \underline{0.8615}\\
        && \ding{56} & \ding{52} & \ding{56} & \ding{56} && 0.5441 & 2.2529 & 1.7811 & 0.7446\\
        && \ding{56} & \ding{56} & \ding{52} & \ding{56} && 0.5236 & 2.303 & 1.8658 & 0.7287\\
        && \ding{52} & \ding{52} & \ding{56} & \ding{56} && \textbf{0.7285} & \textbf{1.7387} & \textbf{1.3739} & \textbf{0.8702}\\
        && \ding{52} & \ding{56} & \ding{52} & \ding{56} && 0.6602 & 1.945 & 1.5216 & 0.827\\
        && \ding{56} & \ding{52} & \ding{52} & \ding{56} && 0.528 & 2.2925 & 1.8631 & 0.7325\\
        && \ding{52} & \ding{52} & \ding{52} & \ding{56} && 0.6593 & 1.9478 & 1.5254 & 0.8261\\
        \midrule
        \multirow{8}{*}{AGILE} && \ding{56} & \ding{56} & \ding{56} & \ding{52} && 0.2655 & 2.86 & 2.3328 & 0.5488\\
        && \ding{52} & \ding{56} & \ding{56} & \ding{52} && \textbf{0.706} & \textbf{1.8091} & \textbf{1.3497} & \textbf{0.8412}\\
        && \ding{56} & \ding{52} & \ding{56} & \ding{52} && 0.2965 & 2.7987 & 2.2763 & 0.5618\\
        && \ding{56} & \ding{56} & \ding{52} & \ding{52} && 0.292 & 2.8076 & 2.2721 & 0.5765\\
        && \ding{52} & \ding{52} & \ding{56} & \ding{52} && 0.5676 & 2.1941 & 1.6587 & 0.7548\\
        && \ding{52} & \ding{56} & \ding{52} & \ding{52} && \underline{0.6705} & \underline{1.9155} & \underline{1.4274} & \underline{0.8203}\\
        && \ding{56} & \ding{52} & \ding{52} & \ding{52} && 0.3355 & 2.7199 & 2.1461 & 0.5903\\
        && \ding{52} & \ding{52} & \ding{52} & \ding{52} && 0.5271 & 2.2945 & 1.7435 & 0.7344\\
        \bottomrule
    \end{tabular}
    % }
    \caption{Feature representation comparison across different models. The table compares the performance of models across various feature sets, including Morgan fingerprints, Expert descriptors, Grover embeddings, and end-to-end learned representations. \textbf{Bold} highlight the best performance, and \underline{underlined} denote the second-best. Overall, Morgan fingerprints consistently achieve the best predictive performance across models. Additionally, AGILE without the inclusion of our added features performs worse than all other models, highlighting the significance of feature representation in transfection efficiency prediction.}
    \label{tab:feature_analysis}
\end{table}

\begin{figure}[t]
	\centering
	
    \includegraphics[width=\linewidth]{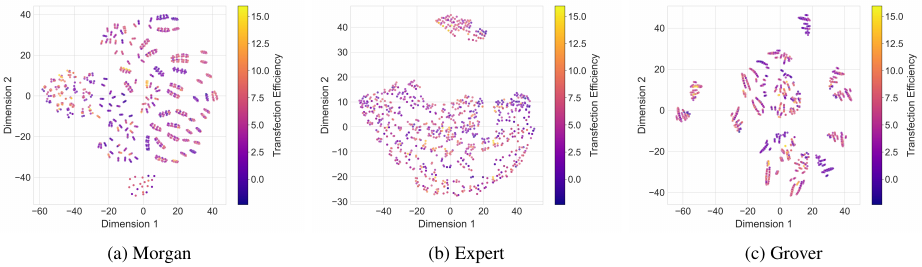}
    \caption{t-SNE visualization of feature representations. Each plot illustrates the distribution of LNP representations in a two-dimensional space, colored by transfection efficiency values. (a) Morgan fingerprints form well-defined clusters, suggesting they effectively capture key molecular substructures, leading to superior predictive performance. (b) Expert descriptors exhibit a more continuous spread, making them less discriminative but still informative. (c) Grover embeddings display a fragmented pattern, indicating less alignment with molecular properties relevant to transfection efficiency, which explains their weaker performance.}
    % \caption{}
	\label{fig:tsne}
\end{figure}

Nonlinear dimensionality reduction using t-distributed stochastic neighbor embedding (t-SNE) \cite{van2008visualizing} further supports these findings (Figure~\ref{fig:tsne}). As shown, Morgan fingerprints exhibit clear and coherent clustering, indicating that they capture substructural features highly relevant to transfection efficiency. Expert descriptors produce a more diffuse distribution, reflecting their ability to encode physicochemical properties while being less discriminative. Grover embeddings form fragmented and weakly structured clusters, consistent with their lower predictive performance and suggesting limited alignment with the molecular determinants of transfection efficiency.

It should be noted that t-SNE was used solely as a qualitative visualization tool to illustrate how different molecular representations organize the dataset in a reduced-dimensional space. The t-SNE parameters followed the original implementation described by van der Maaten and Hinton \cite{van2008visualizing}.

A similar pattern was observed for the AGILE model. In its original configuration, based on end-to-end learned embeddings supplemented with Mordred molecular descriptors, it consistently underperformed relative to all MLP and SVR variants evaluated in this study, irrespective of the feature representation used. This observation highlights the importance of benchmarking sophisticated architectures to ensure that model performance is not constrained by suboptimal molecular representations.

Interestingly, extending AGILE by incorporating additional explicit molecular features, particularly Morgan fingerprints, resulted in a marked improvement in predictive accuracy (Table~\ref{tab:feature_analysis} and Supplementary Figure~4). This finding underscores the critical role of chemically meaningful descriptors in the prediction of LNP transfection and suggests a viable strategy for enhancing the performance of future end-to-end molecular representation models. 

%%%%%%%%%%%%%%%%%%%%%%%%%%%%%%%%%%%%%%%%%%%%%%%%%%%%%%%%%%%%

\subsection{Comparative Analysis of ML Models for LNP Transfection Prediction}
The proposed framework also enables systematic benchmarking across diverse model architectures. Given the superior performance of Morgan fingerprints—either alone or combined with Expert descriptors—in MLP, SVR, and AGILE evaluations, we centered benchmarking on models trained with these feature sets. This design ensures that comparisons are made using the most informative and chemically meaningful representations identified in earlier analyses.

The proposed framework systematically evaluates the predictive performance of a broad range of model architectures, spanning traditional ML models, feedforward neural networks, and transformer-based deep learning methods, while applying multiple evaluation strategies to ensure robust benchmarking. Specifically, we incorporated five representative models: MLP, transformer-based architectures, SVR, random forest (RF), and k-nearest neighbors (kNN), each evaluated using Morgan fingerprints or their combination with Expert descriptors.

MLP and transformer architectures \cite{NIPS23_Attention} were selected to represent feedforward learning approaches. MLPs are well suited for capturing nonlinear relationships and complex feature interactions, while transformers leverage a self-attention mechanism \cite{arXiv19_Bert} that enables learning of hierarchical and context-dependent representations. Among traditional ML models, we included SVR, RF, and kNN, all of which have demonstrated strong performance in molecular property prediction. These models differ fundamentally in their learning strategies: SVR and RF are trainable algorithms—SVR using flexible kernel functions for nonlinear regression, and RF \cite{ML01_RandomForest} employing ensemble decision trees to reduce variance—whereas kNN \cite{kramer2013k} is a non-parametric method that makes predictions based on local similarity in feature space without explicit model training.

As the field advances, increasingly sophisticated architectures for LNP transfection prediction are expected to rely on GNNs and graph transformers, as exemplified by recent models such as AGILE \cite{xu2024agile} and LUMI \cite{cui2025lumi}. Evaluating the performance of these architectures against simpler yet strong baseline models, including those implemented in this study, is essential. Equally important is benchmarking new graph-based approaches against established GNN frameworks to contextualize their performance within the broader landscape of molecular representation learning. To support such comparisons, we incorporated Chemprop \cite{chemprop} and KPGT (Knowledge-Powered Graph Transformer) \cite{li2023knowledge} into the workflow. Chemprop is a graph neural network that processes molecular graphs through message-passing operations to extract structural features, whereas KPGT extends beyond conventional GNNs by integrating transformer-based self-attention mechanisms to capture higher-order structural dependencies. All these graph-based models rely on end-to-end feature extraction, learning molecular representations dynamically during pretraining and supervised training, with feature generation fully integrated into the predictive pipeline. We trained all models on the refined dataset using the same random data split as in Section~\ref{sec:feature_rep} and evaluated performance on the held-out test set to ensure a fair and consistent comparison. AGILE was included in its published configuration \cite{xu2024agile} for reference.

Table~\ref{tab:model_comparison} summarizes the regression metrics for all models, and Supplementary Figure~5 presents prediction-versus-true-value plots using the optimal feature representation for each architecture. Under these conditions, MLP achieved the highest accuracy when trained with Morgan fingerprints combined with Expert descriptors. Traditional ML models such as SVR and RF also performed competitively, demonstrating strong suitability for transfection prediction when paired with chemically meaningful representations. Importantly, all models implemented within the proposed framework outperformed both AGILE and Chemprop. Among the graph-based models, KPGT achieved the highest predictive accuracy, followed by Chemprop, with AGILE performing the worst. The relatively strong performance of KPGT indicates that data-driven feature extraction can be effective for LNP transfection prediction when supported by well-designed architectures and robust pretraining strategies.

The inferior performance of all graph-based approaches relative to models using explicit molecular substructure encoding (MLP, SVR, and RF) is notable and likely reflects the limited ability of GNNs and graph transformers to capture key structural features when trained on a small, domain-specific dataset. These architectures may require substantially larger datasets or pretraining on lipid-rich molecular corpora to learn chemically meaningful representations—resources that are currently scarce for ionizable lipids. Consequently, the development and evaluation of graph-based models should be approached with caution, as inadequate benchmarking can lead to misleading conclusions or inflated performance estimates. One may argue that graph-based models should be compared primarily against architecturally similar approaches; however, given their current limitations in predicting LNP transfection efficiency, these models alone are insufficient to serve as meaningful benchmarks. In contrast, models leveraging explicit molecular representations included in the proposed framework—particularly MLP, SVR, and RF trained on Morgan fingerprints with or without Expert descriptors—consistently provide the most accurate transfection predictions and should serve as essential baselines for future LNP transfection model development. These findings further highlight the need for improved GNN and graph-transformer architectures, or hybrid strategies that integrate domain-informed chemical features, to advance LNP transfection prediction.

\begin{table}[!t]
    \centering
    \small
    % \renewcommand{\arraystretch}{1.1} 
    % \resizebox{0.95\textwidth}{!}{
    \begin{tabular}{c|ccccccccc}
        \toprule
         && \multicolumn{3}{c}{\textbf{Feature Set}} && \multicolumn{4}{c}{\textbf{Metrics}}\\
        \cmidrule{3-5} \cmidrule{7-10}
        \textbf{Method} && \textbf{Morgan} & \textbf{Expert} & \textbf{End-to-End} && $\textbf{R}^\textbf{2}$ & \textbf{RMSE} & \textbf{MAE} & \textbf{r} \\
        \midrule
        \multirow{2}{*}{MLP} && \ding{52} & \ding{56} & \ding{56} && \underline{0.7974} & \underline{1.5018} & \underline{1.1507} & \underline{0.8959}\\
        && \ding{52} & \ding{52} & \ding{56} && \textbf{0.8161} & \textbf{1.4308} & \textbf{1.1003} & \textbf{0.9053}\\
        \midrule
        \multirow{2}{*}{Transformer} && \ding{52} & \ding{56} & \ding{56} && 0.6555 & 1.9585 & 1.5181 & 0.81\\
        && \ding{52} & \ding{52} & \ding{56} && 0.6368 & 2.011 & 1.5381 & 0.8005\\
        \midrule
        \multirow{2}{*}{SVR} && \ding{52} & \ding{56} & \ding{56} && 0.7109 & 1.7942 & 1.4162 & 0.8615\\
        && \ding{52} & \ding{52} & \ding{56} && 0.7285 & 1.7387 & 1.3739 & 0.8702\\
        \midrule
        \multirow{2}{*}{RF} && \ding{52} & \ding{56} & \ding{56} && 0.6976 & 1.835 & 1.3548 & 0.84\\
        && \ding{52} & \ding{52} & \ding{56} && 0.7169 & 1.7753 & 1.3391 & 0.848\\
        \midrule
        \multirow{2}{*}{kNN} && \ding{52} & \ding{56} & \ding{56} && 0.6001 & 2.11 & 1.6109 & 0.7769\\
        && \ding{52} & \ding{52} & \ding{56} && 0.6148 & 2.071 & 1.5969 & 0.7866\\
        \midrule
        KPGT && \ding{56} & \ding{56} & \ding{52} && 0.6655 & 1.9298 & 1.4597 & 0.8211\\
        \midrule
        Chemprop && \ding{56} & \ding{56} & \ding{52} && 0.5215 & 2.3081 & 1.7908 & 0.7248\\
        \midrule
        AGILE && \ding{56} & \ding{56} & \ding{52} && 0.2655 & 2.86 & 2.3328 & 0.5488\\
        \bottomrule
    \end{tabular}
    % }
    \caption{Performance comparison across models and feature sets. The table presents the predictive performance of various ML models trained with different feature representations, including Morgan fingerprints, Expert descriptors, and end-to-end learned embeddings, compared against the AGILE baseline. MLP consistently achieves the best performance, with its highest accuracy observed when using the Morgan fingerprints and Expert descriptors combination, as indicated by the \textbf{bold} values. \underline{Underlined} values represent the second-best performance.}

    \label{tab:model_comparison}
\end{table}

%%%%%%%%%%%%%%%%%%%%%%%%%%%%%%%%%%%%%%%%%%%%%%%%%%%%%%%%%%%%

\subsection{Impact of Murcko Scaffold Splitting on Model Performance}

Generalization capability, beyond predictive accuracy, is a critical property that determines whether a model is suitable and reliable for discovery applications and therefore must be rigorously evaluated during model development. The proposed framework enables such evaluation by supporting performance comparisons across its included models under Murcko scaffold splitting \cite{bemis1996properties}, a more stringent and chemically informed data partitioning strategy than random splitting. This approach groups molecules according to their core scaffolds, ensuring that structurally distinct compounds are assigned to the training, validation, and test sets, thereby allowing assessment of a model’s ability to make predictions on previously unseen chemical structures.

Analysis of pairwise Tanimoto similarities computed from Morgan fingerprints under both random and Murcko scaffold splitting  indicates that, under Murcko splitting, the mean similarity within individual subsets (training, validation, or test) is higher than under random splitting, suggesting that structurally related molecules are preferentially grouped together. In contrast, similarities between subsets are consistently lower under Murcko splitting, reflecting increased structural separation between training and evaluation sets. Collectively, these results confirm that Murcko scaffold splitting more effectively reduces structural overlap between subsets, thereby limiting information leakage and enabling a more stringent and realistic assessment of model generalization. It should be noted, however, that when datasets are relatively small, scaffold-based partitioning can lead to substantial data fragmentation, reducing the number of structurally related training examples and increasing the difficulty of the prediction task.

Table~\ref{tab:murko_split} summarizes the predictive performance of all models under Murcko scaffold splitting, with prediction-versus-true-value plots shown in Supplementary Figure~6. Relative to the random-split setting, most models exhibit reduced accuracy, reflecting the increased difficulty of generalizing to unseen molecular scaffolds. Notably, among deep learning models, Chemprop is the only architecture that maintains performance comparable to the random split, indicating robustness to scaffold-based distribution shifts. The pronounced performance decline observed for other deep learning models suggests a reliance on structural patterns that do not extrapolate well to novel scaffolds, an issue that is exacerbated by limited dataset size.

In this context, the importance of benchmarking generalization capability becomes clearer when examining AGILE’s performance. Although AGILE was originally designed for discovery applications and would therefore be expected to generalize well, it exhibits the lowest predictive accuracy under scaffold splitting. This result, together with the earlier observations, suggests that AGILE’s pretraining strategy and training regime do not yield sufficiently meaningful or transferable molecular representations.

It should be noted that the curated HeLa transfection dataset used in this study exhibits a relatively narrow dynamic range, spanning less than two orders of magnitude, which reflects the inherent variability of the transfection assays reported by Xu et al. \cite{xu2024agile}. This compressed distribution influences the magnitude of absolute error values. Accordingly, our conclusions emphasize comparative model performance rather than absolute error magnitudes: because all models are evaluated on the same dataset and data splits, relative differences in predictive accuracy remain meaningful despite the limited numerical spread of the target variable. We also note that the proposed framework is designed to accommodate broader datasets as they become available, enabling future benchmarking under wider and more diverse response ranges.

Overall, these findings suggest that nonparametric approaches such as kNN may be less sensitive to scaffold-level domain shifts, at least within the context of transfection prediction and the dataset size examined here, whereas deep learning models, despite their higher representational capacity, require more structurally diverse training data to generalize effectively. These results underscore the substantial influence of data-splitting strategies on model performance and highlight the necessity of benchmarking new models against a broad set of baselines to properly assess their suitability for generalization and discovery.

% } 

\begin{table}[!t]
    \centering
    \small
    % \renewcommand{\arraystretch}{1.1} 
    % \resizebox{0.95\textwidth}{!}{
    \begin{tabular}{c|ccccccccc}
        \toprule
         && \multicolumn{3}{c}{\textbf{Feature Set}} && \multicolumn{4}{c}{\textbf{Metrics}}\\
        \cmidrule{3-5} \cmidrule{7-10}
        \textbf{Method} && \textbf{Morgan} & \textbf{Expert} & \textbf{End-to-End} && $\textbf{R}^\textbf{2}$ & \textbf{RMSE} & \textbf{MAE} & \textbf{r} \\
        \midrule
        \multirow{2}{*}{MLP} && \ding{52} & \ding{56} & \ding{56} && 0.5265 & 1.9305 & 1.5465 & 0.7337\\
        && \ding{52} & \ding{52} & \ding{56} && 0.4532 & 2.0746 & 1.7726 & 0.7344\\
        \midrule
        \multirow{2}{*}{Transformer} && \ding{52} & \ding{56} & \ding{56} && 0.3324 & 2.2924 & 1.8597 & 0.6552\\
        && \ding{52} & \ding{52} & \ding{56} && 0.4097 & 2.1555 & 1.7421 & 0.6621\\
        \midrule
        \multirow{2}{*}{SVR} && \ding{52} & \ding{56} & \ding{56} && 0.3082 & 2.3336 & 2.0003 & 0.6374\\
        && \ding{52} & \ding{52} & \ding{56} && 0.3157 & 2.3209 & 1.9835 & 0.6425\\
        \midrule
        \multirow{2}{*}{RF} && \ding{52} & \ding{56} & \ding{56} && 0.3496 & 2.2627 & 1.7943 & 0.6043\\
        && \ding{52} & \ding{52} & \ding{56} && 0.4747 & 2.0334 & 1.5947 & 0.6895\\
        \midrule
        \multirow{2}{*}{kNN} && \ding{52} & \ding{56} & \ding{56} && \textbf{0.6146} & \textbf{1.7419} & \textbf{1.3644} & \textbf{0.7955}\\
        && \ding{52} & \ding{52} & \ding{56} && \underline{0.5919} & \underline{1.7923} & \underline{1.4177} & \underline{0.7892}\\
        \midrule
        KPGT && \ding{56} & \ding{56} & \ding{52} && 0.4878 & 2.0079 & 1.5751 & 0.7173\\
        \midrule
        Chemprop && \ding{56} & \ding{56} & \ding{52} && 0.5129 & 1.9582 & 1.4912 & 0.7188\\
        \midrule
        AGILE && \ding{56} & \ding{56} & \ding{52} && 0.0057 & 2.7976 & 2.3389 & 0.4690\\
        \bottomrule
    \end{tabular}
    % }
    \caption{Performance comparison across models and feature sets under Murcko scaffold split. This table presents the predictive performance of various ML models trained with different feature representations, including Morgan fingerprints, Expert descriptors, and end-to-end learned embeddings. \textbf{Bold} highlight the best performance, and \underline{underlined} denote the second-best. Due to the more challenging scaffold-based data split, overall performance is lower compared to random splitting. While most models show a decline in accuracy, kNN and Chemprop maintain similar performance levels, suggesting their robustness to distribution shifts. kNN achieves the best performance, followed by MLP, while AGILE exhibits the lowest predictive accuracy.
}
    \label{tab:murko_split}
\end{table} 

%%%%%%%%%%%%%%%%%%%%%%%%%%%%%%%%%%%%%%%%%%%%%%%%%%%%%%%%%%%%

\subsection{Error Distribution and Ranking-Based Evaluation of Regression Models}

The proposed framework provides complementary insights into the reliability of LNP transfection prediction beyond standard regression metrics by enabling additional benchmarking through analysis of relative error distributions and ranking accuracy.

Relative error is defined as the absolute difference between the predicted and true transfection efficiency, normalized by the true value. Importantly, relative error distributions do not always correlate directly with traditional regression metrics and can reveal additional nuances related to model uncertainty. Narrower and more concentrated distributions indicate greater reliability and reduced variability in predictions, whereas broader distributions suggest increased uncertainty and diminished predictive consistency.

\begin{figure}[!t]
	\centering
        \includegraphics[width=0.47\textwidth]{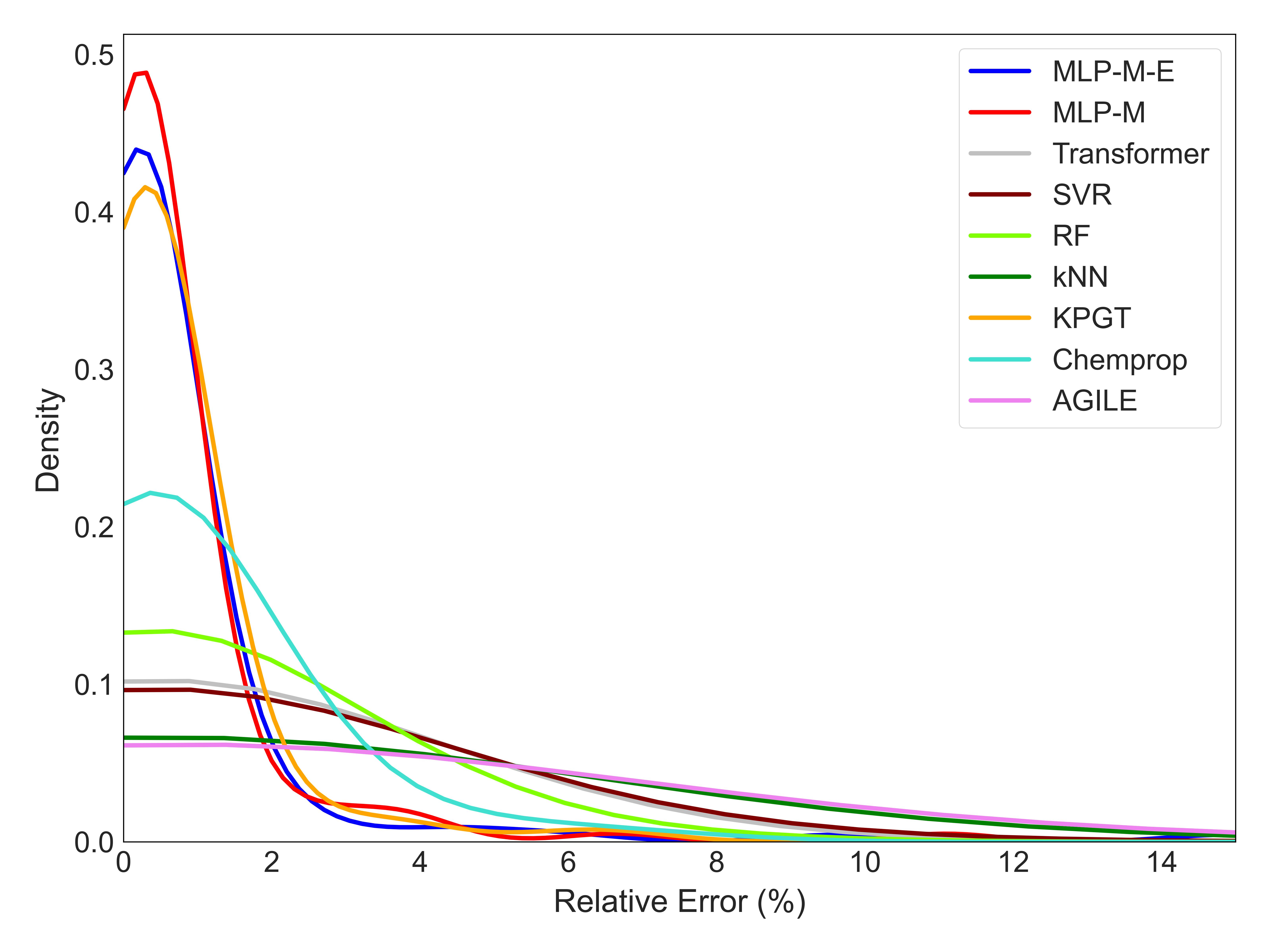}
        \caption{Relative error distribution across different models. This figure presents the relative error histograms for nine models evaluated, including two MLP variants trained with Morgan fingerprints (MLP-M) and Morgan fingerprints combined with Expert descriptors (MLP-M-E). For models utilizing multiple feature sets, the best-performing configuration is selected based on predictive accuracy. The distribution of errors highlights the reliability and precision of each model, with narrower and more concentrated histograms indicating superior predictive performance. }
    % \caption{}
	\label{fig:errors}
\end{figure}

Figure~\ref{fig:errors} shows the relative error distributions for all models evaluated on the test set under random splitting of the curated transfection dataset. As seen, MLP and KPGT exhibit similarly narrow and concentrated distributions, indicating stable performance with fewer extreme mispredictions—even though MLP outperforms KPGT in standard regression metrics. In contrast, transformer-based models and traditional ML models (SVR, RF, kNN), while achieving comparable RMSE and $\text{R}^2$ values to KPGT (Table~\ref{tab:model_comparison}), display broader distribution tails, reflecting greater variability in their predictions. This highlights the importance of assessing model performance from multiple perspectives, as error distributions can substantially influence the interpretation of predictive reliability.

Among all models, AGILE exhibits the widest relative error distribution, further reinforcing its poor predictive performance and lack of reliability in estimating transfection efficiency. Together, these discrepancies underscore the necessity of benchmarking and evaluating models using multiple complementary criteria to gain a comprehensive understanding of model robustness and suitability for downstream discovery applications.

\begin{figure}[!t]
	\centering
    \includegraphics[width=\linewidth]{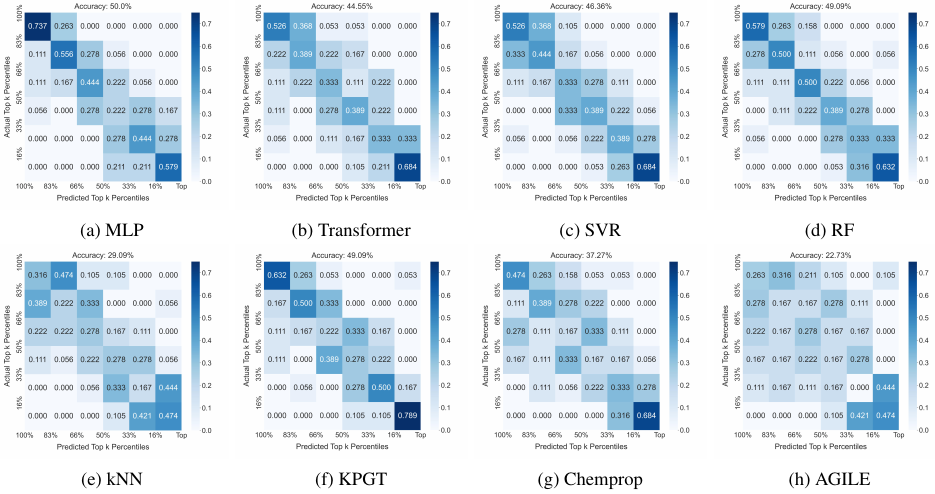}
    \caption{Confusion matrices for different models using the AGILE ranking-based evaluation approach. Each matrix represents the performance of a model on the test set, where the best-performing feature set is selected for each model. (a–h) show the confusion matrix of MLP, Transformer, SVR, RF, kNN, KPGT, Chemprop, and AGILE, respectively. The confusion matrices are constructed based on percentile-based ranking, where true and predicted transfection efficiencies are assigned to discrete percentile bins. The diagonal elements indicate correct rankings, while off-diagonal elements reflect ranking misclassifications. Higher values along the diagonal correspond to better performance. MLP achieves the highest ranking accuracy, while AGILE shows the lowest predictive ranking performance.}
    % \caption{}
	\label{fig:confusion}
\end{figure}

One may argue that regression-based metrics alone are insufficient for evaluating models intended for discovery applications, where ranking candidates by transfection efficiency is often more relevant than predicting exact numerical values. To address this perspective, and to further support benchmarking when ranking accuracy is of interest, the proposed framework incorporates ranking-based evaluation. In this approach, predictions are categorized into percentile bins, following the strategy used by Xu et al. \cite{xu2024agile}, and model performance is illustrated using confusion matrices that show how effectively each model ranks formulations into the correct percentile groups. We applied this evaluation method to all models trained under random splitting (Figure~\ref{fig:confusion}). In these matrices, diagonal entries represent correctly ranked predictions, whereas off-diagonal entries indicate misclassified rankings.

Applying this ranking-based evaluation reveals that MLP achieves the highest ranking accuracy, with the largest proportion of correct assignments along the diagonal. The overall ranking accuracy for MLP is 50.0\%, with particularly strong performance in identifying high-transfection formulations (73.7\%) and low-transfection formulations (57.9\%). Among graph-based and traditional models, KPGT and RF perform comparably to MLP, with KPGT excelling at identifying the highest and lowest transfection groups, while RF demonstrates more uniform accuracy across all percentile bins. AGILE again demonstrates the weakest performance, with an average ranking accuracy of only 22.7\%. It correctly identifies just 26.3\% of high-transfection formulations and 47.4\% of low-transfection formulations. 
This further highlights how insufficient benchmarking can lead to the development of sophisticated transfection prediction models that nonetheless perform poorly across multiple evaluation criteria.

%%%%%%%%%%%%%%%%%%%%%%%%%%%%%%%%%%%%%%%%%%%%%%%%%%%%%%%%%%%%

\section{Discussion}

The present study introduces a robust benchmarking framework for predicting transfection efficiency from ionizable lipid structures. The proposed framework enables systematic evaluation of the predictive performance of diverse ML models on a given dataset, emphasizing the critical roles of feature selection, model architecture, and evaluation strategy when selecting appropriate baselines for reliable LNP transfection prediction.

This framework supports benchmarking across three distinct feature representations: Morgan fingerprints, Expert RDKit-derived molecular descriptors, and Grover embeddings, a graph-based representation learned with transformer architectures. These representations were used to train two representative model architectures: a multi-layer perceptron (MLP), a deep learning model capable of capturing complex molecular relationships, and a support vector regression (SVR), a traditional ML approach that uses kernel functions for nonlinear modeling.

The proposed framework also enables benchmarking across multiple classes of ML models, spanning traditional models, feedforward neural networks, and transformer-based deep learning architectures, using Morgan fingerprints alone or combined with Expert descriptors as the feature set. In addition, it incorporates state-of-the-art GNN and graph-transformer models to provide a comprehensive comparison. A wide range of evaluation metrics, including regression accuracy, relative-error distributions, and ranking-based assessments, are integrated into the framework to support rigorous benchmarking.

Morgan fingerprints encode molecular substructures as bit vectors by fragmenting molecules into atomic neighborhoods and iteratively expanding circular substructures around each atom \cite{rogers2010extended}. They are widely used in molecular similarity-based tasks \cite{maggiora2014molecular}, including virtual screening against protein targets \cite{petrone2012rethinking}, structure–activity relationship modeling for ADME-Tox prediction \cite{orosz2022comparison}, bioactivity assessment of natural compounds \cite{periwal2022bioactivity}, drug combination synergy prediction \cite{zagidullin2021comparative}, and identification of novel protein inhibitors \cite{belenahalli2023development}. Despite their utility in capturing substructural similarity patterns, Morgan fingerprints lack direct interpretability. As demonstrated by Tejera et al. \cite{tejera2019cell}, the choice between similarity-based and ML-based prediction methods is inherently task-specific, with certain compound–cell line interactions better captured by one approach over the other.

We incorporated count-based Morgan fingerprints (C-MF), introduced by Zhong et al. \cite{zhong2023count}, into the framework. C-MF encode not only the presence of molecular substructures but also their frequency, providing a richer representation than binary Morgan fingerprints (B-MF), which capture only substructure presence or absence. Zhong et al. \cite{zhong2023count} demonstrated that C-MF outperformed B-MF in nine out of ten datasets across six ML models (ridge regression \cite{hoerl1970ridge}, SVM \cite{svm}, kNN, RF, XGBoost \cite{chen2016xgboost}, and CatBoost \cite{prokhorenkova2018catboost}), underscoring the predictive value of incorporating substructure counts. Expert molecular descriptors, on the other hand, encode predefined physicochemical and topological properties, such as molecular weight, logP, and topological polar surface area (TPSA), derived from domain knowledge \cite{landrum2013rdkit, rdkitdesc}. These interpretable features help characterize molecular properties relevant to transfection efficiency.

When comparing Morgan fingerprints and Expert descriptors, our findings show that Morgan fingerprints consistently outperformed Expert descriptors in predicting transfection efficiency. However, this is not universally true across all molecular prediction tasks. For example, Tayyebi et al. \cite{tayyebi2023prediction} found that Mordred descriptors \cite{moriwaki2018moesm4} provided superior accuracy over Morgan fingerprints for aqueous solubility prediction using RF and MLR. Similarly, Seal et al. \cite{seal2021comparison} demonstrated that Cell Painting descriptors outperformed both ErG and Morgan fingerprints for cytotoxicity and proliferation assays. These studies highlight that the effectiveness of molecular descriptors is highly dependent on dataset characteristics and the specific predictive task.

In our analysis, Morgan fingerprints, particularly when combined with Expert descriptors, consistently yielded the highest predictive performance, emphasizing the value of explicit molecular substructure encoding in LNP transfection prediction. This improvement aligns with findings from Zhou et al. \cite{zhou2024utility}, who demonstrated that incorporating Morgan fingerprints into the FINDSITE suite enhanced ligand homology modeling, traditional ligand similarity-based approaches, and ML-driven virtual screening, reinforcing the added value of including explicit substructural features. It is important to note, however, that alternative fingerprinting methods—such as MACCS keys, Atom Pair or Topological Torsion fingerprints, and ECFP variants with different radii—may also provide informative representations for LNP transfection prediction. Evaluating these alternatives represents an important direction for future work and can be readily integrated into benchmarking frameworks as the field and available datasets continue to evolve.

While GNNs have demonstrated strong performance in many molecular property prediction tasks, their reliance on large datasets and extensive pretraining makes them less effective in small-data regimes, such as the LNP transfection prediction task considered here. Our results highlight systematic limitations of graph-based representations, including Grover embeddings and AGILE’s end-to-end learned embeddings, when applied to relatively small, domain-specific datasets. Grover embeddings use self-supervised learning with graph-transformer architectures trained on large molecular corpora to extract structural features directly from molecular graphs \cite{rong2020self}. However, this purely data-driven feature extraction limits interpretability and does not incorporate domain-specific chemical knowledge.

Fang et al.~\cite{fang2023mfgb} noted that, although GNNs can automatically learn molecular representations, they often fail to capture deep chemical semantics, and simple molecular graph structures are insufficient for robust representation learning. They also emphasized that GNNs are susceptible to over-smoothing, where excessive message passing renders node representations indistinguishable, hindering meaningful feature extraction. To address these limitations, they introduced a strategy that integrates chemical domain knowledge through Morgan fingerprints, effectively mitigating over-smoothing and reducing dependence on long-distance information propagation. They observed substantial improvements in predicting molecular properties such as toxicity and lipophilicity, emphasizing the utility of explicitly encoded substructural information \cite{fang2023mfgb}. This observation aligns with our findings: incorporating Morgan fingerprints into AGILE led to substantial improvements in predictive accuracy (Table~\ref{tab:feature_analysis}), likely due to the introduction of chemically meaningful representations not captured by the original architecture. Notably, adding Expert descriptors to AGILE did not yield comparable performance gains. It should be noted that AGILE includes a descriptor encoder designed to process Mordred descriptors \cite{moriwaki2018mordred} and update the lipid structure representation within the pretrained graph encoder. Our results suggest that AGILE’s integration strategy for descriptors is not effective for improving accuracy in LNP transfection prediction. A similar pattern was observed when Expert descriptors were combined with Grover embeddings for MLP and SVR: unlike Morgan fingerprints, Expert descriptors did not enhance predictive performance relative to Grover features alone. 
Collectively, these trends highlight the superior utility of Morgan fingerprints as chemically informative representations for LNP transfection prediction, compared with generic physicochemical descriptors in this context. Nevertheless, it is important to recognize that descriptor performance is dataset-dependent. Differences in dataset size, structural diversity, and chemical space coverage may lead to different outcomes. Accordingly, while the presented baselines provide strong reference points for benchmarking, their relative performance may not universally generalize across all datasets in the field.

We further investigated the performance of additional graph-based models for transfection prediction relative to traditional approaches. Li et al. \cite{li2023knowledge} noted that many self-supervised GNN methods suffer from two major limitations: (i) the lack of a well-defined self-supervised learning strategy and (ii) limited capacity to capture chemically meaningful representations. To examine whether more advanced architectures can address these issues, we evaluated KPGT, a knowledge-powered graph transformer designed to generate robust and generalizable molecular embeddings. We also included Chemprop, a widely used GNN model that integrates message-passing and aggregation mechanisms for molecular feature extraction. Our results show that KPGT substantially outperforms both AGILE and Chemprop, demonstrating the critical importance of architecture design and pretraining strategy in graph-based LNP transfection prediction. KPGT leverages a pretrained backbone, the LiGhT (Line Graph Transformer) model, trained on approximately two million molecules from ChEMBL29 \cite{gaulton2017chembl} as part of its knowledge-guided pretraining. Through extensive validation across 63 molecular property prediction datasets, KPGT has been shown to consistently outperform conventional GNN-based approaches \cite{li2023knowledge}, underscoring the value of chemistry-aware pretraining. However, despite its improved performance relative to other graph-based models, KPGT still performed inferiorly to models using explicit molecular substructure encoding across multiple evaluation criteria.

Based on these results, we recommend that newly developed transfection prediction models be benchmarked against models that use Morgan fingerprints—either alone or in combination with Expert descriptors—to ensure rigorous and meaningful comparison. Among all models trained on the curated HeLa transfection dataset, MLP consistently outperforms the others, demonstrating superior regression accuracy, narrower relative-error distributions, and the highest ranking performance. Accordingly, MLP should receive particular emphasis when selecting baseline models for benchmarking. The proposed framework provides the necessary tools and standardized evaluation procedures to support such rigorous and reproducible assessment. 

With the emergence of increasingly sophisticated graph-based models—such as AGILE \cite{xu2024agile} and LUMI \cite{cui2025lumi}—for LNP transfection prediction, there is a growing expectation that these architectures will deliver high accuracy, well-behaved error distributions, and strong generalizability to guide lipid discovery. The increasing reliance on such models underscores the need for robust benchmarking resources, as inadequate benchmarking—illustrated by the cases of AGILE and LUMI—poses a significant risk and may call into question the reliability of these models for discovery applications. AGILE was evaluated against weak baselines that were not competitive with the standards established in this study, leading to an overstated assessment of its performance, while LUMI, although benchmarked more rigorously, still achieves substantially lower accuracy than the best-performing models proposed here (Pearson = 0.79 vs. 0.90 for MLP with Morgan and Expert descriptors). These cases highlight the necessity of using strong, representative baselines to ensure meaningful and reliable evaluation of new architectures for LNP transfection prediction. By incorporating competitive baseline models, diverse feature representations, and multiple evaluation metrics, the current study provides a rigorous and comprehensive framework for determining whether newly developed graph-based models achieve the reliability and generalizability required for discovery applications.

%%%%%%%%%%%%%%%%%%%%%%%%%%%%%%%%%%%%%%%%%%%%%%%%%%%%%%%%%%%%

\section{Conclusion}

This work introduces a standardized benchmarking framework that addresses a critical gap in the development of predictive models for LNP transfection. By integrating multiple chemically informed feature representations, diverse ML models, and complementary evaluation strategies, this framework provides a rigorous foundation for assessing the predictive accuracy, generalizability, and robustness of newly developed LNP transfection prediction methods.

Evaluation of the models included in the proposed framework on a rigorously curated HeLa transfection dataset demonstrates that explicit molecular substructure representations—particularly Morgan fingerprints combined with Expert descriptors—paired with MLP, SVR, or RF architectures consistently outperformed state-of-the-art graph-based methods, including AGILE, Chemprop, and KPGT. While these results do not imply that such descriptor–model combinations should be used directly for discovery without further validation—especially given their potential limitations in extrapolating to novel chemical spaces—they underscore the need for improved graph-based or hybrid architectures that incorporate domain-informed chemical features and can at least match or surpass these strong baseline performances.

By establishing strong, representative baselines and a transparent evaluation pipeline, the proposed framework ensures that future LNP transfection prediction models can be meaningfully benchmarked and reliably interpreted, ultimately enabling more efficient and trustworthy discovery of high-performing ionizable lipids for RNA therapeutics. To support continued progress in data-driven LNP design, we release both the refined dataset and the source code, which together facilitate rapid evaluation and deployment of predictive models for LNP transfection efficiency and related molecular property prediction tasks.

%%%%%%%%%%%%%%%%%%%%%%%%%%%%%%%%%%%%%%%%%%%%%%%%%%%%%%%%%%%%

\section{Methods}\label{sec:method}

\subsection{ML Models and Training Procedure}\label{sec:models}
We evaluated five ML models that have been widely applied across domains \cite{mehradfar2025supervised, an2023comprehensive}, including multi-layer perceptron (MLP), transformer \cite{NIPS23_Attention}, support vector regression (SVR) \cite{awad2015support}, random forest (RF) \cite{ML01_RandomForest}, and k-nearest neighbors (kNN) \cite{kramer2013k}. These models were selected due to their strong performance in regression tasks and their diversity in learning paradigms. In addition, we benchmarked three domain-specific models developed for molecular property prediction: Chemprop, KPGT, and AGILE. All model hyperparameters were tuned to ensure optimal performance, except for AGILE, which was used directly from the original implementation without any modifications. All models were trained using a curated dataset derived from HeLa cell transfection experimental data originally reported by Xu et al. \cite{xu2024agile}. The final dataset was rigorously validated to ensure consistency with the original experimental measurements and to eliminate redundancy by removing duplicate SMILES representations associated with geometric isomers, thereby preserving structural uniqueness and ensuring data integrity.

For the MLP, we used a seven-layer feedforward architecture (including the output layer) with hidden layer sizes of 200, 300, 500, 500, 300, and 200, each followed by ReLU activation. The model was trained for up to 100 epochs using the Adam optimizer \cite{arXiv17_Adam} with a learning rate of 0.0002. Mean squared error (MSE) was used as the loss function, and early stopping was applied to prevent overfitting.

The transformer model was implemented using an encoder-based architecture with five layers, each containing two attention heads. The hidden dimension within each feedforward layer was set to 500, and the model dimension (embedding size) was set to 100. It was trained for 100 epochs using the Adam optimizer with a learning rate of 0.0002, MSE loss, and early stopping.

For traditional models, the SVR model uses a radial basis function (RBF) kernel to capture complex nonlinear relationships in the data. This kernel was selected due to its strong ability to approximate highly nonlinear mappings between input features and target values, making it a powerful choice for small to medium-sized datasets where flexible function fitting is beneficial. The RF model was configured with 100 estimators and used the squared error criterion. For kNN, we set the number of neighbors to 5 and used uniform weighting with mean aggregation.

For domain-specific models, we used the official AGILE implementation with the same training settings reported in the original study \cite{xu2024agile}. In AGILE, molecular descriptors are processed through an MLP and concatenated with the GNN-extracted features. The default configuration uses Mordred descriptors~\cite{moriwaki2018mordred}, while alternative feature sets (e.g., Morgan fingerprints and expert-derived descriptors) were supplied through the same interface without modifying the AGILE architecture. For Chemprop, we trained the graph encoder and regressor on our curated dataset using MSE loss for 100 epochs, with a hidden dimension of 500 in the regressor. For KPGT, we used their pretrained model embeddings and trained a three-layer regressor with a hidden dimension of 512 for 20 epochs.

The dataset was once randomly partitioned into training, validation, and test sets using an 80\%, 10\%, and 10\% split, respectively. All experiments were conducted using a fixed random seed to ensure reproducibility. Unless otherwise stated, all reported metrics are evaluated on the held-out test split.

%%%%%%%%%%%%%%%%%%%%%%%%%%%%%%%%%%%%%%%%%%%%%%%%%%%%%%%%%%%%

\subsection{Feature Representations}
We employed three types of molecular features: Morgan fingerprints, Expert descriptors, and Grover embeddings. Morgan fingerprints were computed using the DeepChem  \cite{Ramsundar-et-al-2019} Python library, generating 2048-dimensional count-based circular fingerprints. Expert descriptors were extracted using RDKit's \cite{landrum2013rdkit} 210-dimensional built-in descriptor module. For Grover embeddings \cite{rong2020self}, we used the pretrained model provided by the original authors, applying the default configuration.

%%%%%%%%%%%%%%%%%%%%%%%%%%%%%%%%%%%%%%%%%%%%%%%%%%%%%%%%%%%%

\subsection{Evaluation Metrics}

Model performance was evaluated using four standard regression metrics: coefficient of determination ($\text{R}^2$), root mean squared error (RMSE), mean absolute error (MAE), and Pearson correlation coefficient ($\text{r}$). 

\begin{itemize}[leftmargin=*]
  \item $\text{R}^2$: Measures the proportion of variance in the target explained by the model. Higher values indicate better performance. It is computed as:
  \begin{equation}
    \text{R}^2 = 1 - \frac{\sum_{i=1}^{n}(y_i - \hat{y}_i)^2}{\sum_{i=1}^{n}(y_i - \bar{y})^2}
  \end{equation}
  where $\hat{y}_i$ is the predicted value, $y_i$ is the true value, and $\bar{y}$ is the mean of the observed data.
  \item RMSE: Quantifies the average squared difference between predicted and true values. Lower values indicate better accuracy and are sensitive to outliers. It is calculated as:
  \begin{equation}
    \text{RMSE} = \sqrt{\frac{1}{n}\sum_{i=1}^{n}(\hat{y}_i - y_i)^2}
  \end{equation}

  \item MAE: Measures the average magnitude of prediction errors. Unlike RMSE, it treats all errors equally. It is defined as:
  \begin{equation}
    \text{MAE} = \frac{1}{n}\sum_{i=1}^{n}|\hat{y}_i - y_i|
  \end{equation}
  
  \item Pearson correlation ($\text{r}$): Captures the linear correlation between predictions and ground truth, ranging from -1 (perfect inverse) to 1 (perfect direct correlation):
  \begin{equation}
    \text{r} = \frac{\sum_{i=1}^{n}(\hat{y}_i - \bar{\hat{y}})(y_i - \bar{y})}{\sqrt{\sum_{i=1}^{n}(\hat{y}_i - \bar{\hat{y}})^2} \cdot \sqrt{\sum_{i=1}^{n}(y_i - \bar{y})^2}}
  \end{equation}
\end{itemize}

We additionally computed the relative error for each sample as:
\begin{equation}
    \text{Relative Error} = \frac{|\hat{y} - y|}{y}
\end{equation}
where $\hat{y}$ is the predicted value and $y$ is the ground truth. Relative error distributions were used to assess model uncertainty and stability.

To complement regression analysis, we adopted the percentile-based classification approach used in the AGILE framework. Ground truth and predicted transfection values were mapped to percentile bins, and confusion matrices were constructed using this discretized format. The diagonal elements of each matrix represent correct predictions within the same percentile bin, while off-diagonal elements indicate misclassification. We computed classification accuracy as the proportion of correct bin assignments:

\begin{equation}
    \text{Accuracy} = \frac{\text{Number of correct bin assignments}}{\text{Total number of samples}}
\end{equation}

Unlike AGILE, which computes accuracy over the entire dataset, we evaluate classification performance strictly on the test set to provide a more realistic estimate of model generalization.

%%%%%%%%%%%%%%%%%%%%%%%%%%%%%%%%%%%%%%%%%%%%%%%%%%%%%%%%%%%%

\section*{Acknowledgment}
The authors would like to acknowledge Owen Antholine and Varun Shankar for their valuable assistance with the dataset curation.

\bibliographystyle{unsrt}
\bibliography{references.bib}

\clearpage
\appendix

\renewcommand{\figurename}{Supplementary Figure}
\setcounter{figure}{0}

\section*{Supplementary Information}\label{sec:sup}

\subsection*{Dataset curation details}

To ensure consistency in model evaluation, we based all the models on the HeLa cell transfection experimental data originally reported by Xu et al., which was also used to fine-tune the AGILE model. However, a detailed inspection revealed several discrepancies between the “source data” file published alongside the article (\href{https://www.nature.com/articles/s41467-024-50619-z}{Nature Communications, 2024}) and the dataset hosted on GitHub (\href{https://github.com/bowang-lab/AGILE}{https://github.com/bowang-lab/AGILE}), which AGILE used for fine-tuning. These inconsistencies needed to be resolved to ensure the dataset's integrity for reliable model training and fair comparative analysis. Two major issues present in the original fine-tuning dataset are as follows:

\textit{i) Label Mismatches in SMILES Representations}: We identified 235 inconsistencies among the 1,200 data points reported and used in AGILE’s fine-tuning. These inconsistencies stem from errors in mapping SMILES representations to transfection efficiency values, leading to label mismatches. Specifically, the transfection values reported in the original experimental "source data" file do not always correspond to those found in the dataset used to fine-tune AGILE. For instance, lipid A17B11C3, synthesized by combining head group A17 with tails B11 and C3, has a reported transfection efficiency of 0.28 in the experimental data source, whereas in the dataset used for AGILE’s fine-tuning, the same lipid is erroneously assigned a value of 5.43. Similarly, lipid A16B11C3 has a transfection efficiency of 7.69 in the experimental data but is recorded as -0.69 in the dataset used for AGILE’s fine-tuning. These discrepancies appear to originate from systematic errors during the conversion of experimental data into machine-readable SMILES representations, predominantly affecting lipids with B1, B6, B11, or B12 tails.

To rectify this, we manually verified and corrected the dataset by ensuring that all SMILES representations corresponded to their correct transfection values, assuming the experimental "source data" as the ground truth. This resulted in a curated dataset, which we make publicly available to enable direct comparison with the dataset originally used to fine-tune AGILE. 

\textit{ii) Duplicate Entries Due to Geometric Isomerism}: We identified 100 duplicate SMILES entries arising from the assignment of identical SMILES representations to cis and trans isomers, despite their distinct experimental transfection measurements. In the lipid library proposed by Xu et al., some reactions yield cis-trans mixtures, while others exclusively produce trans isomers. Although the difference lies solely in geometric isomerism, their transfection efficiencies vary significantly.
For example, lipid A17B4C5, which exists as a cis-trans mixture, has an experimental transfection value of 0.85, whereas A17B5C5, which has the same molecular formula but a pure trans structure, has a transfection efficiency of 1.54. This confirms that geometric isomerism has a substantial effect on transfection efficiency. However, AGILE did not account for this distinction, treating cis and trans isomers as the same molecule by assigning them identical SMILES representations.
This misrepresentation results in two key issues: a)	Chemically incorrect labeling – Cis and trans isomers are distinct molecules with different properties. b)	Computational misrepresentation – The model encounters duplicate SMILES representations with conflicting transfection values, leading to inconsistencies and degraded predictive performance. Although the number of duplicates was relatively small compared to the overall dataset size, their presence introduced ambiguity and inconsistency in the dataset. 

To resolve this issue, we removed the duplicate entries and retained only the trans isomers, resulting in a refined dataset comprising of 1,100 unique lipid molecules. The impact of these corrections on model performance is presented in Supplementary Figure~\ref{fig:motiv}, where we compare results before and after dataset refinement. After dataset refinement, AGILE’s accuracy improved marginally from 24.17\% to 24.55\% on the test set. This reported accuracy may appear to contradict the 28.83\% accuracy reported by Xu et al., but this discrepancy arises from their inclusion of training, validation, and test data in performance evaluation, which artificially inflates accuracy. Our evaluation was based strictly on test-set performance, ensuring a more rigorous assessment of the performance of the model. Unless otherwise specified, this refined dataset was used for the training and evaluation of AGILE and all subsequent models in this study.

\newpage
\subsection*{Additional Figures}

\begin{figure}[!htbp]
	\centering
        \begin{subfigure}[t]{0.32\textwidth}
            \centering
            \includegraphics[width=0.99\textwidth]{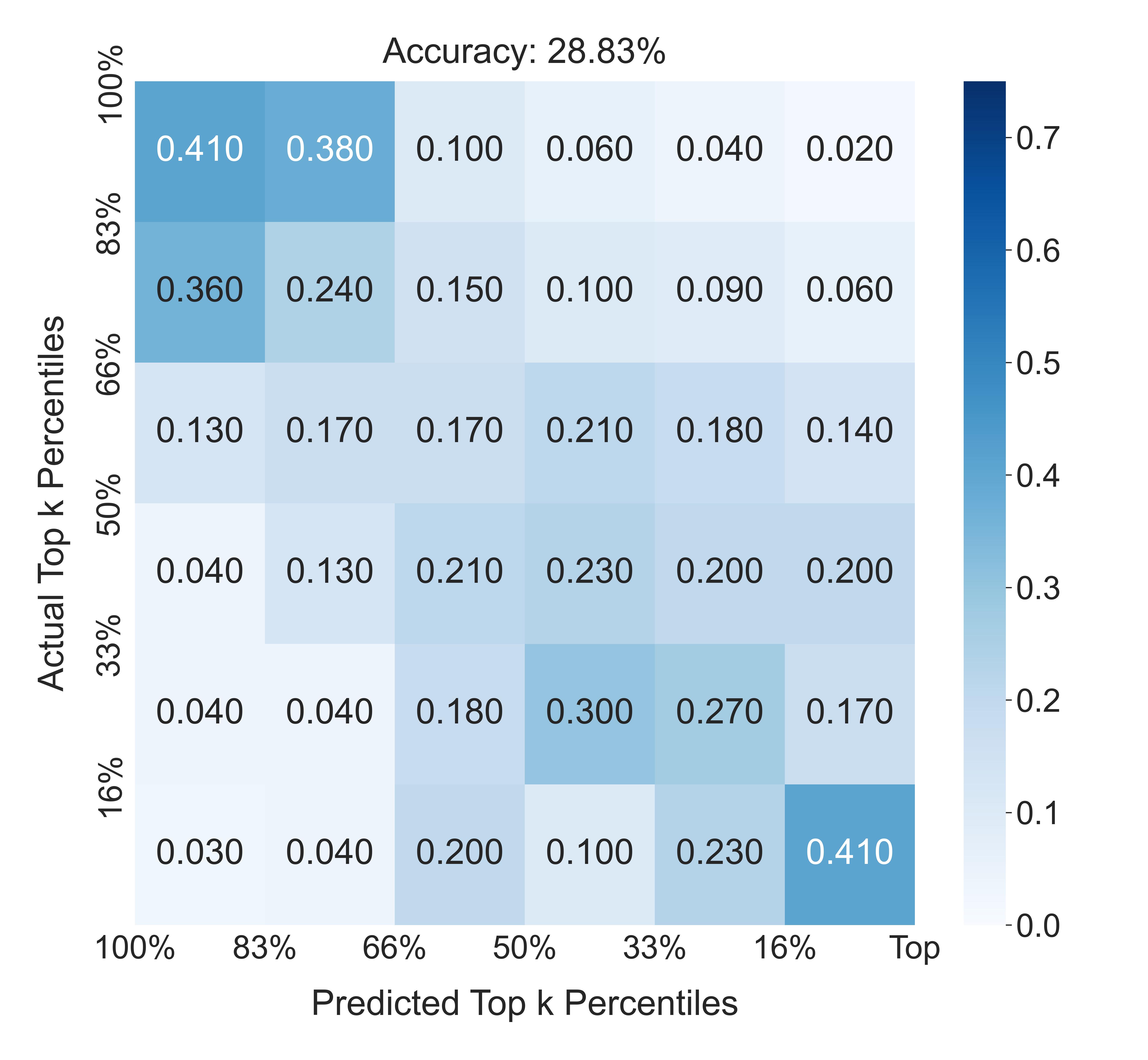}
            \caption{Original data (full set)}
        \end{subfigure}
	\begin{subfigure}[t]{0.32\textwidth}
            \centering
            \includegraphics[width=0.99\textwidth]{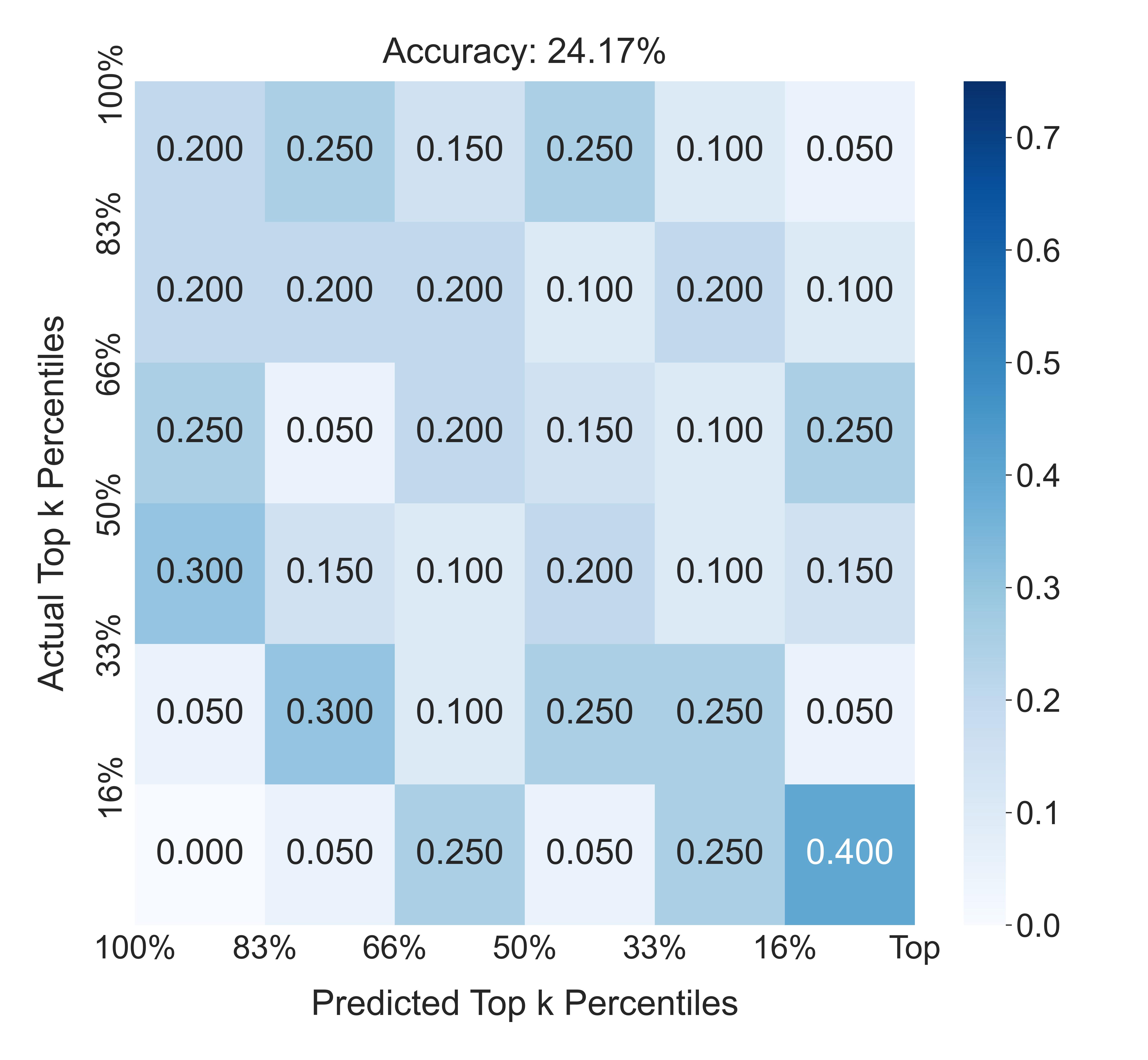}
            \caption{Original data (test split)}
        \end{subfigure}
        % \hfill
	\begin{subfigure}[t]{0.32\textwidth}
            \centering
            \includegraphics[width=0.99\textwidth]{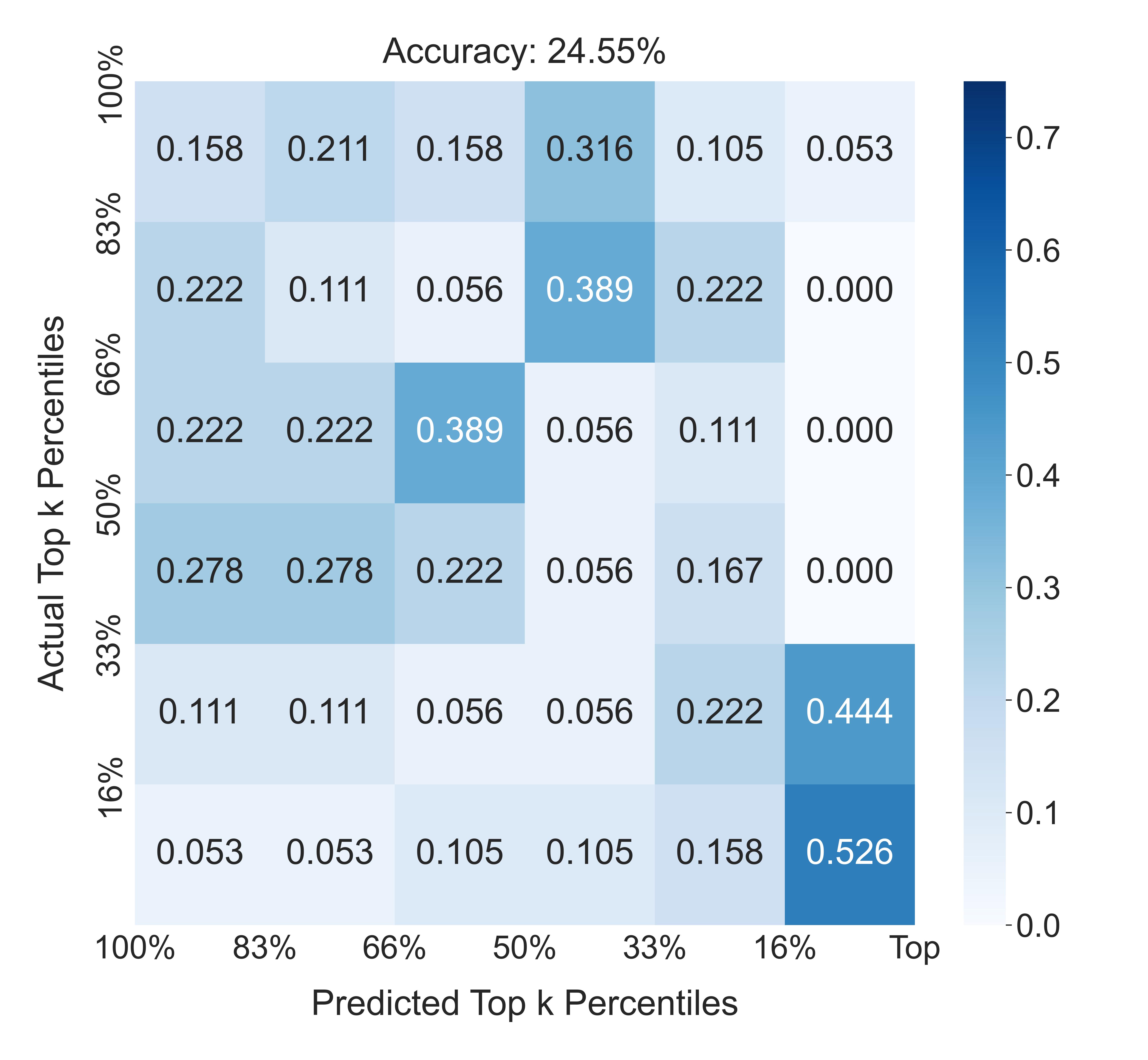}
            \caption{Refined data (test split)}
        \end{subfigure}
        \caption{Confusion matrices for the AGILE model under different data conditions. (a) AGILE trained on the original 1200-point dataset, with the confusion matrix computed over the entire dataset (train, validation, and test), as presented in the AGILE study. (b) The same AGILE model as in (a), but the confusion matrix is computed only on the test split of the original dataset. (c) AGILE trained on the refined 1100-point dataset with Murcko scaffold-based splitting, with the confusion matrix computed on its test split. The differences highlight the impact of dataset refinement on model predictions.}
	\label{fig:motiv}
\end{figure}

\begin{figure}[!htbp]
	\centering
        \begin{subfigure}[t]{0.24\textwidth}
            \centering
            \includegraphics[width=0.99\textwidth]{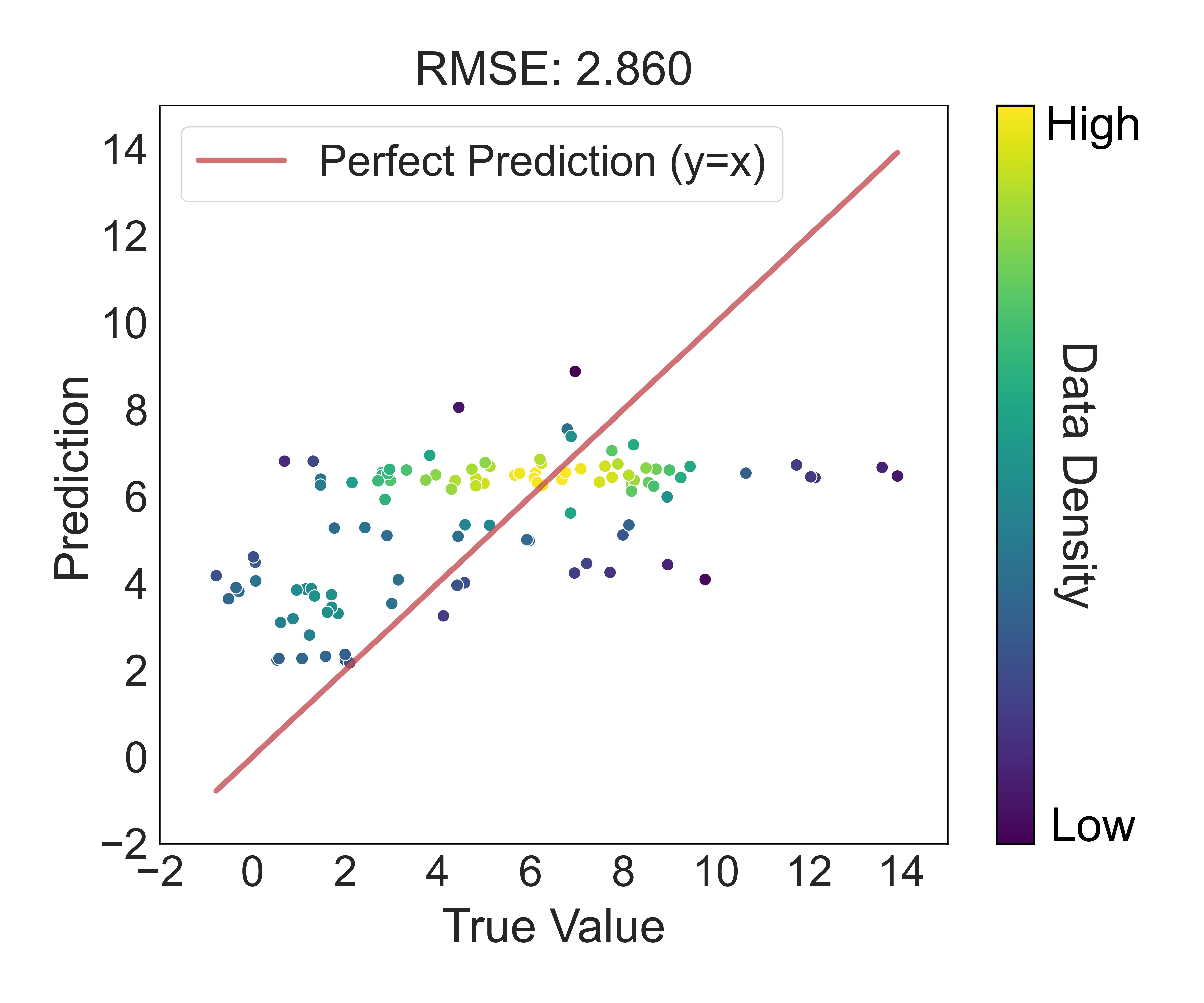}
            \caption{AGILE (End-to-End)}
        \end{subfigure}
	\begin{subfigure}[t]{0.24\textwidth}
            \centering
            \includegraphics[width=0.99\textwidth]{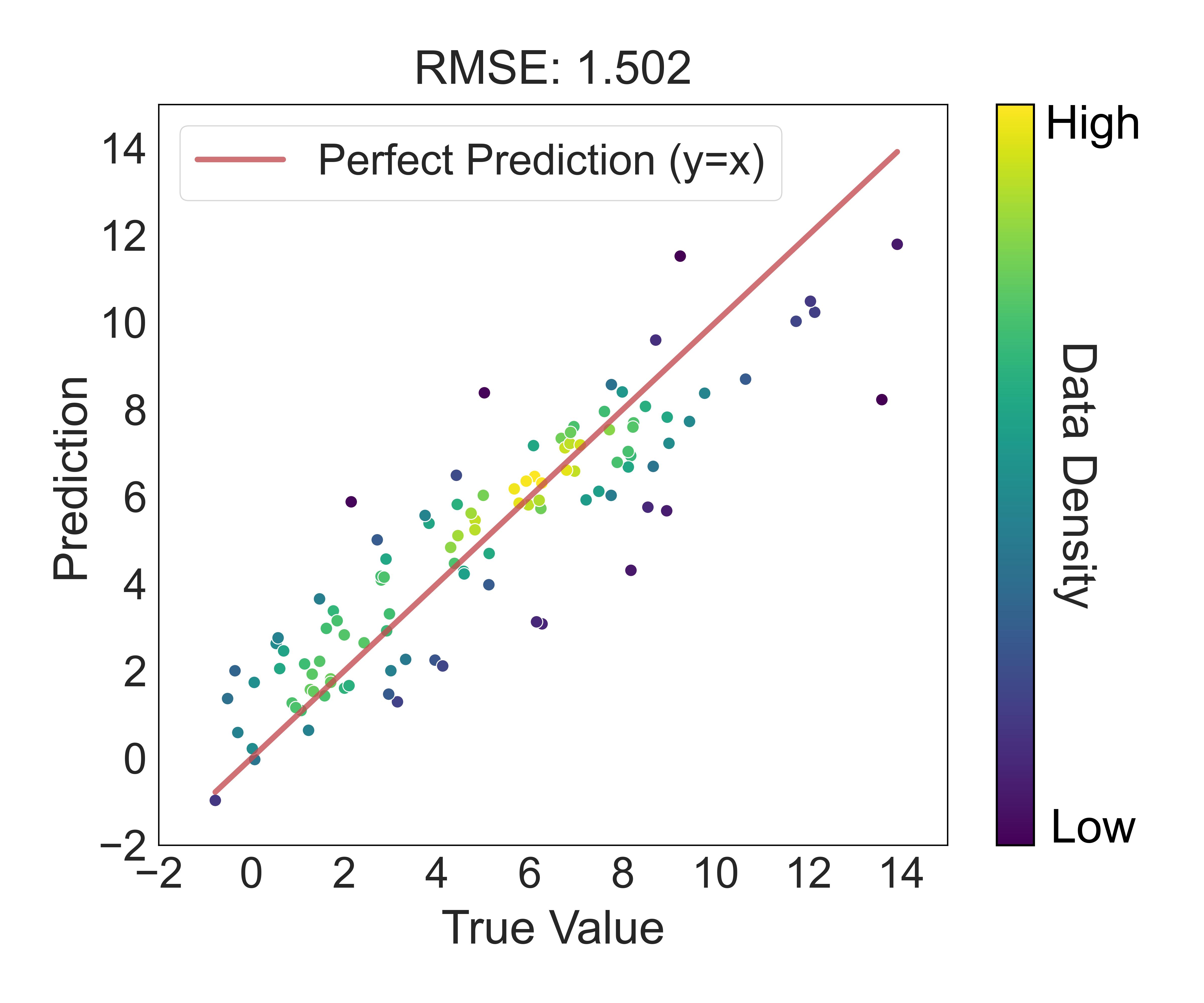}
            \caption{MLP (Morgan)}
        \end{subfigure}
        \begin{subfigure}[t]{0.24\textwidth}
            \centering
            \includegraphics[width=0.99\textwidth]{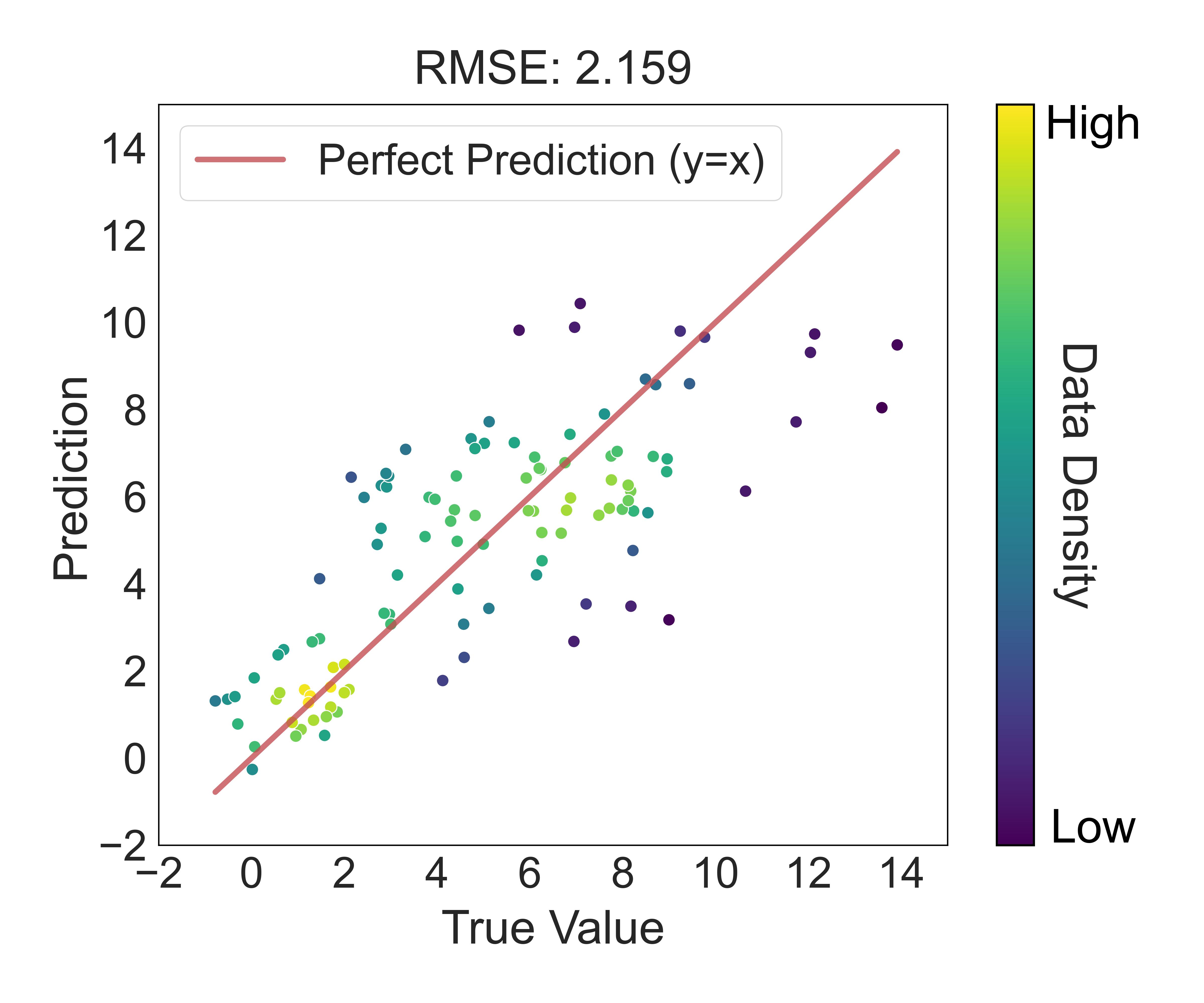}
            \caption{MLP (Expert)}
        \end{subfigure}
        \begin{subfigure}[t]{0.24\textwidth}
            \centering
            \includegraphics[width=0.99\textwidth]{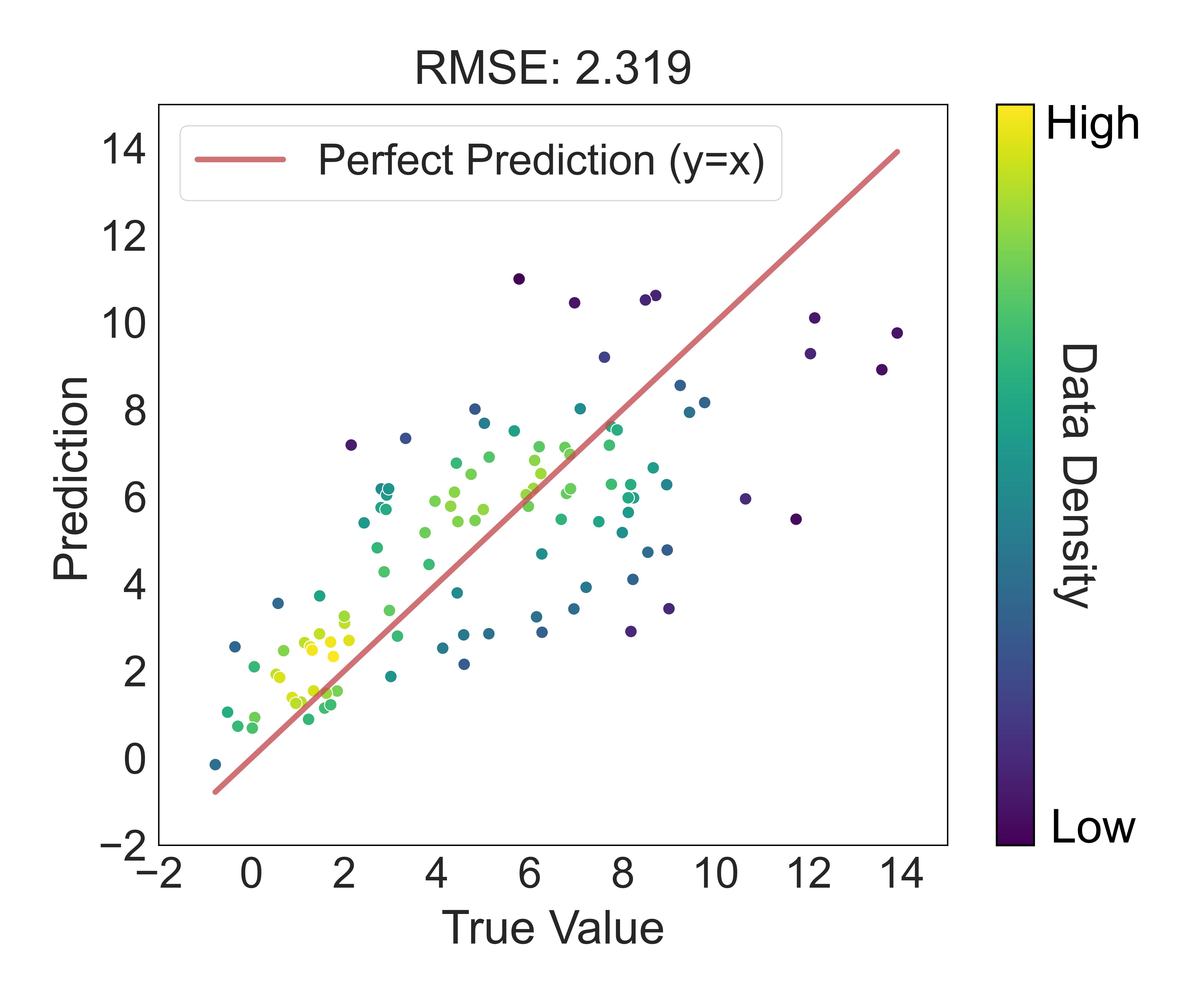}
            \caption{MLP (Grover)}
        \end{subfigure} \\
	\begin{subfigure}[t]{0.24\textwidth}
            \centering
            \includegraphics[width=0.99\textwidth]{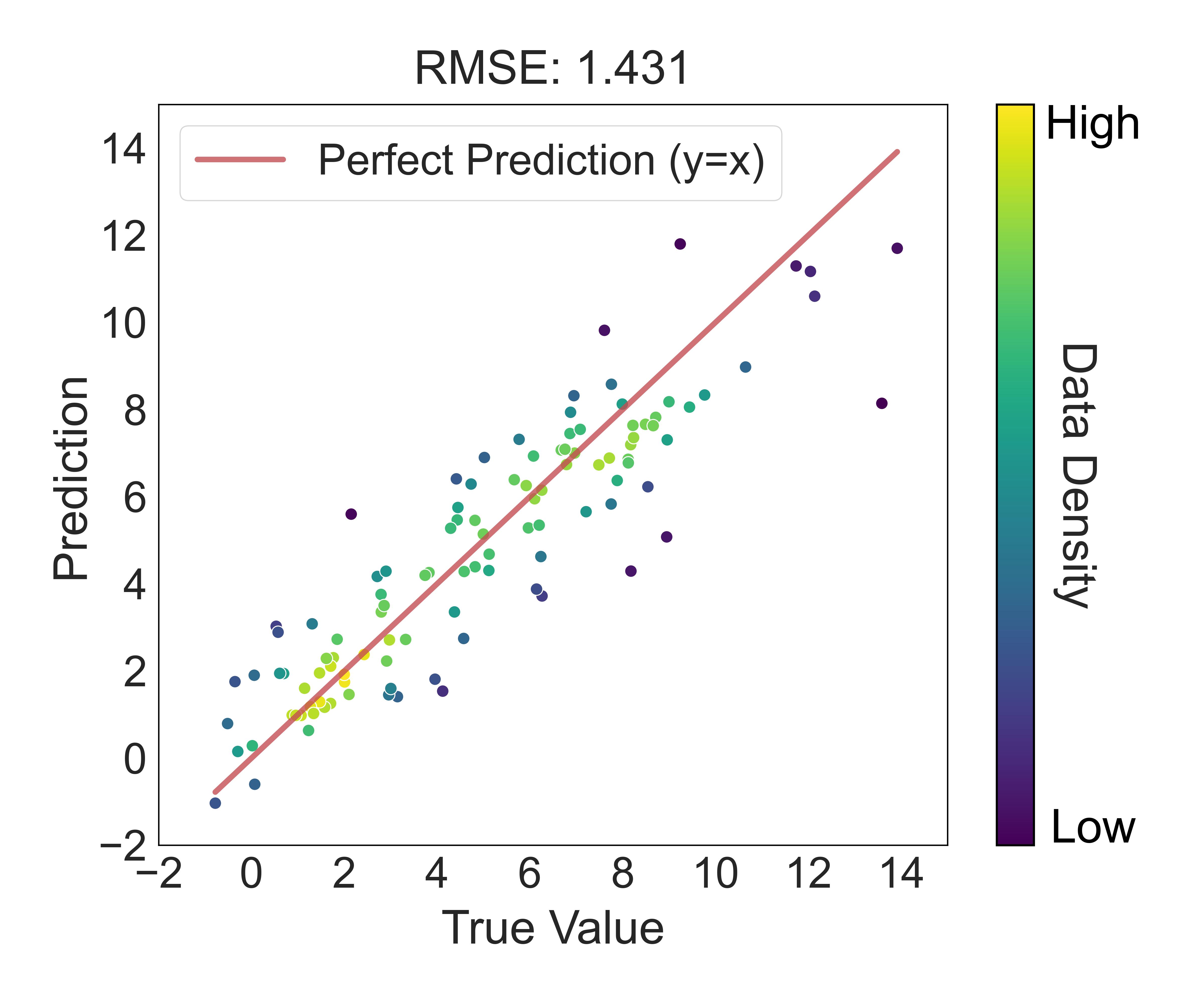}
            \caption{MLP (Morgan + Expert)}
        \end{subfigure}
        \begin{subfigure}[t]{0.24\textwidth}
            \centering
            \includegraphics[width=0.99\textwidth]{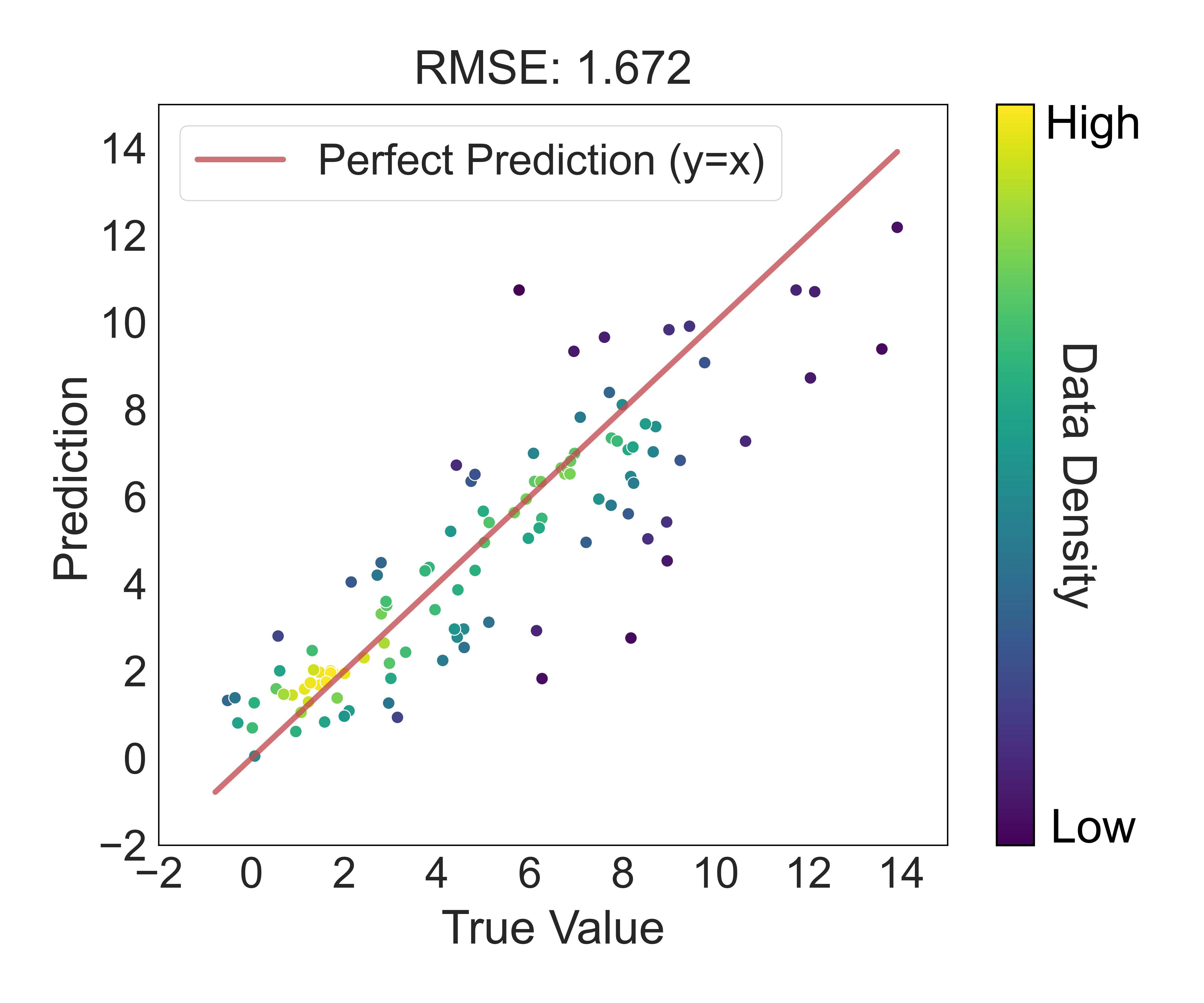}
            \caption{MLP (Morgan + Grover)}
        \end{subfigure}
        \begin{subfigure}[t]{0.24\textwidth}
            \centering
            \includegraphics[width=0.99\textwidth]{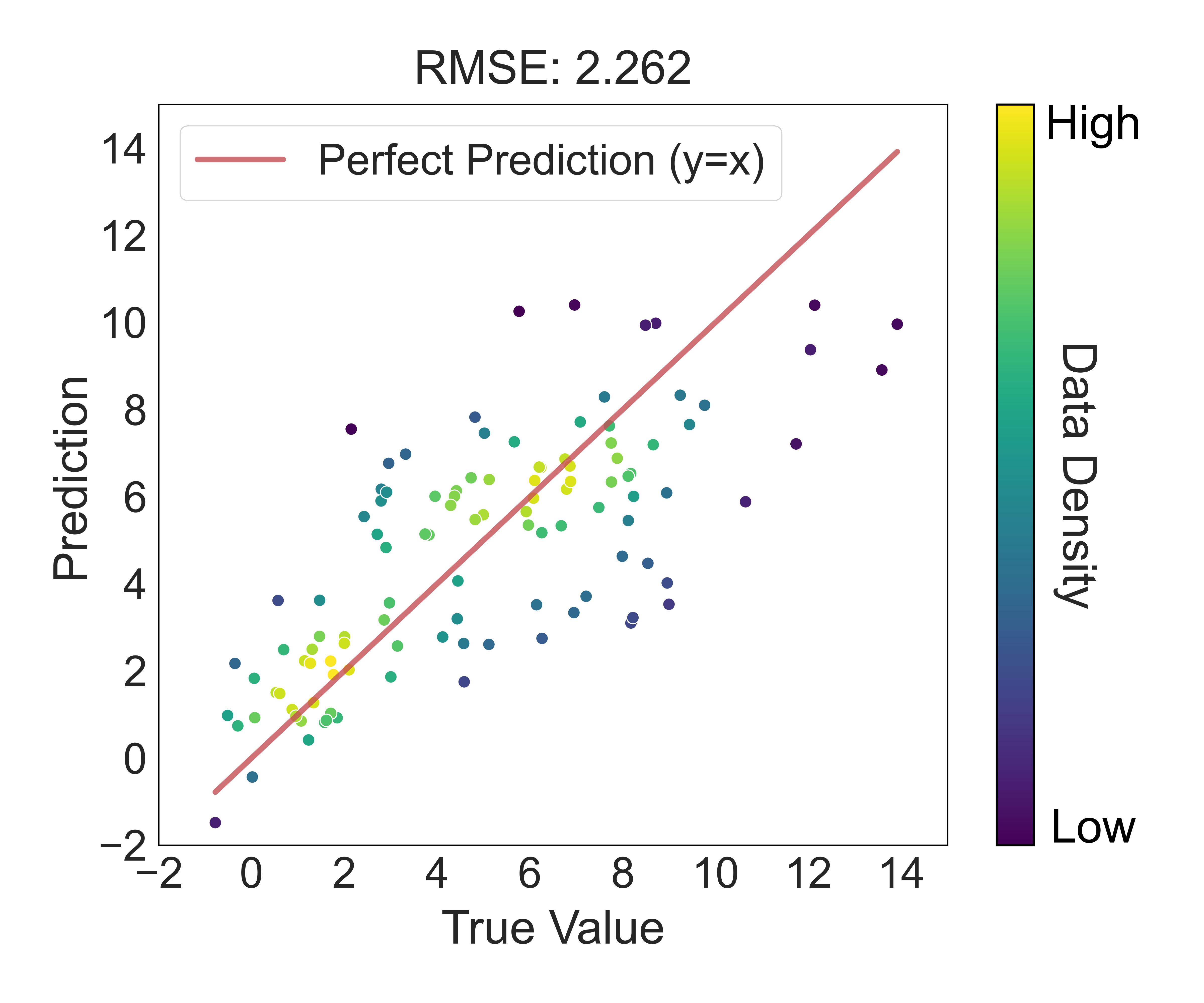}
            \caption{MLP (Expert + Grover)}
        \end{subfigure}
        \begin{subfigure}[t]{0.24\textwidth}
            \centering
            \includegraphics[width=0.99\textwidth]{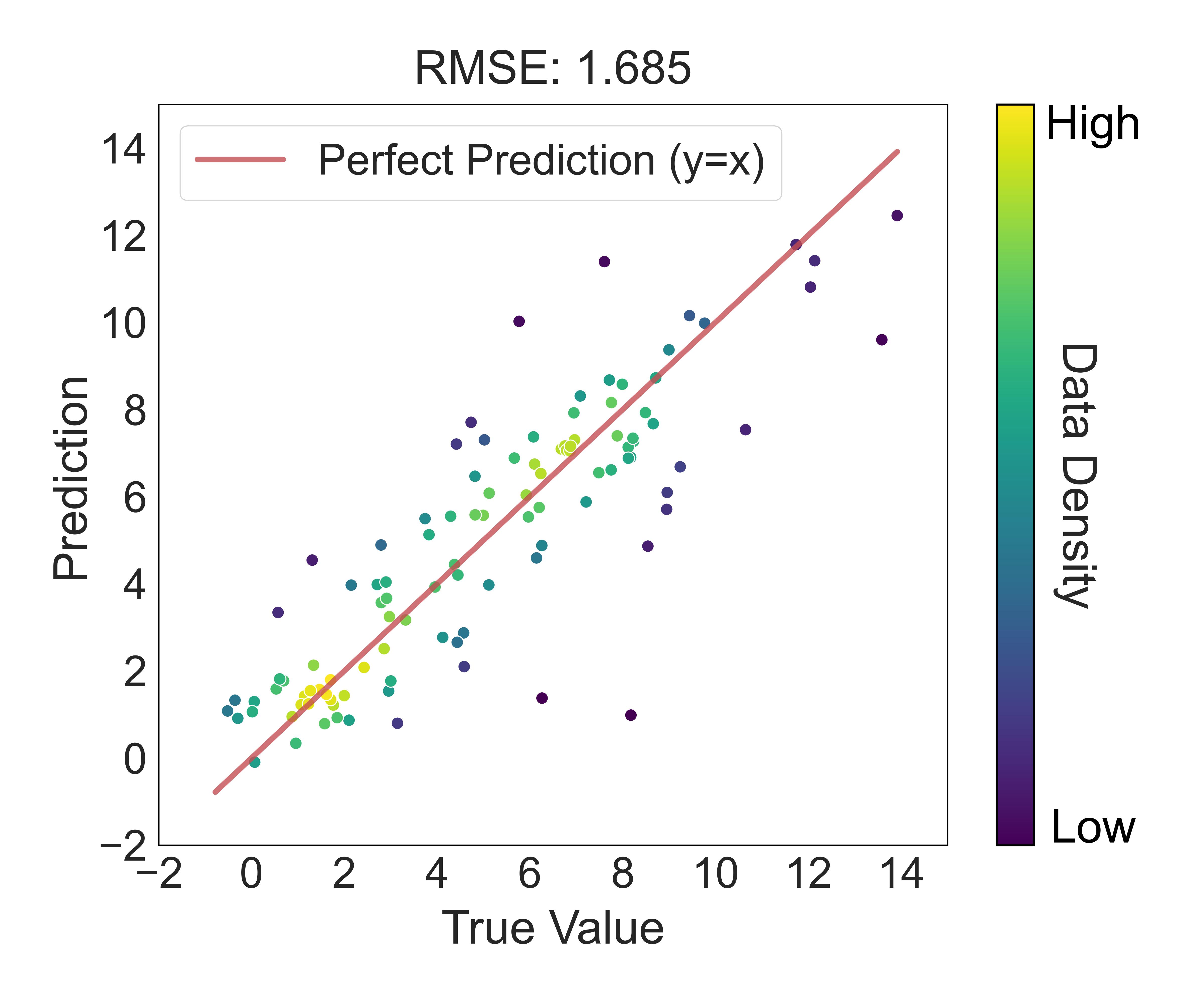}
            \caption{MLP (All Features)}
        \end{subfigure}

        \caption{Prediction vs. true value plots for different feature sets. (a) AGILE model using end-to-end features. (b–h) MLP model trained with different feature sets, including Morgan, Expert, and Grover representations individually and in combination. \textit{Morgan fingerprints} combined with \textit{Expert descriptors} achieve the best performance with the lowest RMSE, demonstrating the advantage of combining domain-specific features with substructure-based Morgan fingerprints.}
	\label{fig:mlp_features}
\end{figure}

\begin{figure}[!htbp]
	\centering
        \begin{subfigure}[t]{0.24\textwidth}
            \centering
            \includegraphics[width=0.99\textwidth]{Figures-supp/Random/AGILE/org.jpg}
            \caption{AGILE (End-to-End)}
        \end{subfigure}
	\begin{subfigure}[t]{0.24\textwidth}
            \centering
            \includegraphics[width=0.99\textwidth]{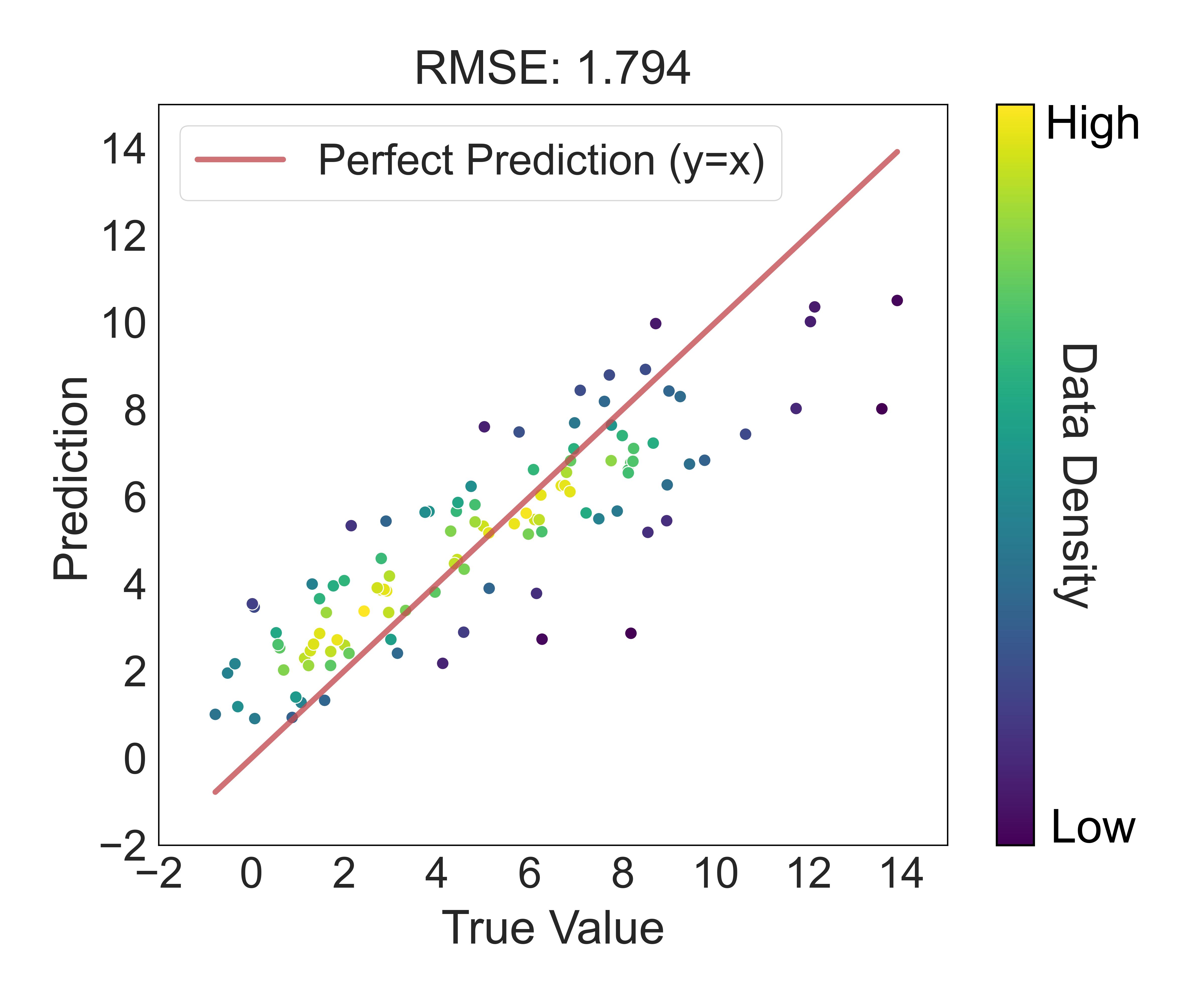}
            \caption{SVR (Morgan)}
        \end{subfigure}
        \begin{subfigure}[t]{0.24\textwidth}
            \centering
            \includegraphics[width=0.99\textwidth]{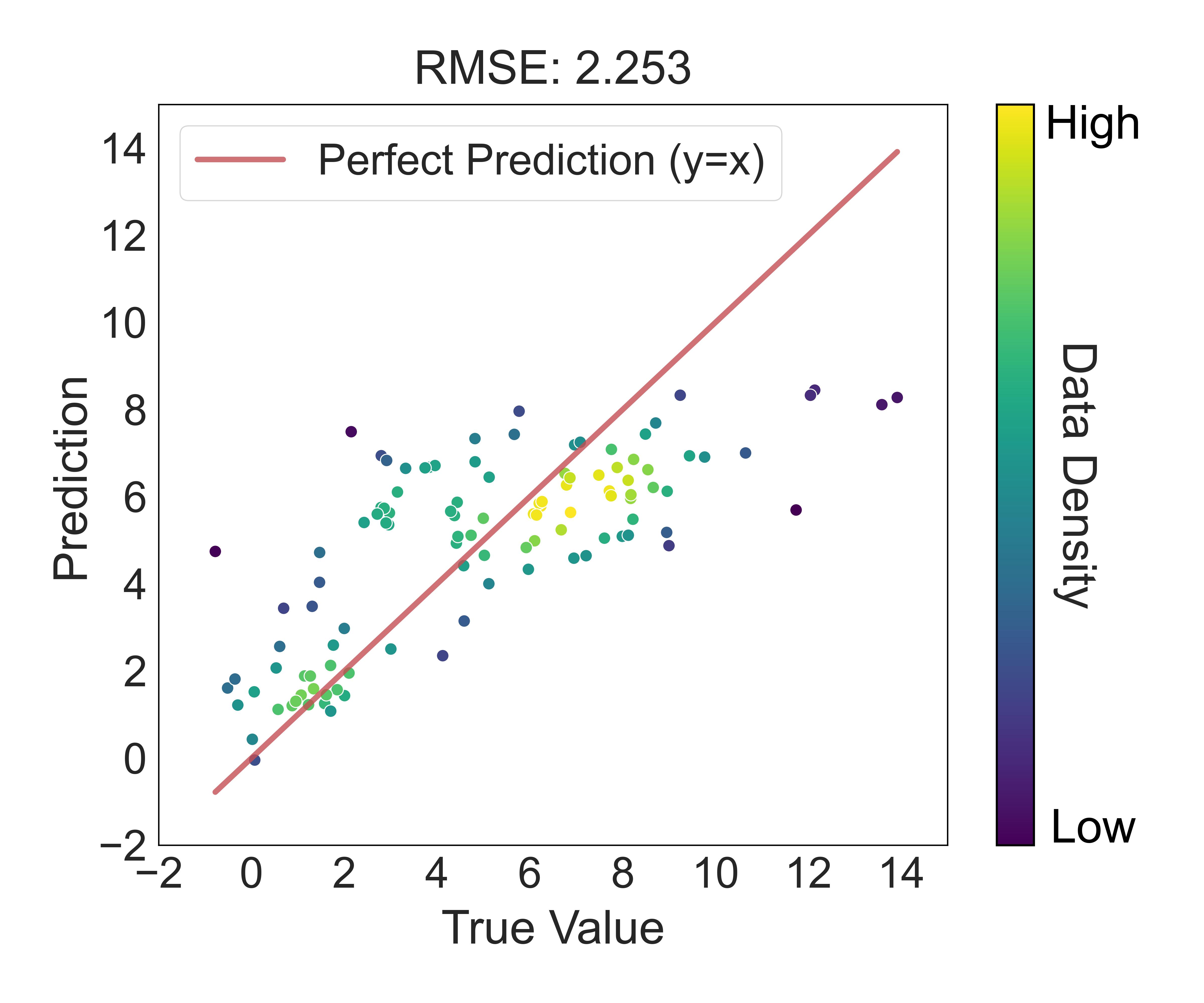}
            \caption{SVR (Expert)}
        \end{subfigure}
        \begin{subfigure}[t]{0.24\textwidth}
            \centering
            \includegraphics[width=0.99\textwidth]{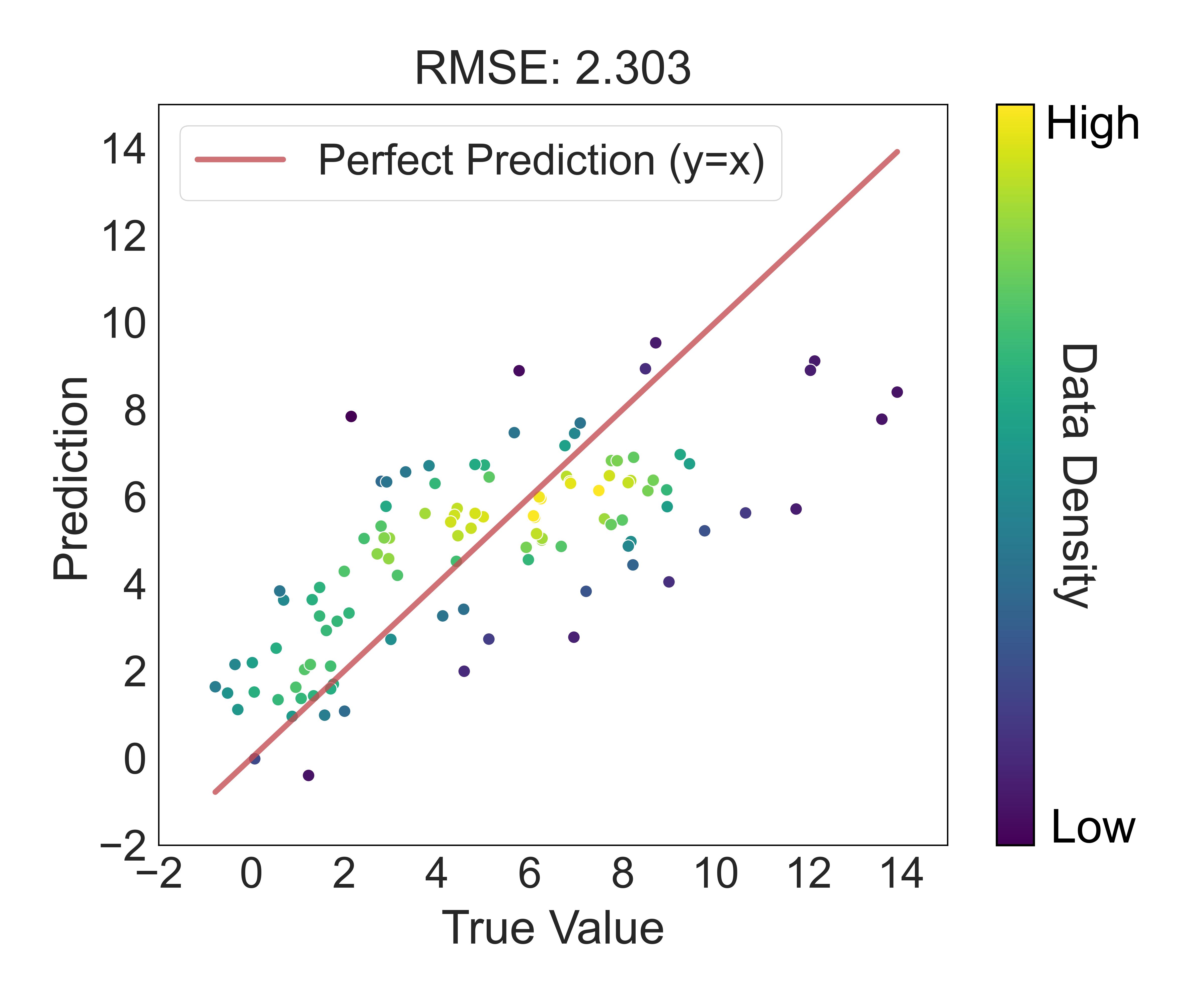}
            \caption{SVR (Grover)}
        \end{subfigure} \\
	\begin{subfigure}[t]{0.24\textwidth}
            \centering
            \includegraphics[width=0.99\textwidth]{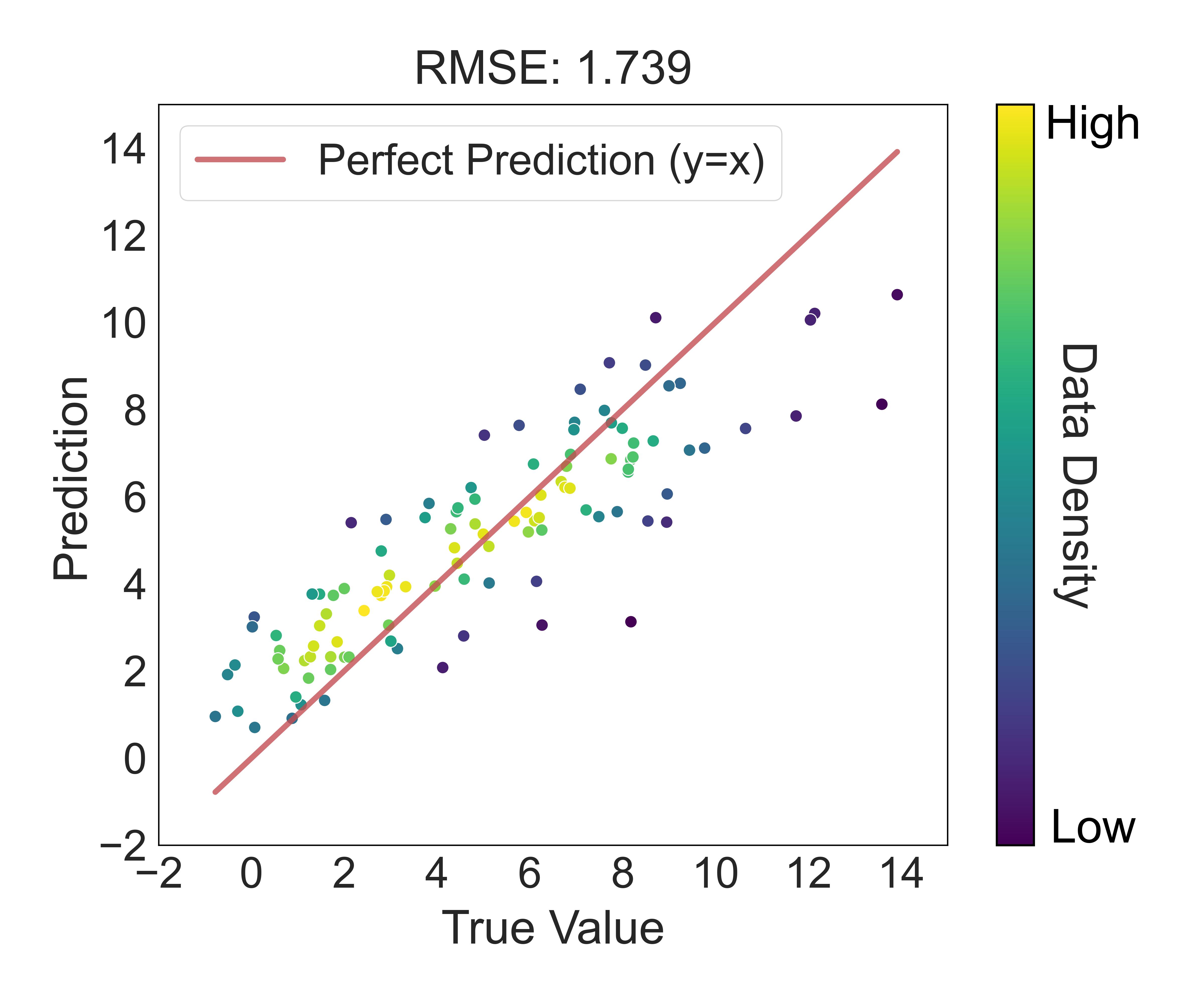}
            \caption{SVR (Morgan + Expert)}
        \end{subfigure}
        \begin{subfigure}[t]{0.24\textwidth}
            \centering
            \includegraphics[width=0.99\textwidth]{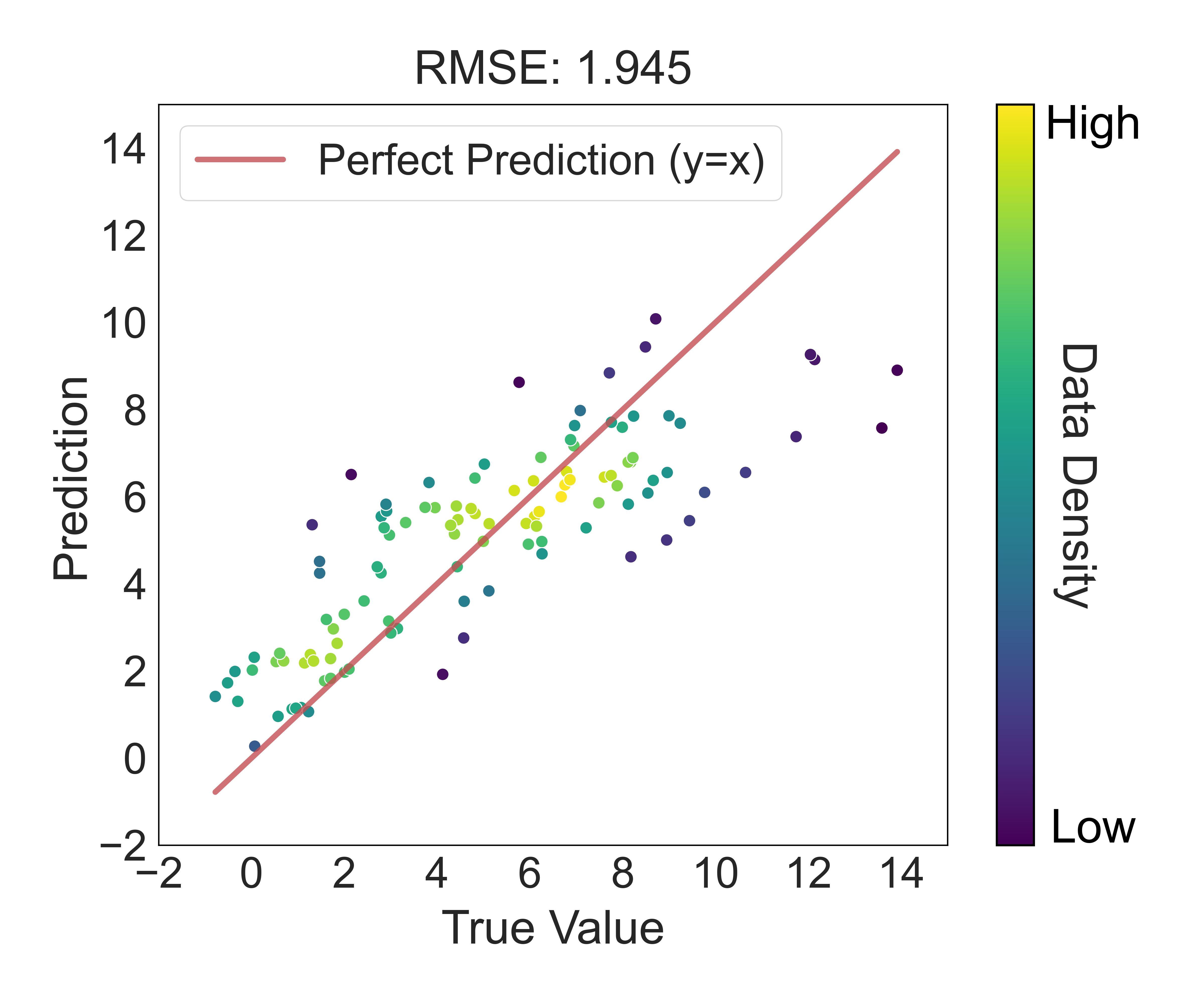}
            \caption{SVR (Morgan + Grover)}
        \end{subfigure}
        \begin{subfigure}[t]{0.24\textwidth}
            \centering
            \includegraphics[width=0.99\textwidth]{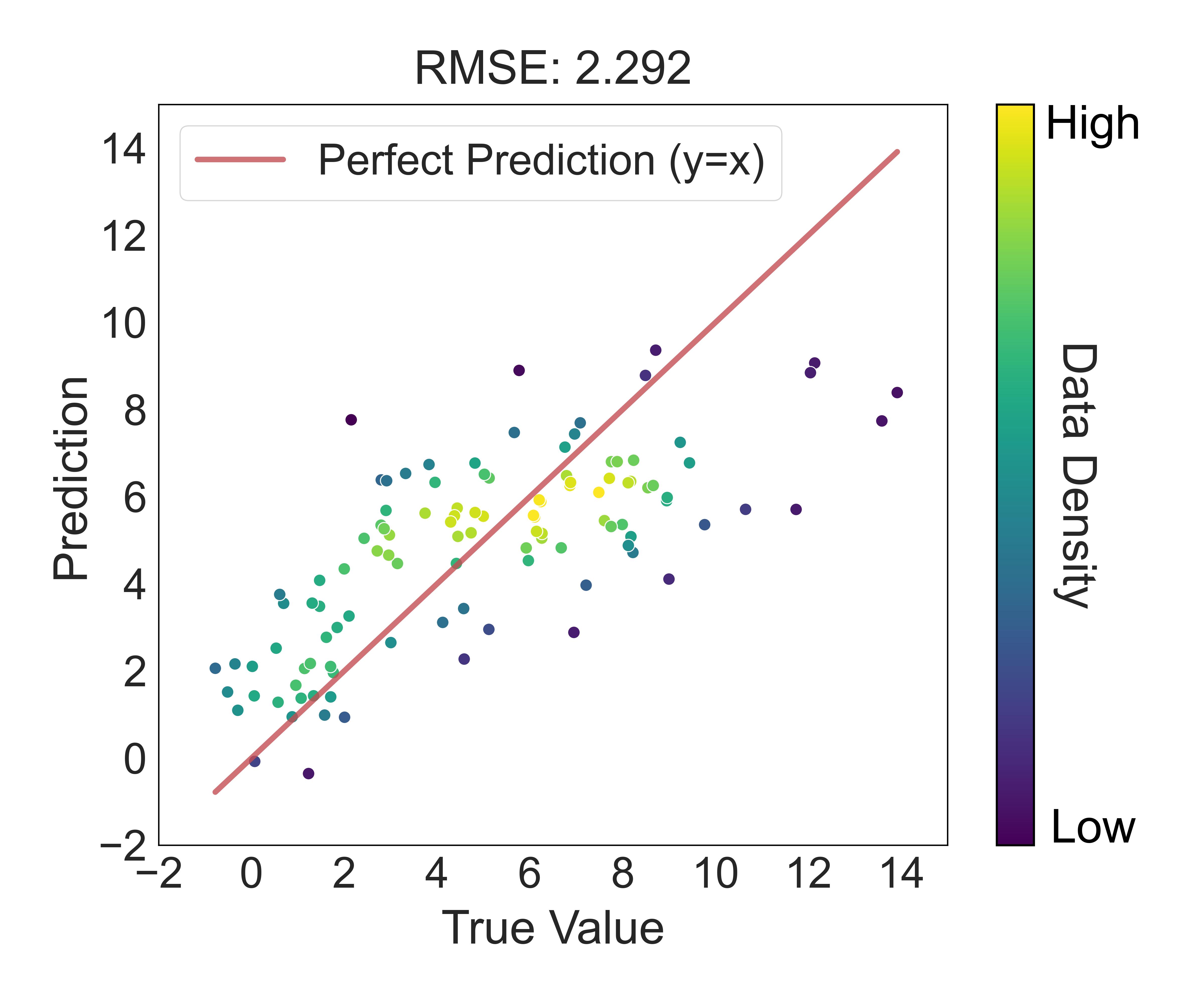}
            \caption{SVR (Expert + Grover)}
        \end{subfigure}
        \begin{subfigure}[t]{0.24\textwidth}
            \centering
            \includegraphics[width=0.99\textwidth]{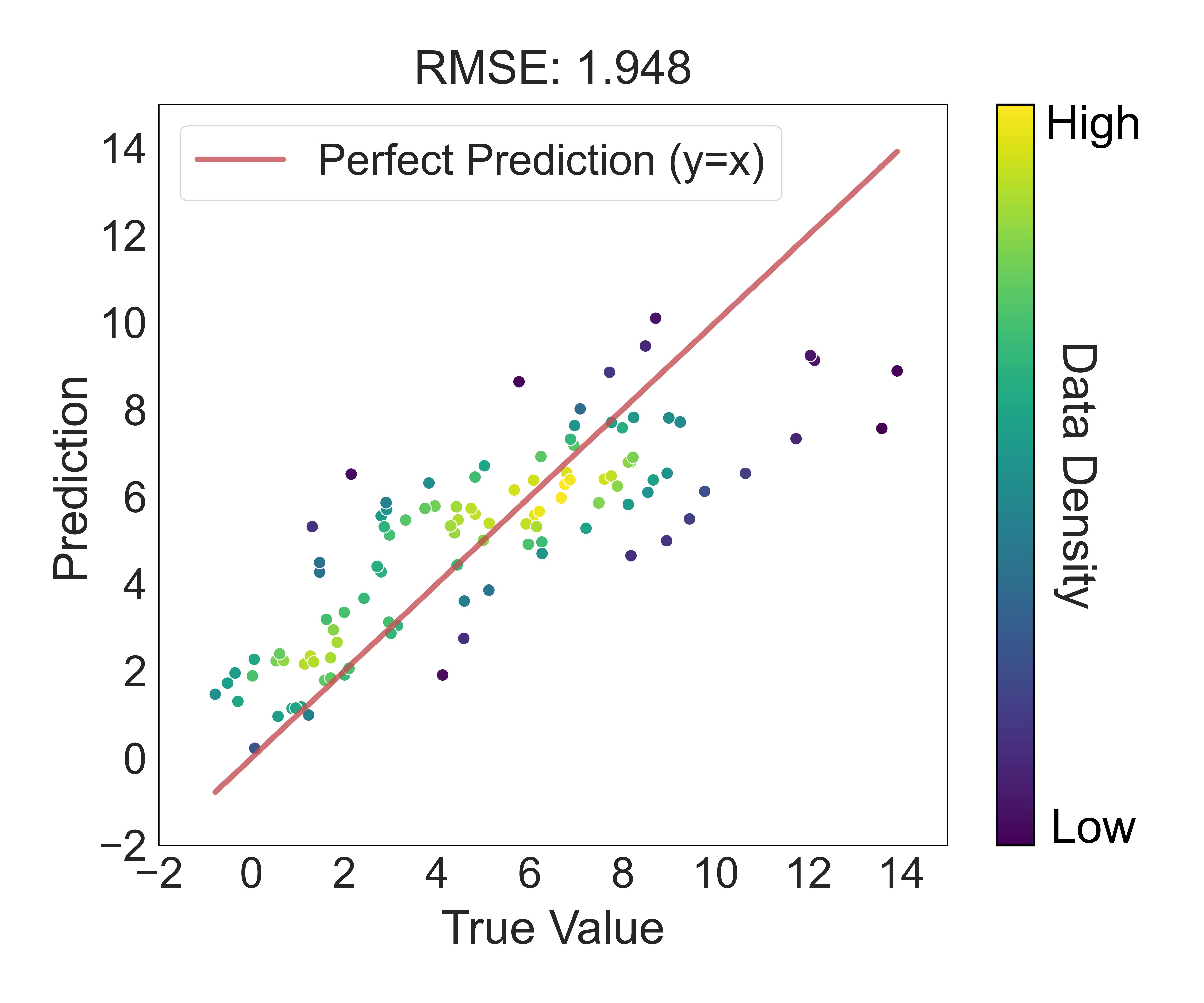}
            \caption{SVR (All Features)}
        \end{subfigure}

        \caption{Prediction vs. true value plots for different feature sets. (a) AGILE model using end-to-end features. (b–h) SVR model trained with different feature sets, including Morgan, Expert, and Grover representations individually and in combination. \textit{Morgan fingerprints} combined with \textit{Expert descriptors} yield the best performance with the lowest RMSE, reinforcing the effectiveness of integrating domain-specific knowledge with substructure-based representations.}
	\label{fig:svr_features}
\end{figure}

\begin{figure}[!htbp]
	\centering
        \begin{subfigure}[t]{0.24\textwidth}
            \centering
            \includegraphics[width=0.99\textwidth]{Figures-supp/Random/AGILE/org.jpg}
            \caption{AGILE (End-to-End)}
        \end{subfigure}
	\begin{subfigure}[t]{0.24\textwidth}
            \centering
            \includegraphics[width=0.99\textwidth]{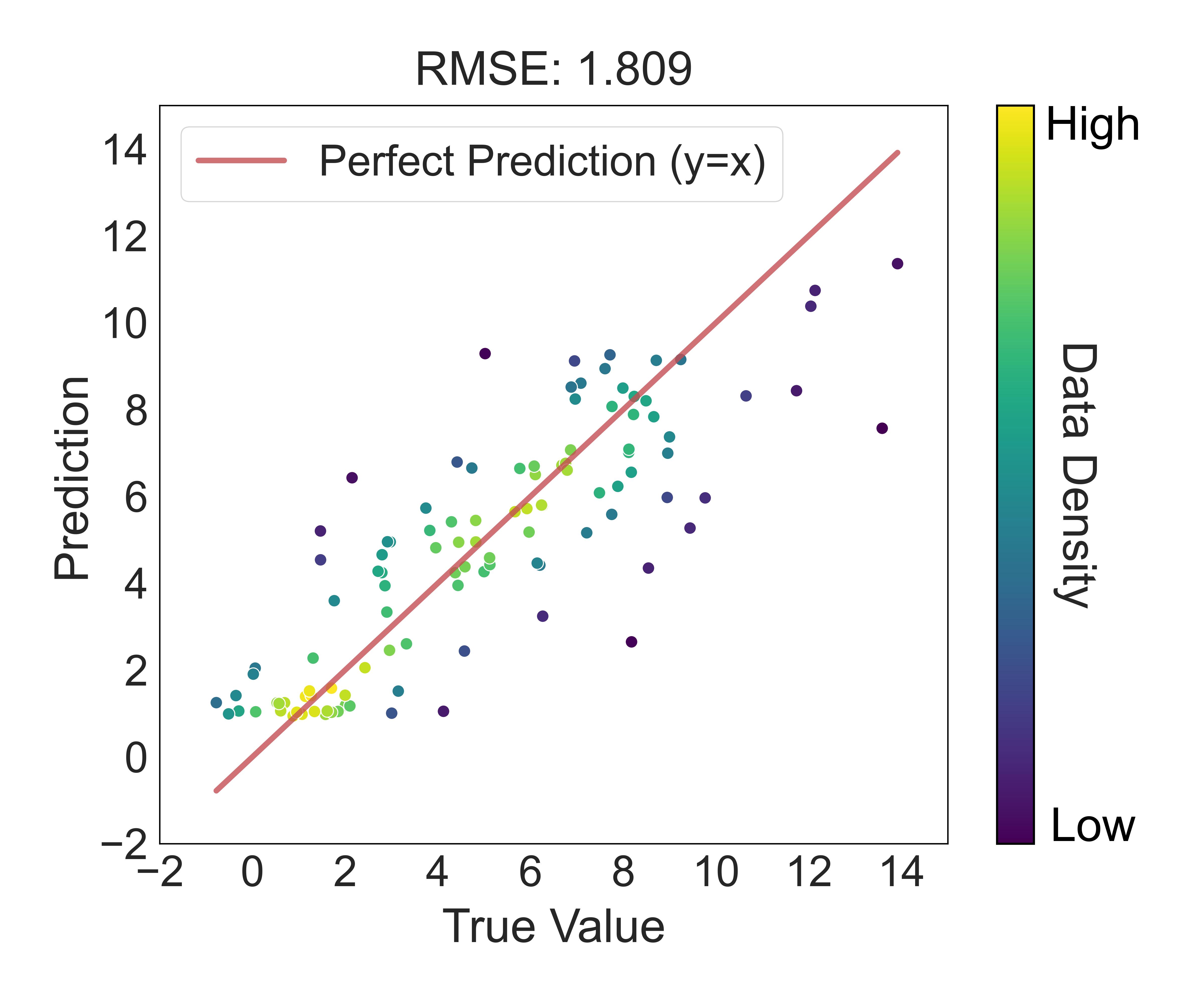}
            \caption{AGILE (Morgan)}
        \end{subfigure}
        \begin{subfigure}[t]{0.24\textwidth}
            \centering
            \includegraphics[width=0.99\textwidth]{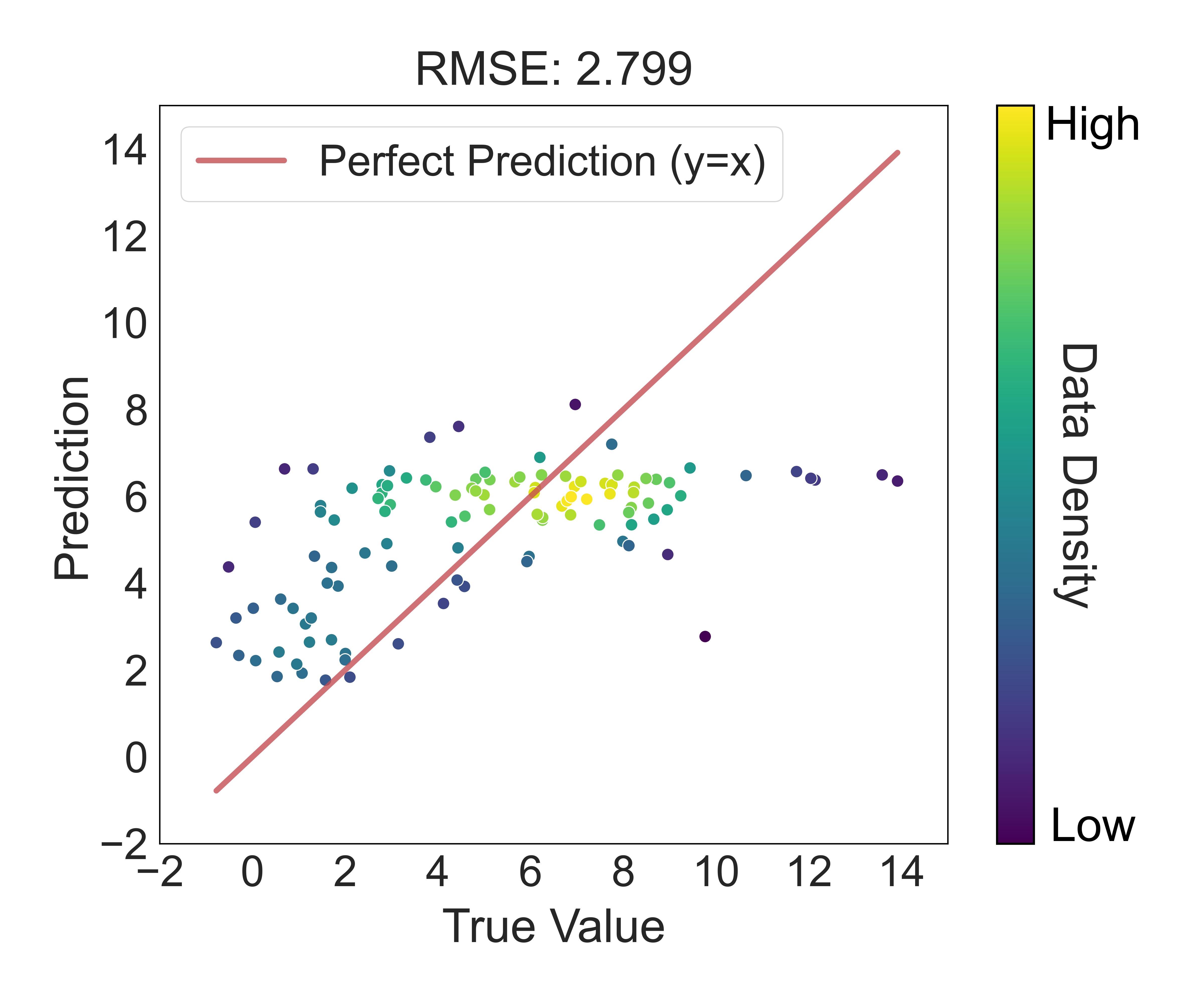}
            \caption{AGILE (Expert)}
        \end{subfigure}
        \begin{subfigure}[t]{0.24\textwidth}
            \centering
            \includegraphics[width=0.99\textwidth]{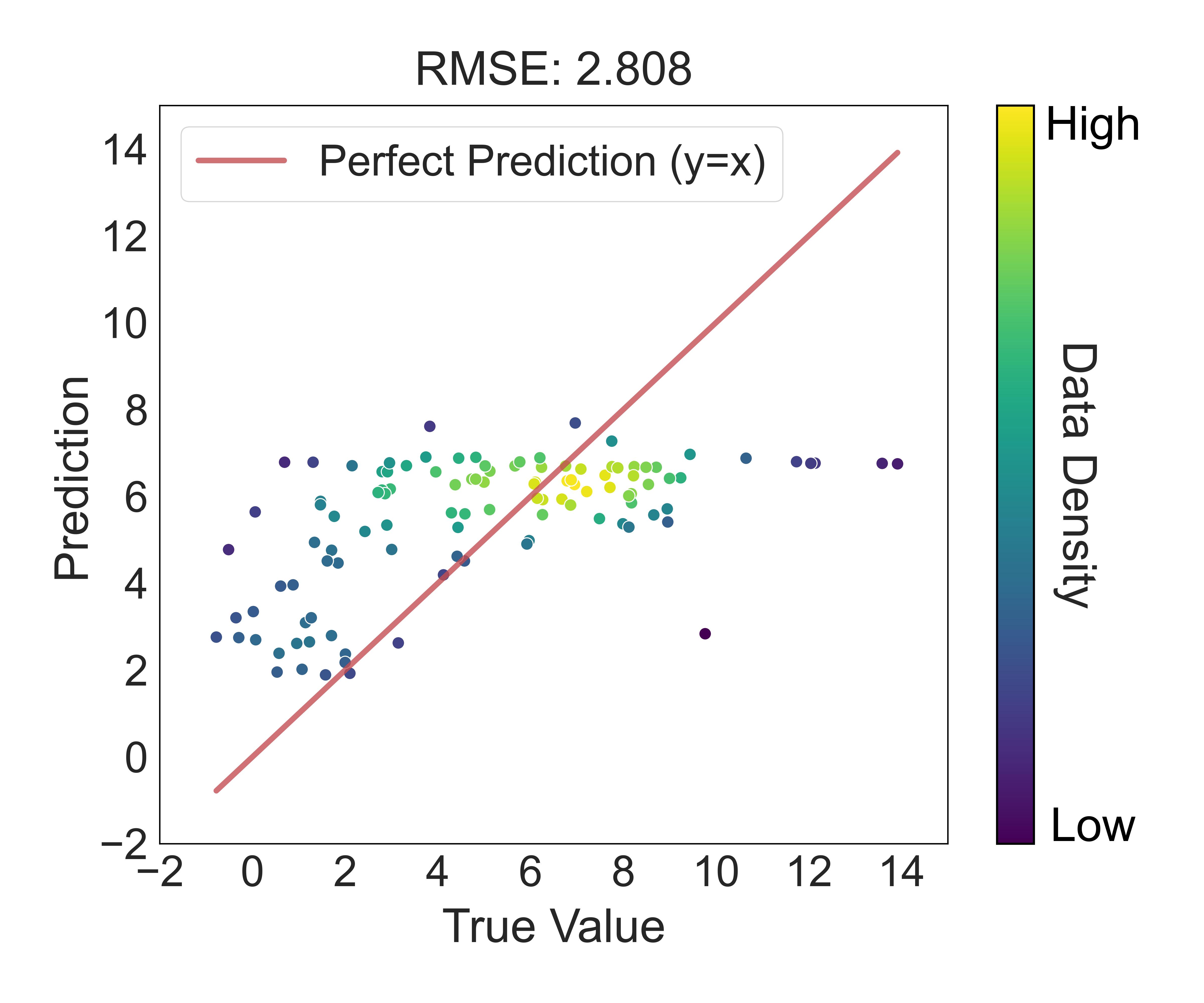}
            \caption{AGILE (Grover)}
        \end{subfigure} \\
	\begin{subfigure}[t]{0.24\textwidth}
            \centering
            \includegraphics[width=0.99\textwidth]{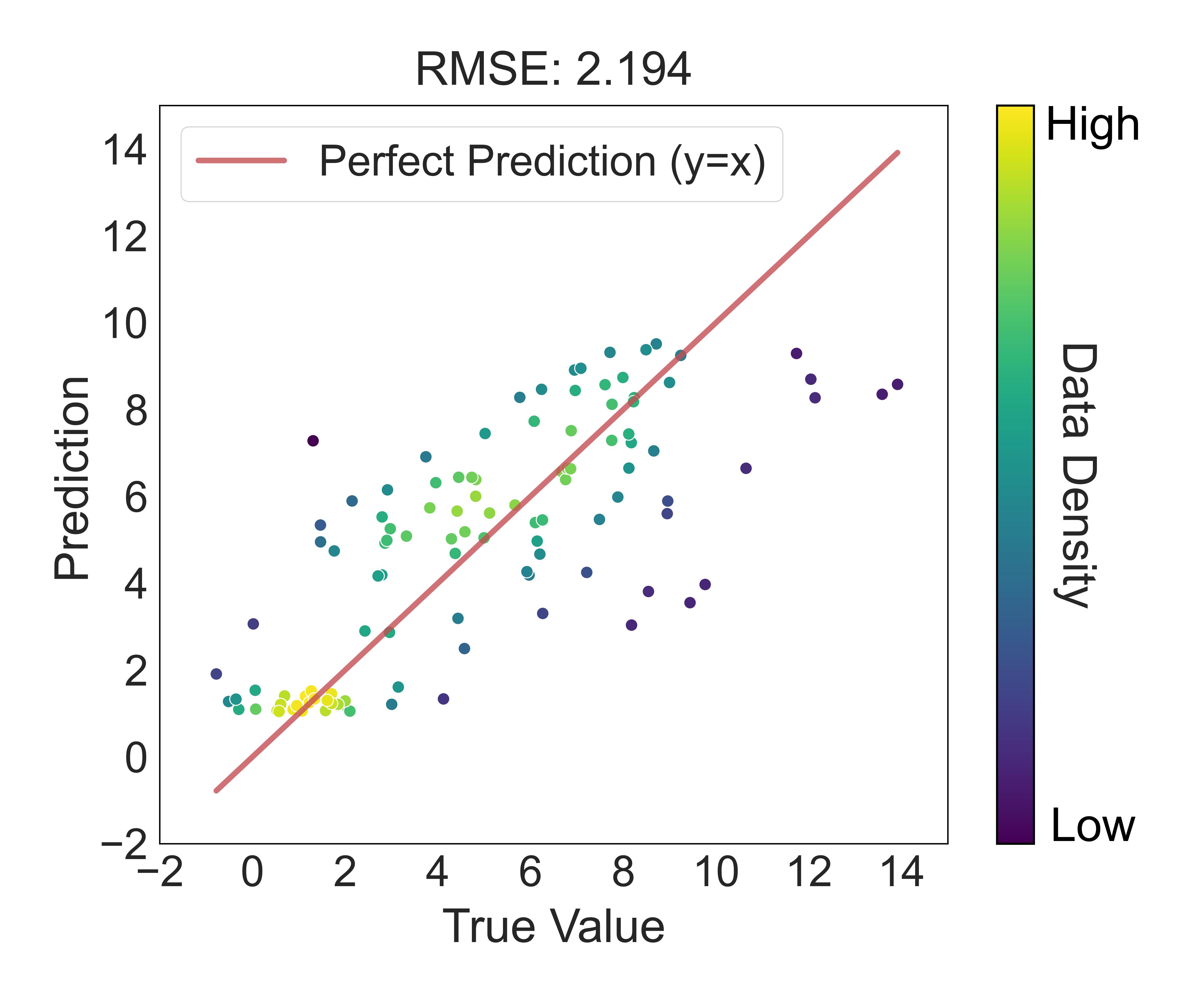}
            \caption{AGILE (Morgan + Expert)}
        \end{subfigure}
        \begin{subfigure}[t]{0.24\textwidth}
            \centering
            \includegraphics[width=0.99\textwidth]{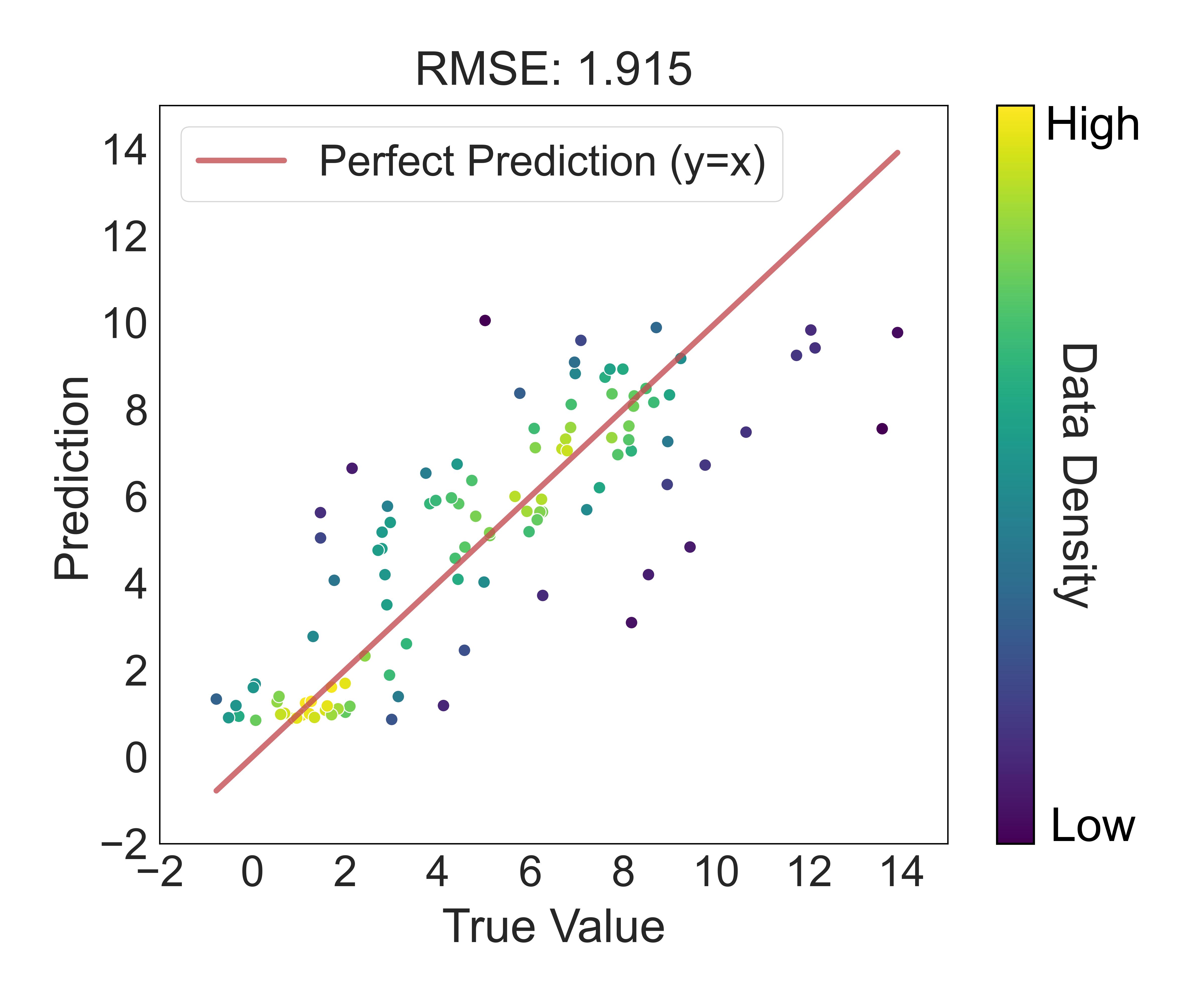}
            \caption{AGILE (Morgan + Grover)}
        \end{subfigure}
        \begin{subfigure}[t]{0.24\textwidth}
            \centering
            \includegraphics[width=0.99\textwidth]{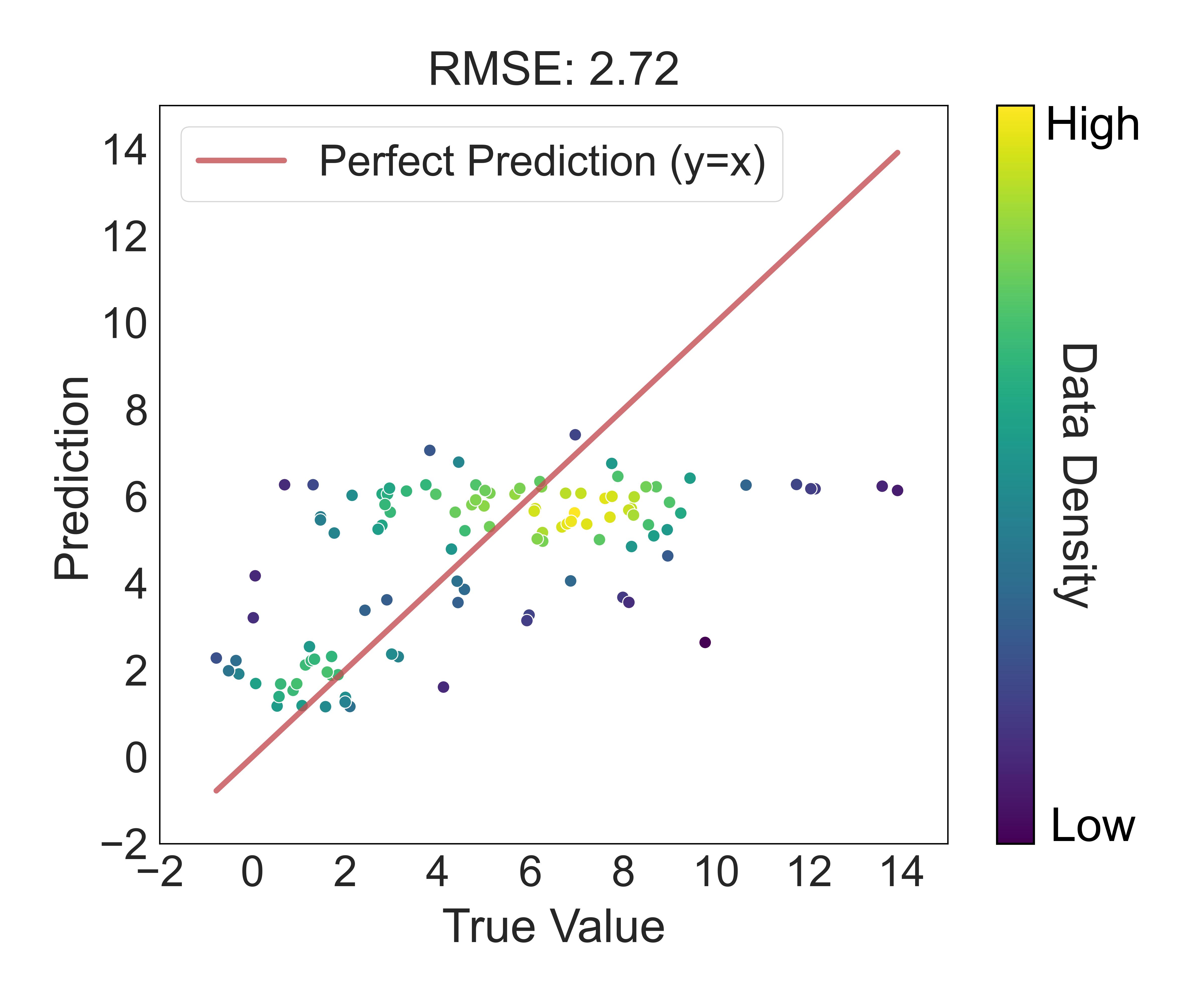}
            \caption{AGILE (Expert + Grover)}
        \end{subfigure}
        \begin{subfigure}[t]{0.24\textwidth}
            \centering
            \includegraphics[width=0.99\textwidth]{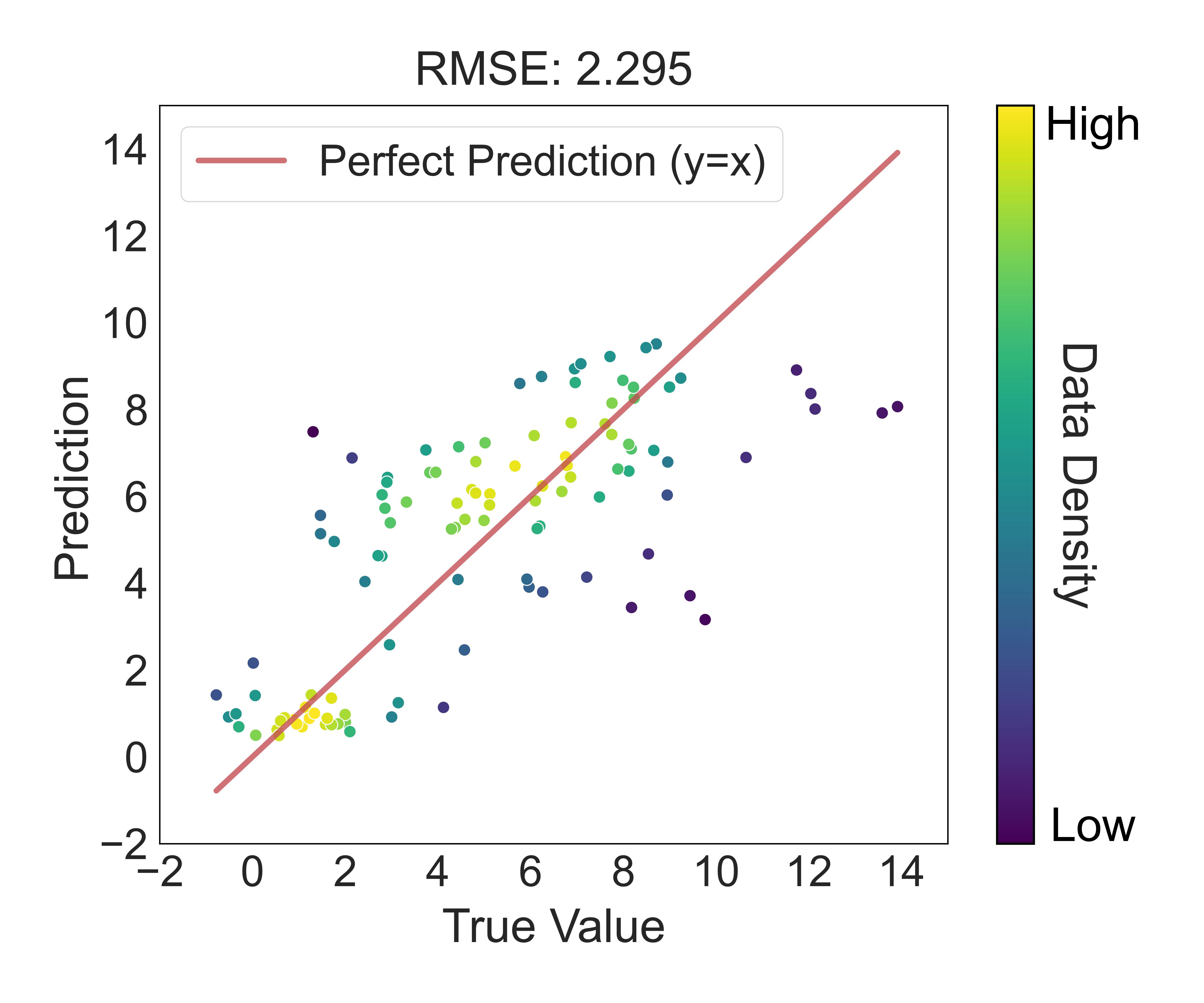}
            \caption{AGILE (All Features)}
        \end{subfigure}

        \caption{Prediction vs. true value plots for different feature sets in AGILE. (a) AGILE model using only its end-to-end learned features. (b–h) AGILE model incorporating additional feature sets alongside its end-to-end representations, including Morgan, Expert, and Grover embeddings individually and in various combinations. \textit{Morgan fingerprints} provide the most significant improvement in predictive accuracy, demonstrating their effectiveness in enhancing AGILE’s performance.
}
	\label{fig:agile_features}
\end{figure}

\begin{figure}[!htbp]
	\centering
        \begin{subfigure}[t]{0.24\textwidth}
            \centering
            \includegraphics[width=0.99\textwidth]{Figures-supp/Random/MLP/circular-expert.jpg}
            \caption{MLP}
        \end{subfigure}
	\begin{subfigure}[t]{0.24\textwidth}
            \centering
            \includegraphics[width=0.99\textwidth]{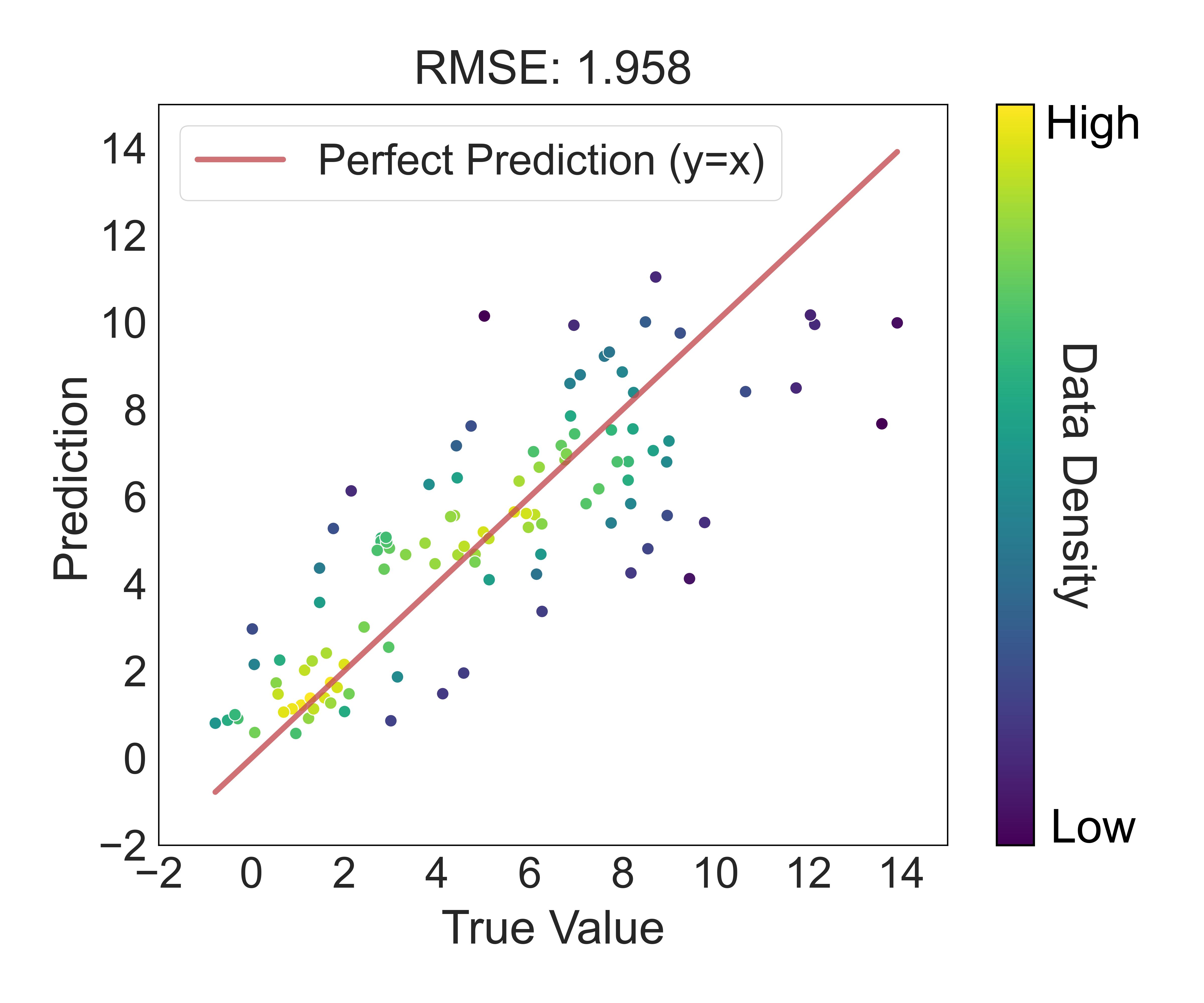}
            \caption{Transformer}
        \end{subfigure}
        \begin{subfigure}[t]{0.24\textwidth}
            \centering
            \includegraphics[width=0.99\textwidth]{Figures-supp/Random/SVR/circular-expert.jpg}
            \caption{\centering SVR}
        \end{subfigure}
        \begin{subfigure}[t]{0.24\textwidth}
            \centering
            \includegraphics[width=0.99\textwidth]{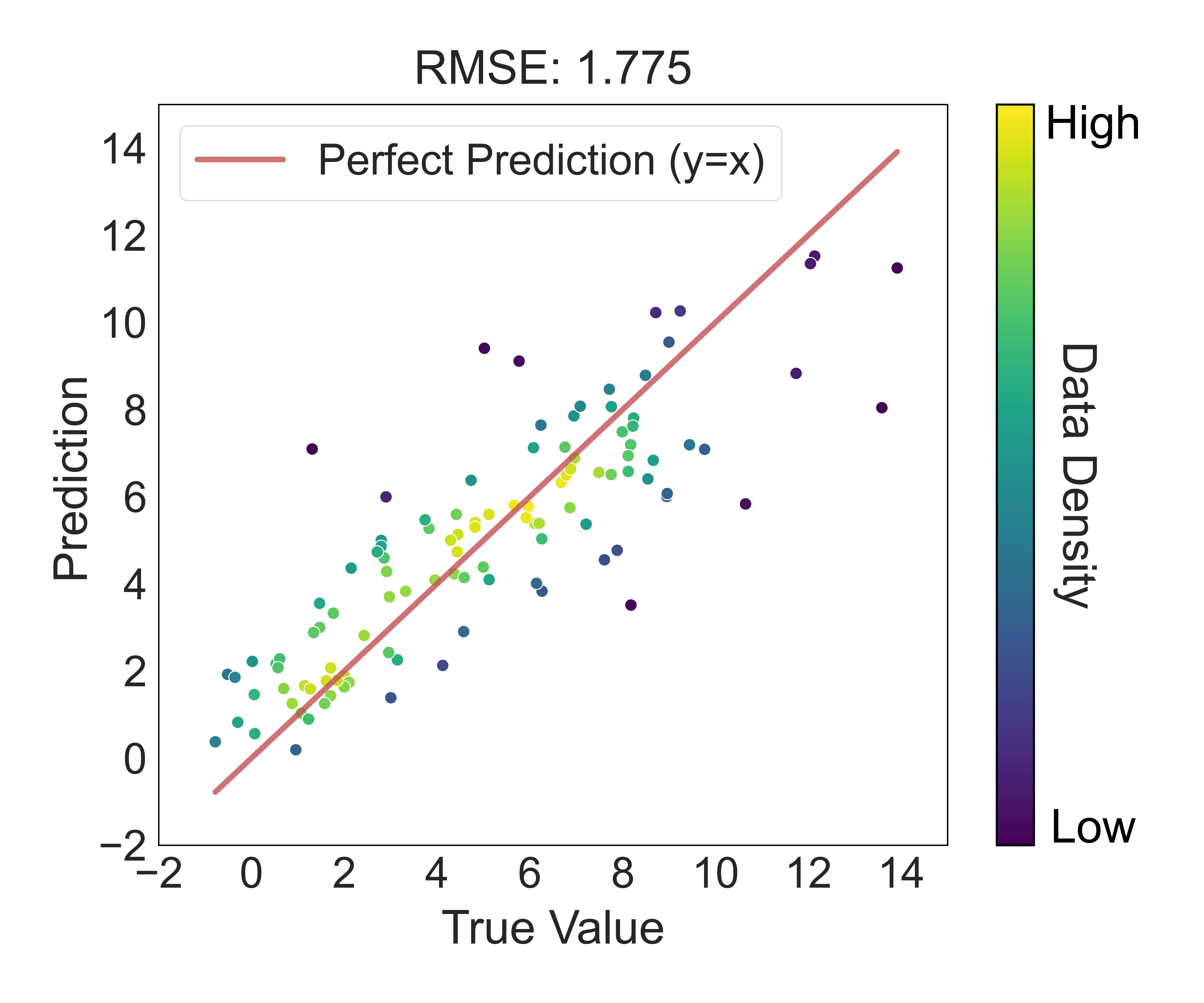}
            \caption{\centering RF}
        \end{subfigure} \\
	\begin{subfigure}[t]{0.24\textwidth}
            \centering
            \includegraphics[width=0.99\textwidth]{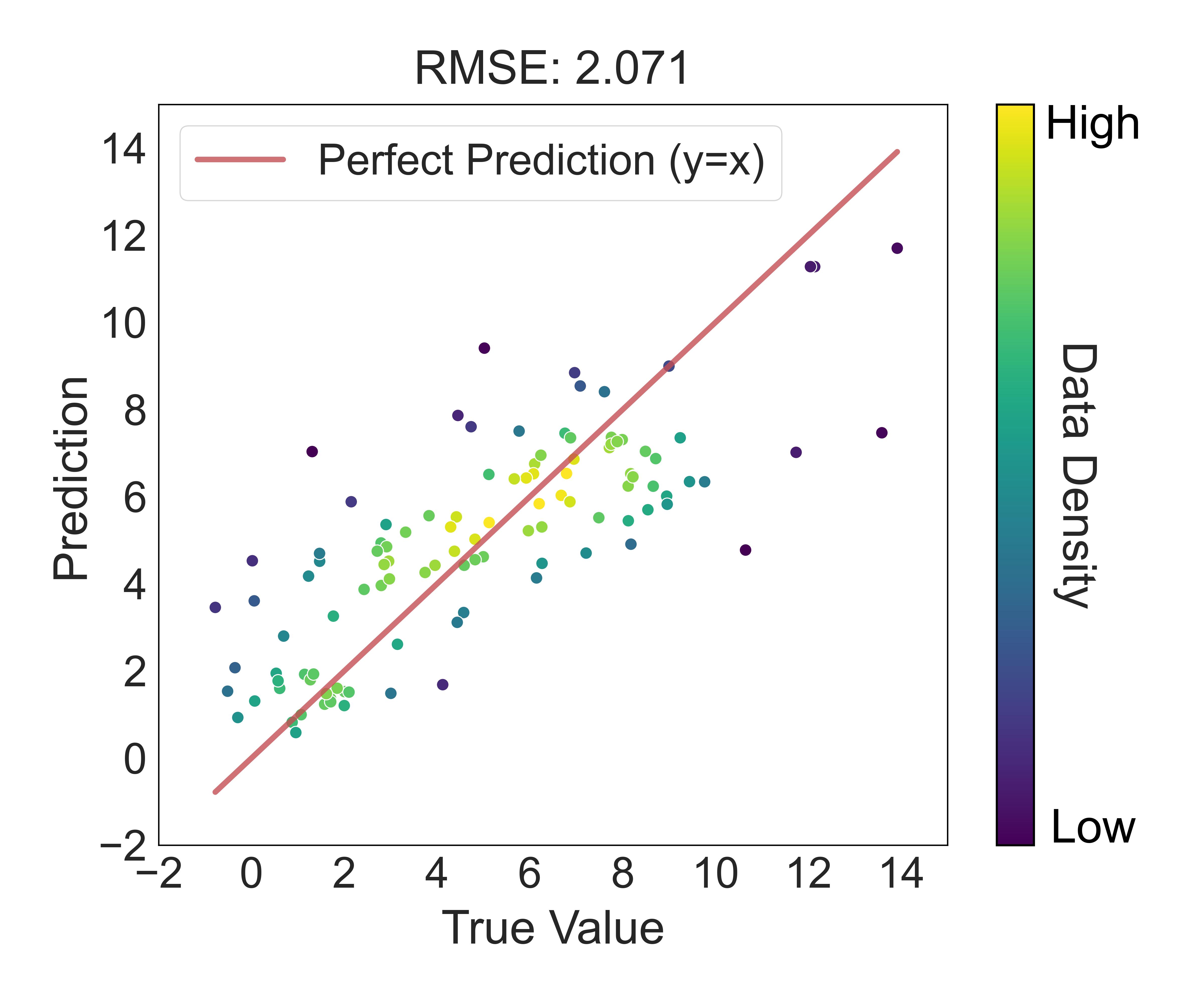}
            \caption{kNN}
        \end{subfigure}
        \begin{subfigure}[t]{0.24\textwidth}
            \centering
            \includegraphics[width=0.99\textwidth]{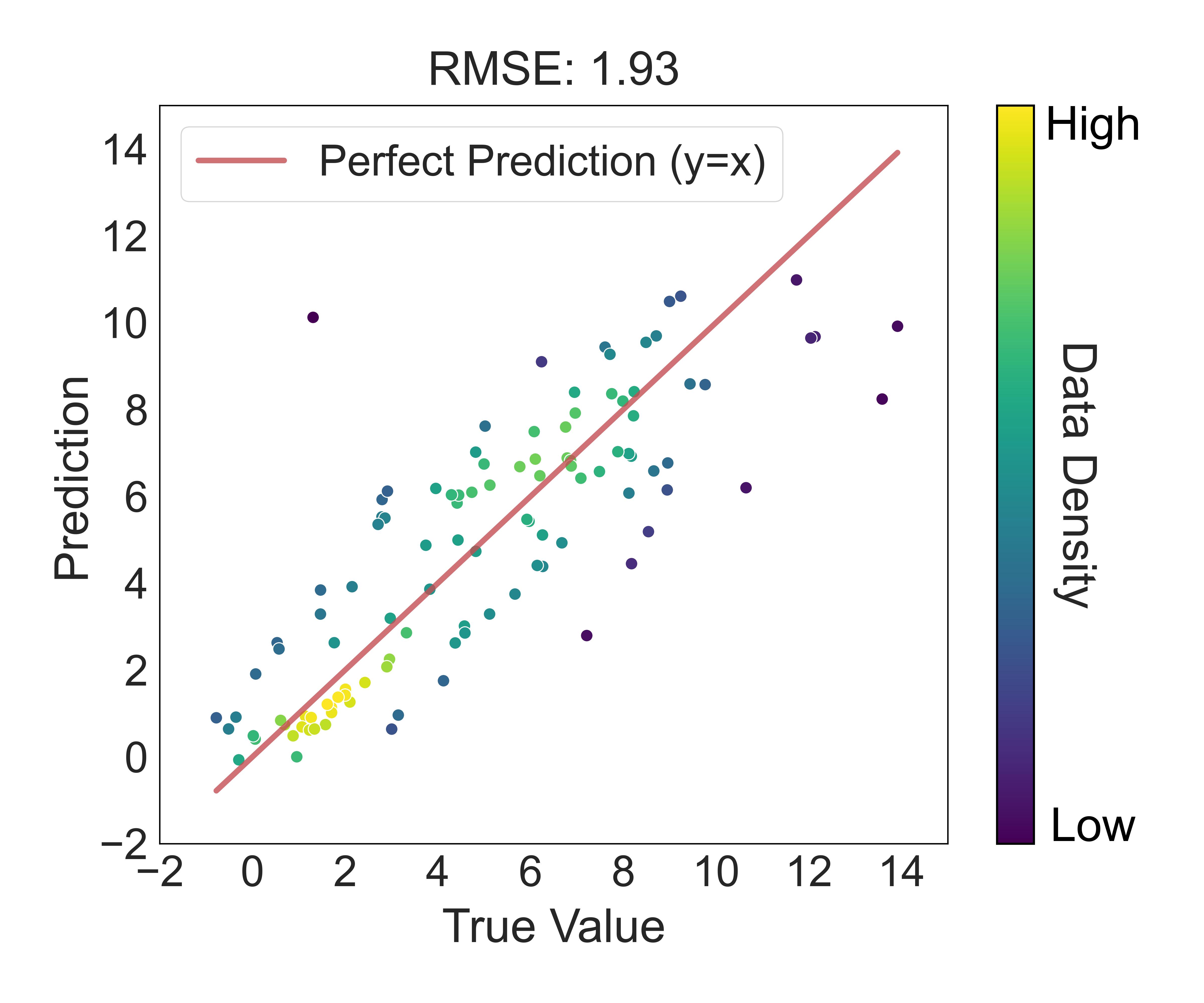}
            \caption{KPGT}
        \end{subfigure}
        \begin{subfigure}[t]{0.24\textwidth}
            \centering
            \includegraphics[width=0.99\textwidth]{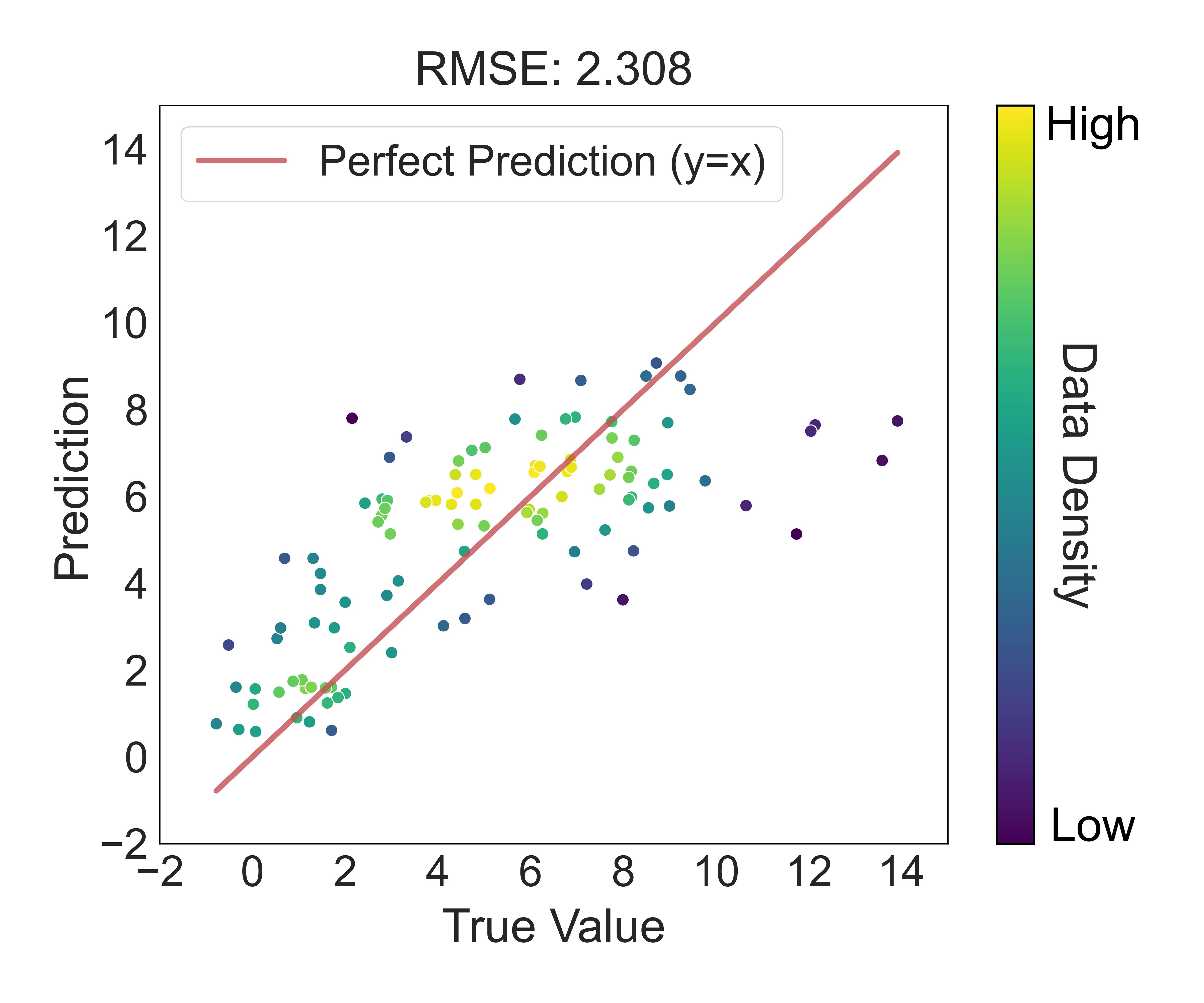}
            \caption{\centering Chemprop}
        \end{subfigure}
        \begin{subfigure}[t]{0.24\textwidth}
            \centering
            \includegraphics[width=0.99\textwidth]{Figures-supp/Random/AGILE/org.jpg}
            \caption{\centering AGILE}
        \end{subfigure}

        \caption{Prediction vs. true value plots for different models. Each model is evaluated using its optimal feature representation, as determined by previous analyses. (a–h) show the predictive performance of MLP, Transformer, SVR, RF, kNN, KPGT, Chemprop, and AGILE, respectively. MLP achieves the highest accuracy among all models, highlighting its strong capacity for capturing complex relationships in the data. Traditional ML models such as SVR and RF perform competitively, while graph models like AGILE and Chemprop exhibit lower predictive accuracy.
}
	\label{fig:models}
\end{figure}

\begin{figure}[!htbp]
	\centering
        \begin{subfigure}[t]{0.24\textwidth}
            \centering
            \includegraphics[width=0.99\textwidth]{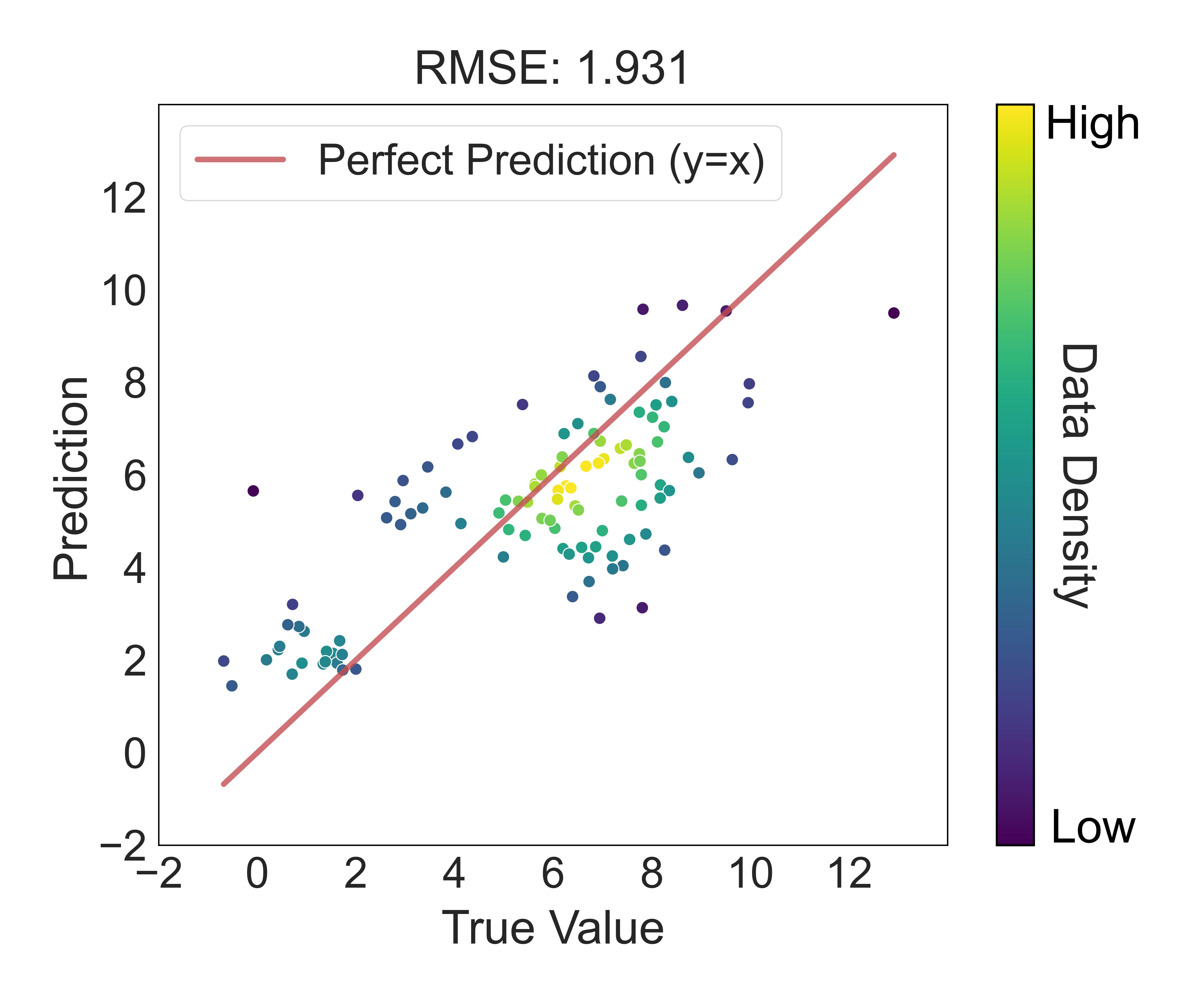}
            \caption{MLP}
        \end{subfigure}
	\begin{subfigure}[t]{0.24\textwidth}
            \centering
            \includegraphics[width=0.99\textwidth]{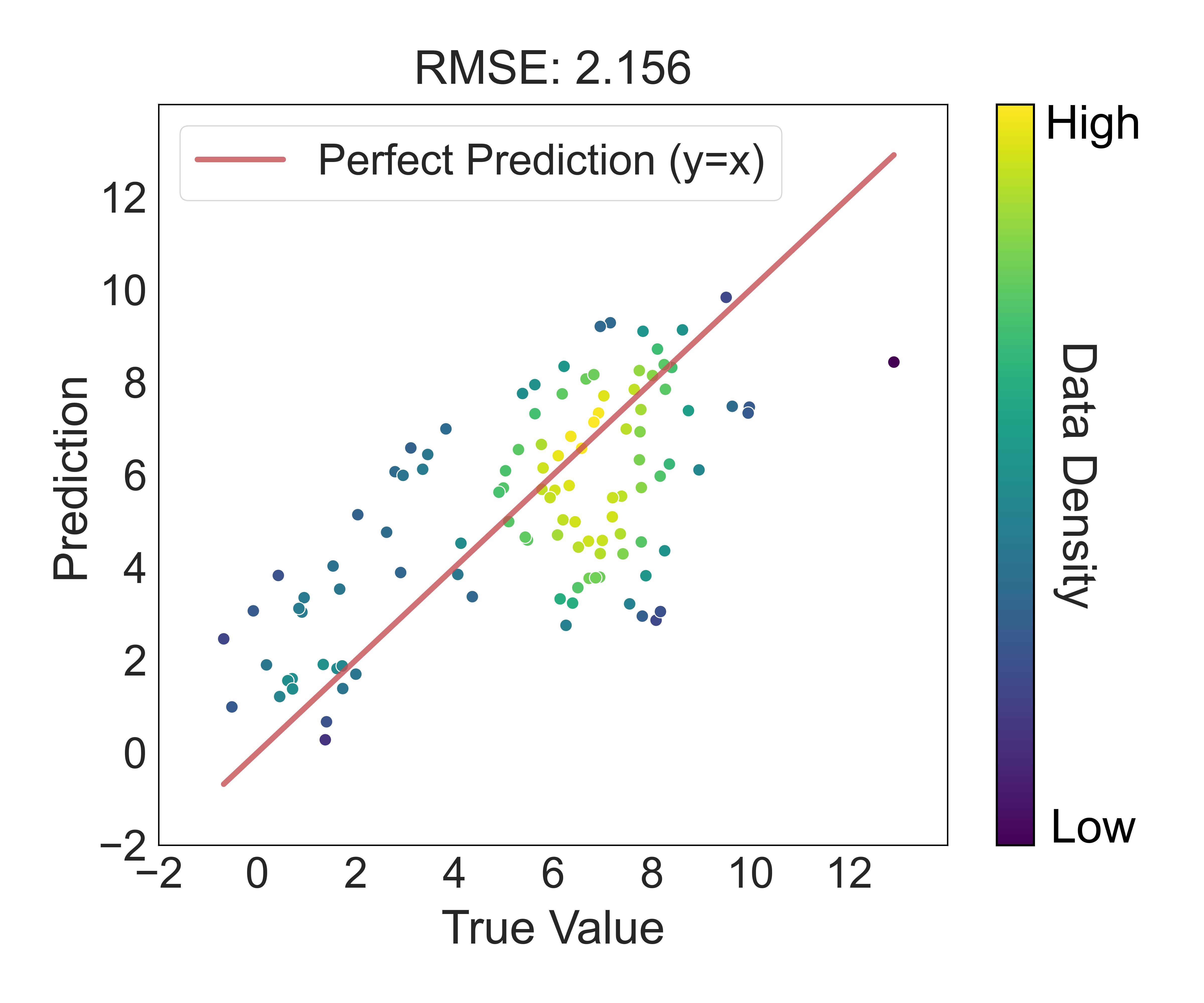}
            \caption{Transformer}
        \end{subfigure}
        \begin{subfigure}[t]{0.24\textwidth}
            \centering
            \includegraphics[width=0.99\textwidth]{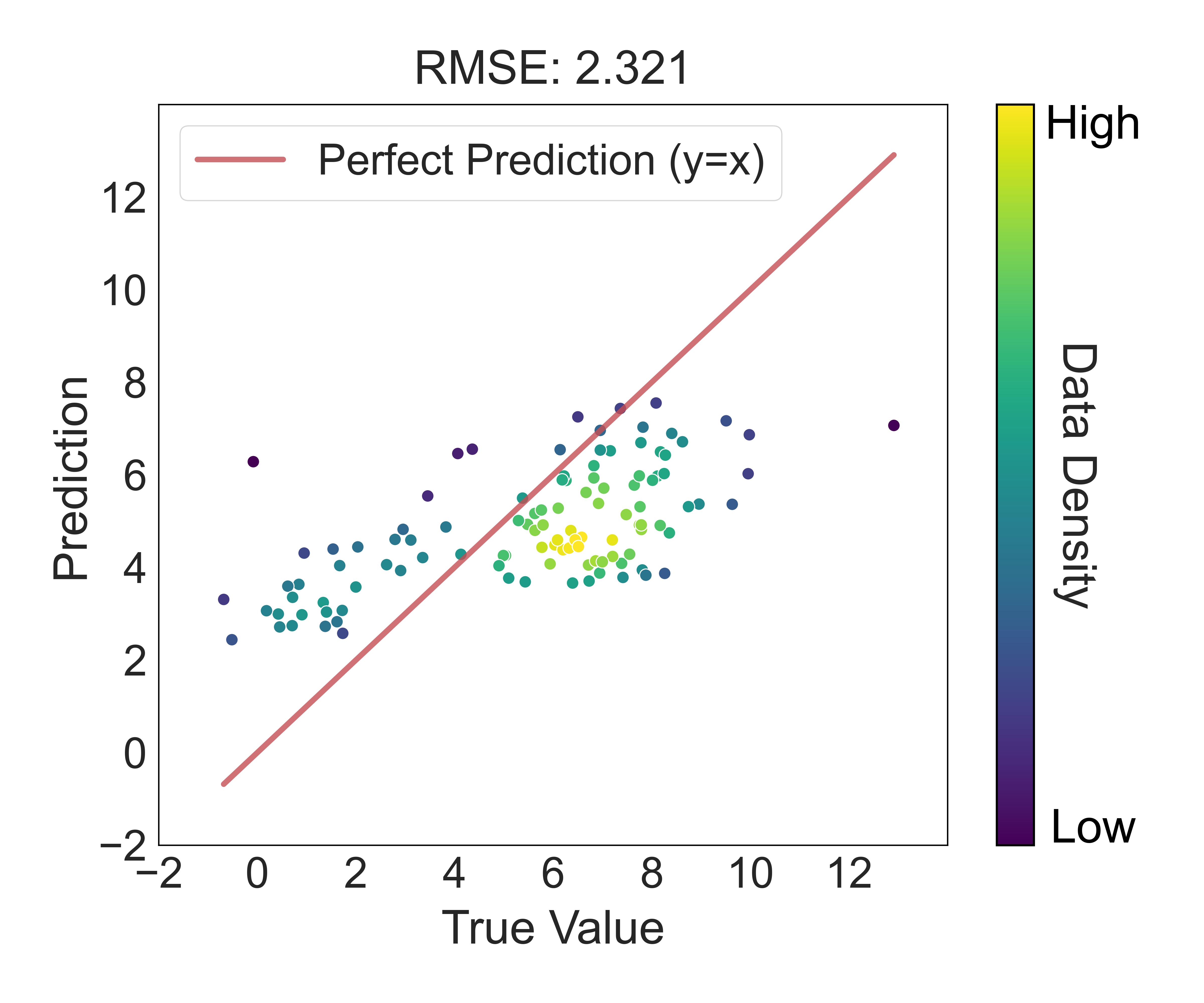}
            \caption{\centering SVR}
        \end{subfigure}
        \begin{subfigure}[t]{0.24\textwidth}
            \centering
            \includegraphics[width=0.99\textwidth]{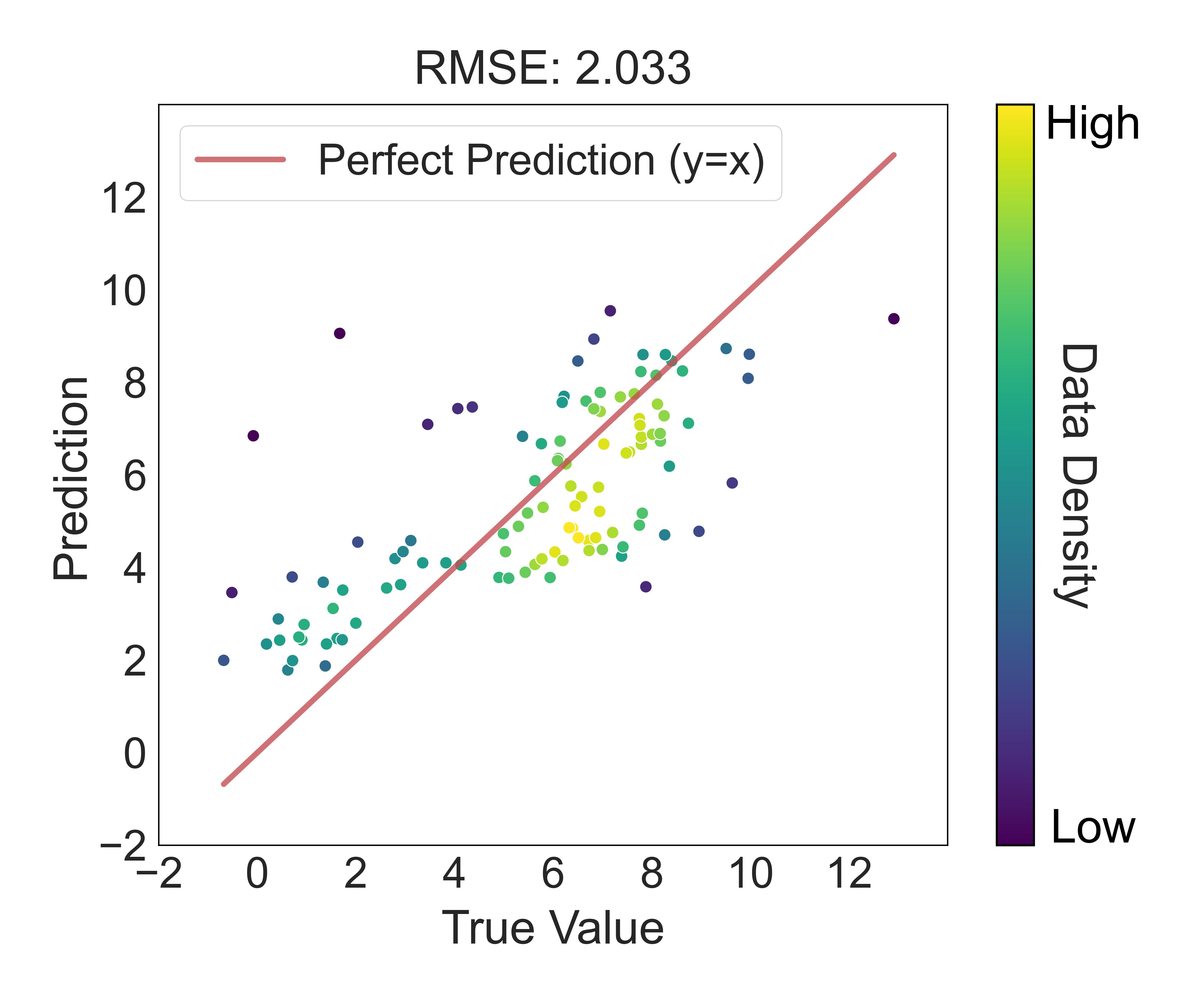}
            \caption{\centering RF}
        \end{subfigure} \\
	\begin{subfigure}[t]{0.24\textwidth}
            \centering
            \includegraphics[width=0.99\textwidth]{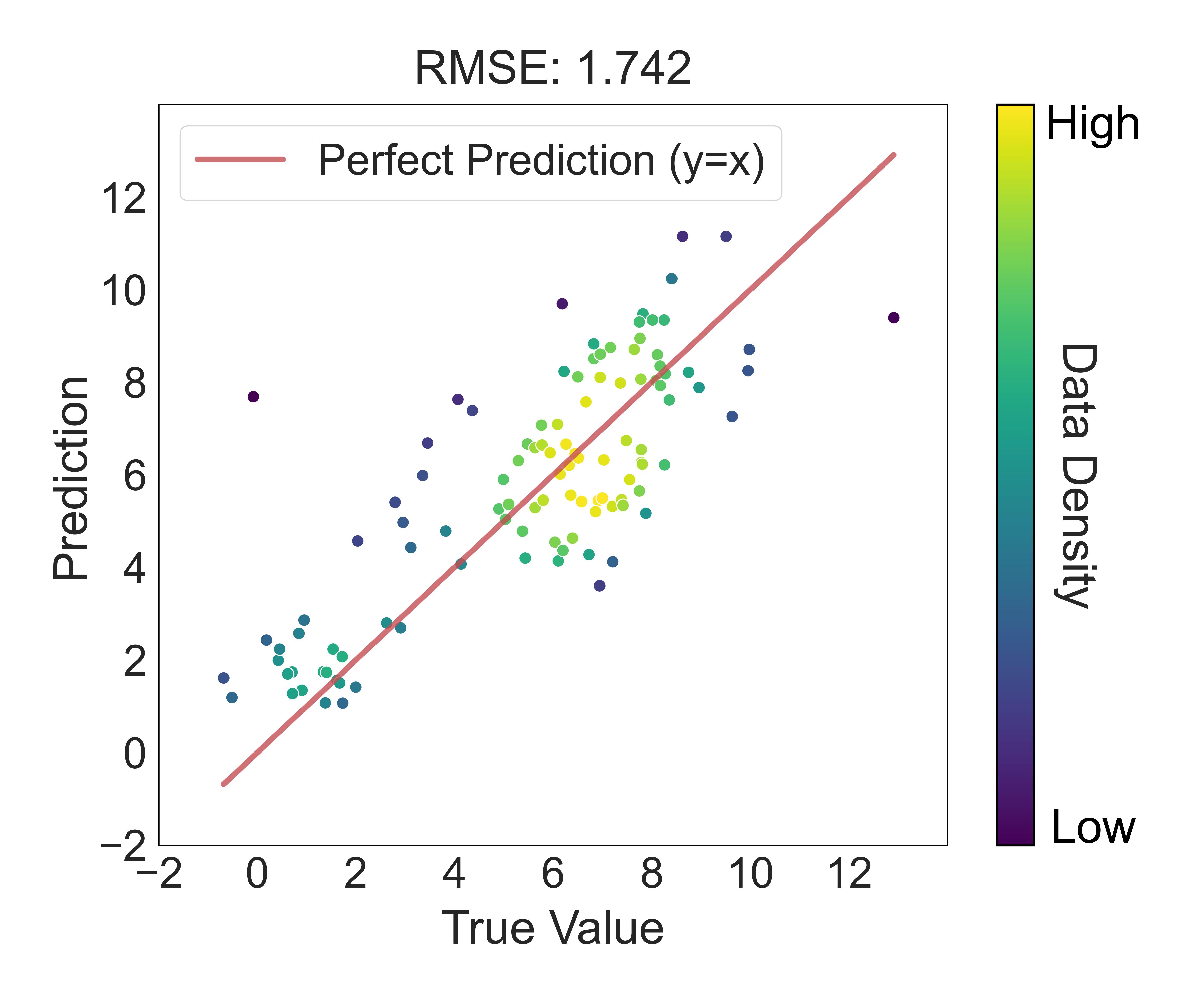}
            \caption{kNN}
        \end{subfigure}
        \begin{subfigure}[t]{0.24\textwidth}
            \centering
            \includegraphics[width=0.99\textwidth]{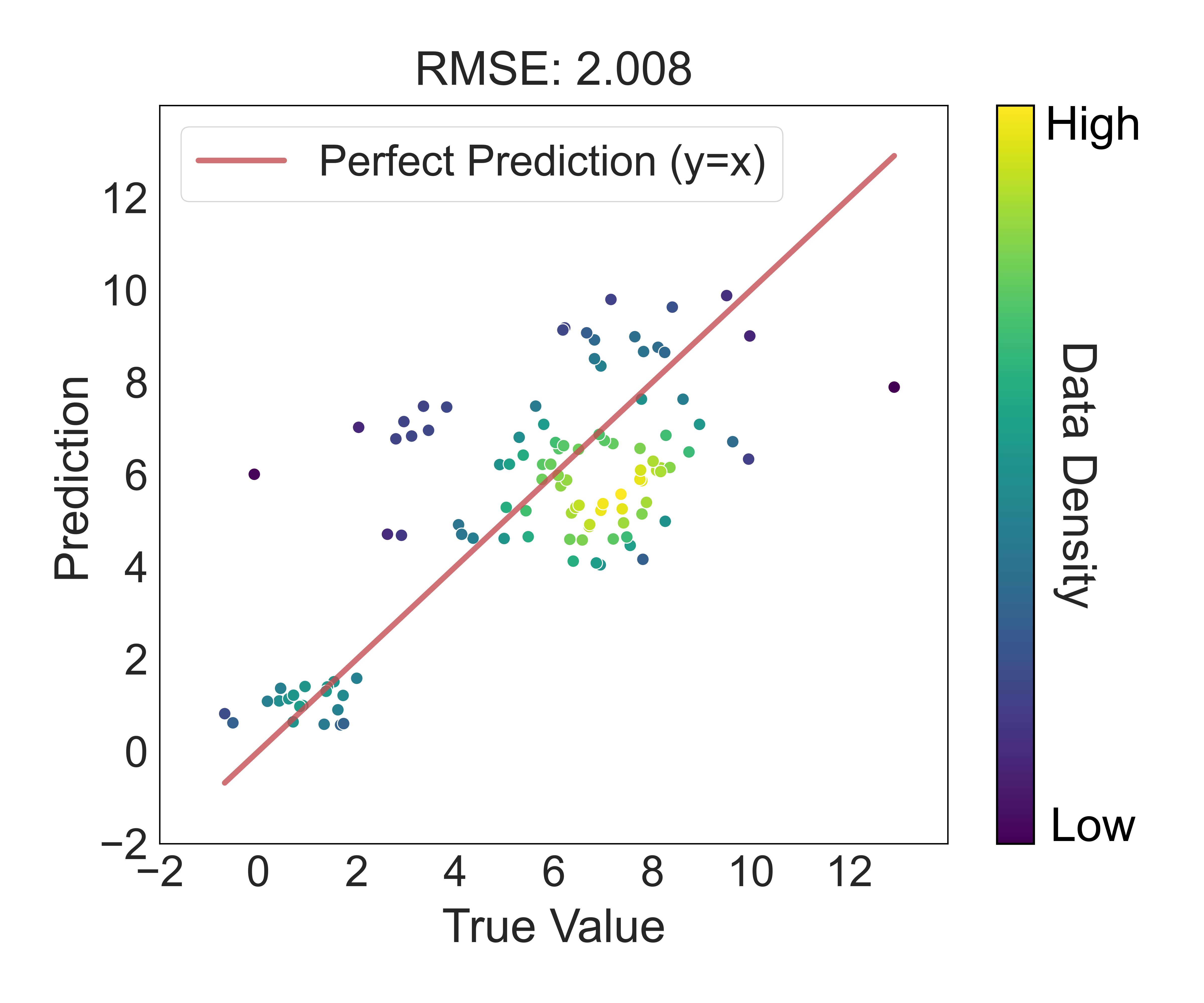}
            \caption{KPGT}
        \end{subfigure}
        \begin{subfigure}[t]{0.24\textwidth}
            \centering
            \includegraphics[width=0.99\textwidth]{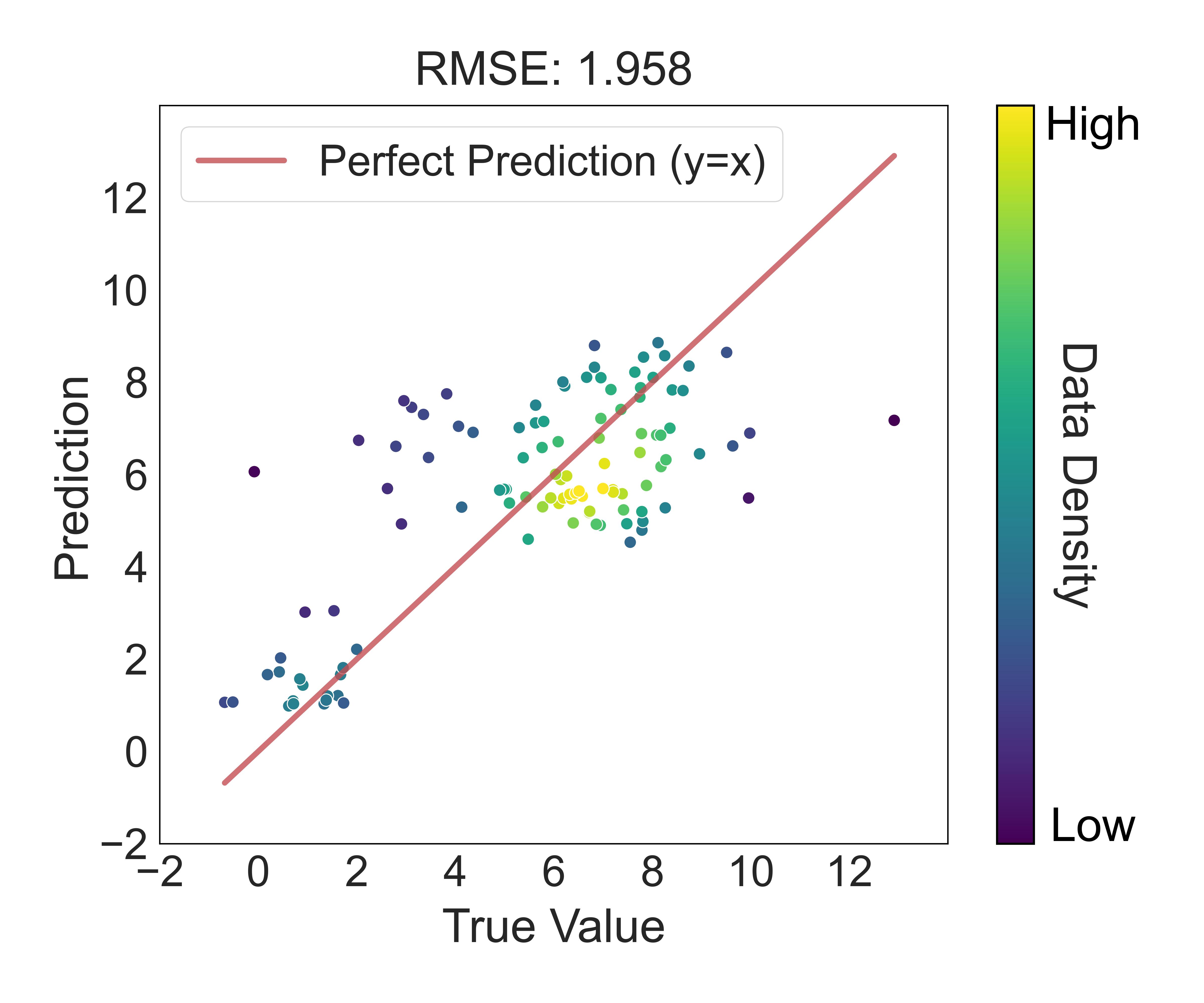}
            \caption{\centering Chemprop}
        \end{subfigure}
        \begin{subfigure}[t]{0.24\textwidth}
            \centering
            \includegraphics[width=0.99\textwidth]{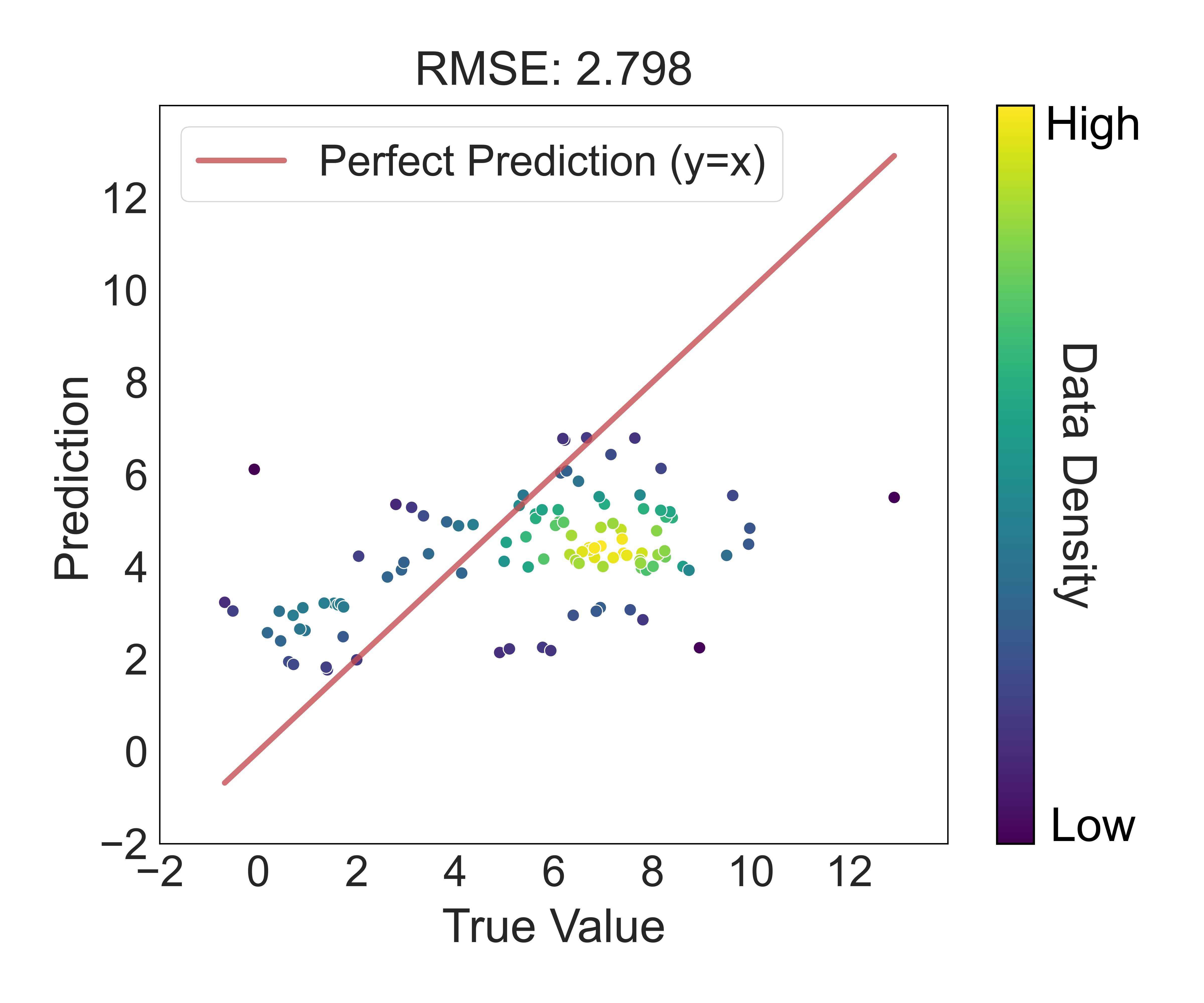}
            \caption{\centering AGILE}
        \end{subfigure}

        \caption{Prediction vs. true value plots for different models using Murcko scaffold split. Each plot presents the best-performing feature set for the corresponding model, selected based on Table~3. (a–h) show the predictive performance of MLP, Transformer, SVR, RF, kNN, KPGT, Chemprop, and AGILE, respectively. Compared to the random split case, most models exhibit a drop in predictive accuracy due to the more stringent generalization requirements of the scaffold-based split. However, kNN and Chemprop maintain similar performance levels, indicating their robustness to distribution shifts, while other models, including MLP and SVR, experience a noticeable decline. AGILE continues to show the lowest accuracy.
}
	\label{fig:models_scaff}
\end{figure}

\end{document}